\documentclass[a4paper,12pt]{article}
\pdfoutput=1
\usepackage{amsfonts,amsmath,amssymb}
\usepackage{graphicx}
\usepackage{color,here}
\usepackage{cite}
\usepackage{hyperref}
\usepackage{mixing_top}
\usepackage{simplewick}
\usepackage{enumitem}
\usepackage{mathrsfs}

\graphicspath{{./paper_figures/}{./npt_ribbon/}{./npt_permutation/}}

\newif\ifusetikz
\usetikzfalse

\ifusetikz
\usepackage{pgf}
\usepackage{xparse}
\usepackage{tikz}
\usetikzlibrary{arrows,arrows.meta,backgrounds,calc,cd,external,graphs,matrix,positioning}
\pgfdeclarelayer{myback}
\pgfsetlayers{myback,background,main}

\fi

\def\chalpha{\check \alpha}
\def\chA{\check A}

\def\hL{L} 

\def\halpha{\hat \alpha}
\def\homega{\omega} 
\def\hmu{\hat\mu}
\def\olcO{\overline{\cal O}}

\numberwithin{equation}{section}

\let\OLDthebibliography\thebibliography
\renewcommand\thebibliography[1]{
  \OLDthebibliography{#1}
  \setlength{\parskip}{2pt}
  \setlength{\itemsep}{2pt plus 2pt}
}

\begin{document}
\thispagestyle{empty}

{\ }
\vspace{-10mm}

\begin{flushright}
{\small \tt KIAS-P18085}
\end{flushright}

\ \vskip 30mm

{\LARGE 
\centerline{\bf Multi-trace Correlators from}

\vspace{3mm}
\centerline{\bf Permutations as Moduli Space}

}

\vskip 25mm

\centerline{
{\large \bf Ryo Suzuki}
}

{\let\thefootnote\relax\footnotetext{{\tt rsuzuki.mp\_at\_gmail.com}}}

\vskip 15mm

\centerline{{\it School of Physics, Korea Institute for Advanced Study},}
\centerline{{\it 85 Hoegiro, Dongdaemun-Gu, Seoul 02455, Korea}}

\vskip 25mm


\centerline{\bf Abstract}

\vskip 6mm

We study the $n$-point functions of scalar multi-trace operators in the $U(N_c)$ gauge theory with adjacent scalars, such as ${\cal N}=4$ super Yang-Mills, at tree-level by using finite group methods.
We derive a set of formulae of the general $n$-point functions, valid for general $n$ and to all orders of $1/N_c$.
In one formula, the sum over Feynman graphs becomes a topological partition function on $\Sigma_{0,n}$ with a discrete gauge group, which resembles closed string interactions.
In another formula, a new skeleton reduction of Feynman graphs generates connected ribbon graphs, which resembles open string interaction.
We define the moduli space ${\cal M}_{g,n}^{\rm gauge}$ from the space of skeleton-reduced graphs in the connected $n$-point function of gauge theory.
This moduli space is a proper subset of ${\cal M}_{g,n}$ stratified by the genus, and its top component gives a simple triangulation of $\Sigma_{g,n}$.

\newpage
\tableofcontents

\section{Introduction}

In the large $N_c$ gauge theories, Riemann surfaces are carved out by the Feynman diagrams in the double-line notation \cite{tHooft:1973alw}.
This observation was a precursor of the AdS/CFT correspondence, that is the conjectured duality between gauge and string theories \cite{Maldacena:1997re}.
The AdS/CFT correspondence can be checked explicitly at any values of the 't Hooft coupling $\lambda = N_c \, g_{\rm YM}^2$ in the integrable setup, such as $\cN=4$ super Yang-Mills (SYM) in four dimensions and superstring on \AdSxS\ \cite{Minahan:2002ve,Bena:2003wd}.

The AdS/CFT correspondence beyond the large $N_c$ limit is not well-understood, including the question of integrability.
It is known that the non-planar integrability, even if exists, does not protect the spectrum, 
because the $1/N_c$ corrections lift the spectral degeneracy of the planar $\cN=4$ SYM \cite{Beisert:2003tq}.
One of the promising ideas in this direction is based on the hexagon form factor \cite{Basso:2015zoa}, which can capture $1/N_c$ corrections \cite{Eden:2017ozn,Bargheer:2017nne,Bargheer:2018jvq}.

Integrability predictions are often formulated and refined with the help of extensive perturbative data.
At weak coupling, the $1/N_c$ corrections can be computed by summing over the graphs at a fixed genus \cite{Beisert:2002bb}.
However, this is tedious, and the finite group method is more efficient.
There one expresses multi-trace operators in terms of permutations, and applies finite-group Fourier-transform to obtain a new basis of gauge-invariant operators labeled by representations or Young tableaux \cite{Corley:2001zk}.
The representation basis diagonalizes tree-level two-point functions at any $N_c$, and solve finite $N_c$ constraints \cite{Brown:2007xh,Bhattacharyya:2008rb}.
The finite group methods are used to compute various quantities, such as partition functions, two-point functions, and recently extremal $n$-point functions \cite{Pasukonis:2013ts}.

We ask two questions in this paper. 
How do we find general non-extremal $n$-point functions in the finite group methods, at any $n$ and to any orders of $1/N_c$\,?
And how do these correlators describe Riemann surfaces?

A similar problem was studied in the Hermitian matrix model, which is a simpler version of large $N_c$ gauge theory \cite{Brezin:1977sv}.
This matrix model describes two-dimensional gravity in the continuum limit, and its exact free energy is given by the $\tau$-function of the KdV hierarchy \cite{Witten:1990hr,Kontsevich:1992ti}.

We will study the tree-level $n$-point functions of scalar operators in $\cN=4$ SYM with $U(N_c)$ gauge group. 
We express Feynman graphs in terms of permutations, and describe the space of Wick-contractions in an algebraic manner.
We obtain a formula which naturally factorizes into the product of pairs of pants, i.e. three-point functions. 
This formula is invariant different pants decomposition, and resembles the interaction of $n$ closed strings.

Then we perform a skeleton reduction to the Feynman graphs in the $n$-point functions.
Under the skeleton reduction, Feynman graphs become connected metric ribbon graphs.
It is known that there is an isomorphism between the space of connected metric ribbon graphs and the decorated moduli space of Riemann surfaces \cite{Jenkins57,Strebel84,Harer86,HarerBook}.
Thus we define the moduli space of Riemann surfaces in gauge theory $\cM_{g,n}^{\rm gauge}$ as the space of Wick-contractions in the skeleton-reduced Feynman graphs.
These graphs resemble the interaction of open strings which triangulates $\Sigma_{g,n}$.

The gauge theory moduli space $\cM_{g,n}^{\rm gauge}$ exhibits two properties.
First, it is a proper subset of the moduli space of the decorated arithmetic Riemann surfaces, equivalent to the connected integral ribbon graphs \cite{MP98,Mulase:2010gw}.
Second, our definition of the skeleton reduction stratifies $\cM_{g,n}^{\rm gauge}$ by genus, meaning that the diagrams with smaller genera contribute to higher powers of $1/N_c$\,.\footnote{This is different from the stratification arising in different situations, such as the Deligne-Mumford compactification of the moduli space \cite{DM69} (see also \cite{DHoker:1988pdl,Witten:1990hr}), hermitian matrix model \cite{Chekhov:1995cq} and hexagonization program \cite{Bargheer:2017nne,Bargheer:2018jvq}.}

In \cite{Gopakumar:2005fx} the correlators of gauge theory are used to define an effective two-dimensional worldsheet theory, by rewriting propagators in the Schwinger parameterization.
This ``CFT'' has been studied in detail, and its unusual properties have been found \cite{Furuuchi:2005qm,Aharony:2006th,David:2006qc,Aharony:2007fs,Aharony:2007rq,Razamat:2008zr,Razamat:2009mc}, including $\cM_{g,n}^{\rm gauge} \subsetneq \cM_{g,n}$ mentioned above.\footnote{Matrix models cover the full moduli space, and describe CFT's in the continuum limit \cite{Charbonnier:2017lom}.}
Another idea is to interpret the correlators of Hermitian matrix model  as the counting of holomorphic maps \cite{Koch:2010zza}. It is argued that the Gaussian matrix model is dual to the A-model topological string on $\bb{P}^1$ \cite{Gopakumar:2011ev,Gopakumar:2012ny,Koch:2014hsa}, similar to the Eguchi-Yang model \cite{Eguchi:1994in}.

Our results may be regarded as a variation of the open/closed duality, exploring the $n$-point functions of a gauge theory at tree-level from various aspects.
Correspondingly, this paper consists of two parts. 
In Sections \ref{sec:graphology} and \ref{sec:PF pants}, closed-string-like formulae of the $n$-point functions are discussed, where one of the formulae has natural pants decomposition.
In Sections \ref{sec:reduction} and \ref{sec:geometry}, open-string-like formulae of the $n$-point functions are discussed, where the skeleton-reduced graphs define the moduli space $\cM_{g,n}^{\rm gauge}$ through triangulation.

\subsection{Outline of ideas}

Our objectives in this paper are to compute $n$-point functions by using permutations, and to define $\cM_{g,n}^{\rm gauge}$ by using permutations.

Let $\cO_i$ be a general multi-trace scalar operator of length $L_i$ in $\cN=4$ SYM,\footnote{More generally multi-trace operators which consist of scalars in the adjoint representation of $U(N_c)$.} whose color structure is specified by the permutation $\alpha_i \in S_{L_i}$ and flavor structure unspecified.
The tree-level $n$-point function is schematically written as the sum over all possible sets of Wick-contractions,
\begin{equation}
G_n \equiv \Vev{ \cO_1 \, \cO_2 \, \dots \, \cO_n }
= \sum_{{\rm Wick} \, \in \, \cW} \ \prod_{i<j}^n {\rm Wick} \( \cO_i , \cO_j \) .
\label{def:Gn Wick sum}
\end{equation}
The space of Wick-contractions $\cW$ decomposes as
\begin{equation}
\cW = \bigsqcup_{\{\ell_{ij} \}} \cW (\{\ell_{ij} \}) 
\end{equation}
where $\ell_{ij}$ is the number of all Wick-contractions between $\cO_i$ and $\cO_j$, also called bridge length. The bridge lengths satisfy
\begin{equation}
\ell_{ii} = 0, \qquad
\ell_{ij} = \ell_{ji} \,, \qquad
L_i = \sum_{j \neq i} \ell_{ij} \,, \qquad
\hL \equiv \sum_{1 \le i < j \le n} l_{ij} = \frac12 \, \sum_{i=1}^n L_i \,.
\label{Wicknum total n-pt}
\end{equation}

The space of Wick-contractions is equivalent to the sum over Feynman diagrams.
In order to study graphs, a permutation is a powerful tool.
Feynman diagram can be interpreted as the Cayley graph generated by a finite group.
The graph data (vertices, edges and faces) are translated into a triple of permutations.
Hence, there should be permutation-based formulae of $G_n$\,.
Once we start looking for such formulae, we encounter the following questions.
\begin{itemize}[nosep,leftmargin=12mm]
\item[i)] How to multiply the elements of different permutation groups?
\item[ii)] How to sum over $\{ \ell_{ij} \}$?
\end{itemize}
Our answer to both is by {\it embedding}, 
\begin{itemize}[nosep,leftmargin=12mm]
\item[i)] to embed $\{ \alpha_i \}$ into a big permutation group to make the group multiplication well-defined.
\item[ii)] to embed $\cW$ into a larger but simpler space by adding unphysical Wick-contractions.
\end{itemize}
These tricks simplify \eqref{def:Gn Wick sum} enormously.

We will obtain three different expressions of $G_n$ in Section \ref{sec:graphology}, which we call vertex-based, edge-based and face-based formulae. 
Let us explain the edge-based formula, which reads
\begin{equation}
G_n = \frac{1}{\prod_{i=1}^n (\hL - L_i)!} \ \frac{1}{\hL!} \ \sum_{ \{ U_i \} \in S_{\hL}^{\otimes n} }
\( \prod_{p=1}^{\hL} \, h^{\hat A^{(1)}_{U_1 (p)} \hat A^{(2)}_{U_2 (p)} \dots \hat A^{(n)}_{U_n (p)} } \) 
N_c^{C \( U_1^{-1} \, \hat \alpha_1 \, U_1 \, 
U_2^{-1} \, \hat \alpha_2 \, U_2 \, 
\dots \, U_n^{-1} \, \hat \alpha_n \, U_n \)} 
\tag{\ref{SL tree n-pt}}
\end{equation}
where $C(\sigma)$ counts the number of cycles in $\sigma$, $\hat \alpha_i$ specifies the color structure of the $i$-th multi-trace operator, and $h^{A_1 A_2 \dots A_n}$ is related to the flavor inner-product.
This formula follows from the idea of extending the operators by adding identity fields,
\begin{equation}
\cO_i \ \sim \ 
(\Phi^{\hat A_1^{(i)}})^{a_1}_{a_{\hat \alpha_i(1)}}
(\Phi^{\hat A_2^{(i)}})^{a_2}_{a_{\hat \alpha_i(2)}}
\dots
(\Phi^{\hat A_{\hL}^{(i)}})^{a_{\hL}}_{a_{\hat \alpha_i(\hL)}} \,,\qquad
( \hat \alpha_i \in S_{L_i} \times {\bf 1}^{\hL -L_i} \subset S_{\hL} )
\end{equation}
where $\Phi^{\hat A_p^{(i)}} = {\bf 1}$ for $L_i < p \le \hL$.

We decompose the formula \eqref{SL tree n-pt} into the product of pairs of pants in Section \ref{sec:PF pants}
\begin{equation}
\begin{gathered}
G_n \sim {\rm Glue} \( \cX [ \cS_1 ] , \cX [ \cS_2 ] , \dots , \cX [ \cS_{n-2} ] \) ,
\qquad
\Sigma_{0,n} = \cS_1 \amalg \cS_2 \amalg \dots \amalg \cS_{n-2} \,.
\end{gathered}
\tag{\ref{correlator as TFT}}
\end{equation}
where $\{ \cS_f \}$ is a local three-point interaction.
The RHS of \eqref{correlator as TFT} depends only on the topology of $\Sigma_{0,n}$. 
We introduce a defect on $\Sigma_{0,n}$ which carries all information on the powers of $N_c$\,.
As a result, no higher-genus surfaces show up.

In Section \ref{sec:reduction}, we study $G_n$ in the opposite way by {\it reduction}, and take the sum over $\{ \ell_{ij} \}$ literally.
This sum is related to the moduli space of Riemann surfaces, which becomes transparent after a skeleton reduction.
The skeleton reduction can be defined by taking the Wick-contractions in two steps.
Let us define a partition of $L_i$ by $\bsl_i \equiv ( \bsl_{i|1} \,, \bsl_{i|2} \,, \dots \,, \bsl_{i|\bL_i} ) \vdash L_i$\,.
We denote a sequence of consecutive $\bsl_{i|r}$ fields in $\cO_i$ by $\bsPhi^{(i)}_r$\,, and define the reduced single-trace operator by
\begin{equation}
\olcO_i = \tr \( \bsPhi^{(i)}_1 \,  \bsPhi^{(i)}_2 \, \dots \, \bsPhi^{(i)}_{\bL_i} \) .
\end{equation}
There are many ways to create $\olcO_i$ from $\cO_i$, which we call partitions of $\cO_i$.
We define the space of the external Wick-contractions by all possible pairings of $\pare{ (i,r),(j,s) }$ in $\bsPhi^{(i)}_r$ and $\bsPhi^{(j)}_s$.
Then we impose the internal Wick-contraction rule between $\bsPhi^{(i)}_r$ and $\bsPhi^{(j)}_s$, which is non-zero only if $\bsl_{i|r} = \bsl_{j|s}$ and expressed as a sum over the internal pairing map $\tau \in S_{\bsl_{i|r}}$.
Since $\tau$ contains non-planar Wick-contractions, the topology of the skeleton-reduced graphs is greatly simplified.

We will obtain two more permutation-based formulae for $G_n$\,, which we call vertex-based and face-based skeleton formulae. 
The face-based skeleton formula for the $n$-point function reads
\begin{equation}
G_n \sim \sum_{ \bsl \, \vdash \hL }
\frac{1}{|{\rm Aut} \, V \! (\bsl)|}
\sum_{\nu \, \in (S_{2 \bL}^{\times \! \times})_{\rm phys} }  \, 
\sum_{\bstau \in S_{\bstau} } \bb{F} ( \bstau |\, \nu ) \ \Big|_{\rm constraints}
\tag{\ref{n-pt Sred}} 
\end{equation}
where $\bsl = (\bsl_1 \,, \bsl_2 \,, \dots \,, \bsl_{\bL}) \vdash L$, and $S_{2 \bL}^{\times \! \times}$ consists of the elements of $S_{2\bL}$ without one- or two-cycles.
The permutation $\nu \in S_{2 \bL}^{\times \! \times}$ defines a skeleton graph.
The function $\bb{F}$ maps a set of skeleton graphs to a polynomial of color and flavor factors, and takes a complicated form
\begin{equation}
\bb{F} ( \bstau |\, \nu ) = 
N_c^{\bZ (\bstau) + C(\bar\omega)} \,
\sum_{\bar\beta \in {\bf CR} (\bar\alpha_\bullet)} \ 
\sum_{ \bsz \in \bb{Z}_\alpha}  \ 
\prod_{s=1}^{\bL} \bsz \cdot 
h \( \bstau \, \Big| \, \bar\mu_{1,s} \ \bar\mu_{1, s'} \ \dots \, \bar\mu_{n,s} \ \bar\mu_{n,s'} \) 
\tag{\ref{gen: construct red F}}
\end{equation}
where $\bar\omega$ is related to $\nu$.
The set $(S_{2 \bL}^{\times \! \times} )_{\rm phys}$ is defined so that it corresponds to a skeleton graph
\begin{equation}
\pare{ \text{Skeleton graphs with $2\bL$ unlabeled vertices} } \quad \leftrightarrow \quad 
\nu \in \( S_{2 \bL}^{\times \! \times} / {\rm Aut} \, V \)_{\rm phys} \,.
\tag{\ref{skeleton graph coset equivalence}}
\end{equation}
From this correspondence, we translate the formula \eqref{n-pt Sred} into a sum over graphs as
\begin{equation}
( G_n )_{\rm connected} = \sum_{g \ge 0} \ 
\sum_{ \Gamma ( \bsl ) \in \cM_{g,n}^{\rm gauge} (\{ L_i \}) } \ 
\frac{1}{|{\rm Aut} \, V (\bsl) |}
\sum_{\bstau \in S_{\bstau} } \mathscr{F} \(\bstau | \Gamma \) 
\tag{\ref{npt reduced formula face}}
\end{equation}
where $\mathscr{F} \(\bstau | \Gamma \)$ is rewriting of $\bb{F} ( \bstau |\, \nu )$ in the language of graphs.

We call $\cM_{g,n}^{\rm gauge} (\{L_i\})$ the moduli space of gauge theory. 
This is a proper subset of the connected metric ribbon graphs, or equivalently of the decorated moduli space of Riemann surfaces. 
\begin{multline}
\cM_{g,n}^{\rm gauge} (\{L_i\}) \subsetneq \cM_{g,n}
\equiv
\Bigl( \text{Choice of } \{ \ell_E > 0 \} \Bigr) \times
\\
\Bigl( \text{Number of $k$-valent vertices} \Bigr) \times
\Bigl( \text{Cyclic ordering of edges at each vertex} \Bigr) .
\label{Mgn gauge intro}
\end{multline}
We give its alternative definition in Section \ref{sec:geometry}, and study examples.

In Section \ref{sec:string} we briefly discuss how metric ribbon graphs appear in string theory.

\subsection{Why many formulae?}

Let us add some words to help the reader understand why many formulae are discussed in this paper.

If one is interested in computing specific $n$-point functions by {\tt Mathematica}, the standard Wick rule is the most efficient operation.
However, the Wick rule does not give an insight into the structure behind the $n$-point functions.
In Section \ref{sec:graphology}, we look for a concise formula, which expresses the sum over the bridge lengths $\{ \ell_{ij} \}$ as a sum over permutations. We introduce extra degrees of freedom to make the formula concise, which makes it practically less efficient for computing concrete examples.
The readers interested in comparing the efficiency can take a look at the attached {\tt Mathematica} files.\footnote{The file {\tt Tree n-pt formula examples.nb} implements the formulae in Section \ref{sec:graphology}, and the file {\tt Tree n-pt formula for graphs.nb} explains the skeleton reduction in Section \ref{sec:reduction}.}

The formulae in Section \ref{sec:reduction} are neither practical nor concise. Our goal is to take the $n$-point functions, and find the geometric structure behind, namely the moduli space. This is a generalization of the work of \cite{Gopakumar:2005fx} when there are multiple scalar fields. 
We notice that the $n$-point function splits into the flavor part and the color part. Since the moduli space carries the information about the color indices, what we need to do is to find the function $\mathscr{F}$ which reconstruct the information about the flavor indices.
It turns out that the techniques developed in Section \ref{sec:graphology} are useful for finding $\mathscr{F}$.

Our results are valid to all orders of the $1/N_c$ expansion, but not at finite $N_c$\,.
We do not take into account the fact that some operators become linearly dependent at finite $N_c$\,.
In order to get the results exact in $N_c$\,, one should use the representation basis of operators, which is a finite-group Fourier transform of the permutation basis, explained in Appendix \ref{app:perm basis}.
The representation basis enables us to reproduce the matrix model results in \cite{Kristjansen:2002bb,Beisert:2002bb}, which will be an important application of our method.\footnote{The author thanks the referee of JHEP for bringing attention to these results.}

\section{Correlators from permutations}\label{sec:graphology}

We express tree-level $n$-point functions by permutations in various ways, and call them vertex-based, edge-based and face-based formulae.
Basic techniques to study Wick-contractions by permutations will be explained below.

We define the permutation basis of scalar multi-trace operators by
\begin{equation}
\cO_i \equiv
\cO^{A_1^{(i)} A_2^{(i)} \dots A_{L_i}^{(i)}}_{\alpha_i} = 
\sum_{a_1, a_2 \dots, a_{L_i} = 1}^{N_c} 
(\Phi^{A_1^{(i)}})^{a_1}_{a_{\alpha_i (1)}}
(\Phi^{A_2^{(i)}})^{a_2}_{a_{\alpha_i (2)}} \dots
(\Phi^{A_{L_i}^{(i)}})^{a_{L_i}}_{a_{\alpha_i (L_i)}} \qquad
(\alpha_i \in S_{L_i} )
\label{def:pm basis}
\end{equation}
and impose the $U(N_c)$ Wick rule \eqref{UNc Wick} between scalar fields.\footnote{Our notation will be explained further in Appendix \ref{app:notation}.}
Then we evaluate the correlator by the repeated application of the pairwise Wick-contractions as
\begin{equation}
G_n = \Vev{ \cO_1 \, \cO_2 \, \dots \, \cO_n }
= \sum_{ \{ \ell_{ij} \} } \prod_{i<j}^n \, \sum_{ W_{ij} } \, 
\prod_{p=1}^{\ell_{ij}} \[ \, g^{A^{(i)}_p A^{(j)}_q } \,
\delta_{b_{\alpha_j (q)}}^{a_p} \,
\delta_{a_{\alpha_i(p)}}^{b_q} \]_{q= W_{ij}^{-1} (p)} 
\label{npt Wick original}
\end{equation}
where $W_{ij}$ runs over the Wick-contractions at fixed bridge lengths. 
We rewrite RHS by permutations.

\subsection{$S_{2 \hL}$ (or vertex-based) formula}\label{sec:S2L formalsim}

Let us label the fields by $P=1,2,\dots, 2\hL$ and forget which index comes from which operator,
\begin{equation}
\Vev{ \cO_1 \, \cO_2 \, \dots \, \cO_n }
= \Vev{ \prod_{i=1}^n \prod_{p=1}^{L_i} (\Phi^{A_p^{(i)}})^{a_p}_{a_{\alpha_i(p)}} (x_i) }
\ \ \to \ \ 
\Vev{ \cO_{\rm all} }
= \Vev{ \prod_{P=1}^{2 \hL} (\Phi^{A_P})^{a_P}_{a_{\alpha(P)}} (x_P) } .
\label{1pt labeling}
\end{equation}
Here $\alpha = \prod_{i=1}^n \alpha_i \in S_{2 \hL}$ is the permutation for all external operators, parameterized by
\begin{equation}
\alpha_k \ \ \leftrightarrow \ \ \text{Permutation of } \ \{ L'_k + 1, L'_k + 2, \dots , L'_{k+1} \}, \qquad
L'_k \equiv \sum_{j=1}^{k-1} L_j \,.
\label{parametrize alpha_k}
\end{equation}
Denote the cycle type of $\alpha$ by $\lambda = [1^{\lambda_1} \, 2^{\lambda_2} \, \dots ] \vdash (2 \hL)$, where $\lambda_K$ is the number of single-trace operator of length $K$ in $\cO_{\rm all}$\,.
Next we introduce a pairing map
\begin{equation}
W_0 \, : \, \pare{ 1, 2, \dots, 2 \hL } 
\ \to \ 
\Bigl\{ (a_1 \, a_2) (a_3 \, a_4) \dots (a_{2\hL - 1} \, a_{2\hL}) \ \Big|\ 
a_P \neq a_Q \ (P \neq Q) \Bigr\}.
\label{def:pairing map}
\end{equation}
This $W_0$ can be regarded as (the top element of )$\bb{Z}_2^{\otimes \hL}$\,,
\begin{equation}
\Bigl\{ \prod_{k=1}^{\hL} \tau_k^{s_k} \, \Big| \,
\tau_k = (a_{2k-1} \,, a_{2k}), \ s_k = \{0,1\} \Bigr\} 
\ \simeq \ \bb{Z}_2^{\otimes \hL} 
\end{equation}
and we write $a_{2k-1} = W_0(a_{2k})$ with $W_0^2=1$. 
By using \eqref{perm identities}, we permute $W_0$ by the adjoint action of $\gamma \in S_{2\hL}$ to create all other choices of the subgroup $\bb{Z}_2^{\otimes \hL} \subset S_{2\hL}$ as\footnote{In the literature on graph theory and combinatorics, the adjoint action of $\gamma$ on edges is called an edge permutation.}
\begin{equation}
W = \gamma \, W_0 \, \gamma^{-1} = \prod_{k=1}^{\hL} \( \gamma(a_{2k-1}) \, \gamma(a_{2k}) \) ,
\qquad (\gamma \in S_{2 \hL}).
\label{def:gWginv}
\end{equation}

\begin{figure}[t]
\begin{center}
\includegraphics[scale=1.3]{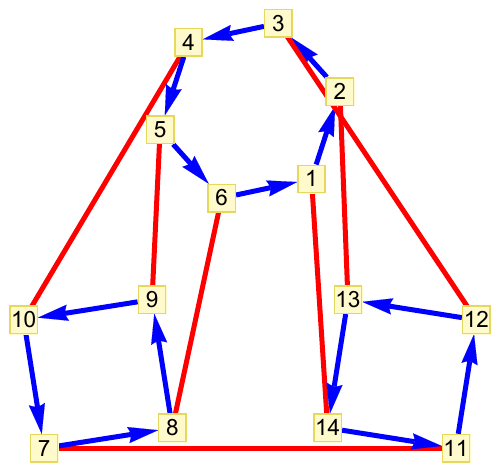}
\hspace{10mm}
\includegraphics[scale=1.3]{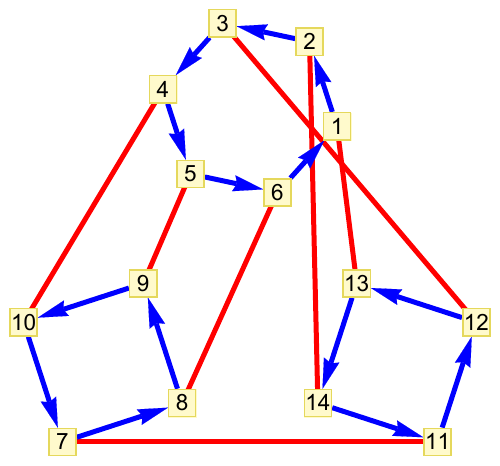}
\caption{The Cayley graphs generated by $(\alpha, W)$ for $\Vev{\cO_1\cO_2\cO_3}$ with $(L_1,L_2,L_3)=(6,4,4)$. The left figure corresponds to a planar term, and the right figure to a non-planar term.}
\label{fig:Cayley644}
\end{center}
\end{figure}

The above definition of $W$ is redundant, because it remains invariant under some relabeling of $\{a_i\}$. This redundancy is generated by the wreath-product group $S_{\hL} [\bb{Z}_2]$ as explained in Appendix \ref{app:notation}.
In contrast, our parameterization of $\cO_{\rm all}$ using $\alpha \in S_{2\hL}$ has no redundancy, because we fix the order of the flavor indices $(A_1 \,, A_2 \,, \dots \,, A_{2\hL})$.
We can identify Feynman graphs with the Cayley graph generated by $\alpha \in S_{2\hL}$ and $W \in \bb{Z}_2^{\otimes \hL}$\,, as shown in Figure \ref{fig:Cayley644}.\footnote{Strictly speaking, in Cayley graphs different colors should be used for different generators of the finite group. Our graphs are Cayley-like, because we use only two colors.}

In the above notation, the original $n$-point formula \eqref{npt Wick original} becomes
\begin{equation}
G_n = \sum_{ \{ \ell_{ij} \} } \sum_{W} 
\prod_{P=1}^{2 \hL} \[  g^{A_P A_{W(P)} } 
\delta_{a_{\alpha W (P)}}^{a_P} \,
\delta_{a_{\alpha (P)}}^{a_{W (P)}} \]^{1/2} 
\end{equation}
where we take the square root of $2 \hL$ factors, and used $W^2=1$. The $\delta$-functions can be simplified as
\begin{equation}
\prod_{P=1}^{2 \hL} \delta_{a_{\alpha W (P)}}^{a_P} \,
\prod_{Q=1}^{2 \hL} \delta_{a_{\alpha (Q)}}^{a_{W (Q)}} 
= N_c^{C(\alpha W) + C(\alpha W^{-1})}
= N_c^{2 \, C(\alpha W)} 
\end{equation}
where $C(\omega)$ is defined in \eqref{def:cycle type}. It follows that
\begin{equation}
G_n = \sum_{ \{ \ell_{ij} \} } \sum_{W} 
\( \prod_{P=1}^{2 \hL} \[  g^{A_P A_{W(P)} } \]^{1/2} \)
N_c^{C(\alpha W)} \,.
\label{npt Wick 2hL}
\end{equation}
This expression looks like a {\it one}-point formula on the permutation basis for any $n$.

Let $\cW (\{ \ell_{ij} \})$ be the space of all pairing maps $W$ in \eqref{npt Wick 2hL}, which satisfies the conditions 
\begin{itemize}[nosep,leftmargin=16mm]
\item[i)] there are $\ell_{ij}$ contractions between $\cO_i$ and $\cO_j$\,.
\item[ii)] there are no self-contractions; $\ell_{ii}=0$ \,.
\end{itemize}
Suppose $W_0$ is one such pairing, for example
\begin{equation}
W_0 \equiv W_0 (\{ \ell_{ij} \}) = \prod_{i<j} \prod_{p=1}^{\ell_{ij}} (L_{i,j}' + p, L_{j,i}'+p), \qquad
L'_{i,j} = \sum_{k=1}^{i-1} L_k + \sum_{m=1}^{j-1} \ell_{im} \,.
\label{explicit W0}
\end{equation}
We can show that all other pairings are generated by the group action
\begin{equation}
\cW (\{ \ell_{ij} \}) = \pare{ \gamma \, W_0 \, \gamma^{-1} \, \Big| \, \gamma \in \prod_{i=1}^n S_{L_i} }
\Big/ S_{\hL} [\bb{Z}_2] .
\label{def:cW ell}
\end{equation}
To see the equivalence, first we prove that RHS does not change $\{ \ell_{ij} \}$. 
Generally $W$ takes the form \eqref{def:gWginv} under the constraint $\gamma \in \prod_{i=1}^n S_{L_i}$\,.
Since any elements of $S_L$ is a product of transpositions, we consider $\gamma = (aa')$. The case of general $\gamma$ can be studied in a similar way. 
Because $\gamma \in \prod_{i=1}^n S_{L_i}$\,, both $a$ and $a'$ must come from the same operator, say $\cO_a$\,. Now
\begin{equation}
\gamma = (aa') \, : \, W_0 = (a c)(a' d) \ \mapsto \ \gamma W_0 \gamma^{-1} = (a'c)(ad), \qquad 
\( a, a' \in \cO_a \,, b \in \cO_b \,, c \in \cO_c \) 
\end{equation}
which does not change any of $(\ell_{ab} \,, \ell_{bc} \,, \ell_{ca})$.
Secondly, we argue that any $\gamma$ which preserves $\{ \ell_{ij} \}$ should belong to $\prod_{i=1}^n S_{L_i}$\,. To see this, suppose that $\cO_i$ and $\cO_j$ are paired by $W_0$ and $W$ as
\begin{equation}
W_0 \ni \prod_{p=1}^{\ell_{ij}} (x_p \, y_p), \qquad
W \ni \prod_{p=1}^{\ell_{ij}} (x'_p \, y'_p).
\end{equation}
The two permutations are related by $W \sim \gamma_{ij} \, W_0 \, \gamma_{ij}^{-1}$, where
\begin{equation}
\gamma_{ij} = \gamma_i \gamma_j \in S_{L_i} \times S_{L_j} \,, \qquad
x'_p = \gamma_i (x_p), \qquad y'_p = \gamma_j (y_p).
\end{equation}
By repeating this argument to all $(i,j)$, one finds that $W = \gamma \, W_0 \, \gamma^{-1}$ with $\gamma \in \prod_{i=1}^n S_{L_i}$\,.

Since $S_{\hL} [\bb{Z}_2]$ is not part of the group $\prod_{i=1}^n S_{L_i}$\,, the space $\cW (\{ \ell_{ij} \})$ in \eqref{def:cW ell} is not a group coset.
For this reason, it is difficult to compute the sum over $W$ in \eqref{npt Wick 2hL}.

\bigskip
In order to simplify the above formula further, we introduce the new inner-product,
\begin{equation}
h^{AB} = \begin{cases}
0 &\qquad (\text{self-contraction within} \ \cO_i) \\
g^{AB} &\qquad (\text{otherwise}).
\end{cases}
\label{def:hAB}
\end{equation}
We extend the space of all pairing maps $W$ as
\begin{equation}
\bigsqcup_{i<j} \cW (\{ \ell_{ij} \}) \ \to \ 
\cW = \pare{ \gamma \, W_0 \, \gamma^{-1} \, \Big| \, \gamma \in S_{2 \hL} / S_{\hL} [\bb{Z}_2] }
\label{def:moduli cW}
\end{equation}
where $W_0 \in \bb{Z}_2^{\otimes \hL}$ is any set of $\hL$ pairs out of $2\hL$ numbers without overlap like \eqref{def:pairing map}.
Now the space $\cW$ becomes a group coset.
It follows that
\begin{equation}
\begin{aligned}
G_n &= \sum_{\gamma \in S_{2 \hL} / S_{\hL} [\bb{Z}_2]} 
\( \prod_{P=1}^{2 \hL} \[  h^{A_P A_{\gamma W_0 \gamma^{-1} (P)} } \]^{1/2} \)
N_c^{C(\alpha \gamma W_0 \gamma^{-1})} 
\\
&= \frac{1}{|{\rm Aut} \, W_0 | } \, \sum_{\gamma \in S_{2 \hL}} 
\( \prod_{P=1}^{2 \hL} \[  h^{A_P A_{\gamma W_0 \gamma^{-1} (P)} } \]^{1/2} \)
N_c^{C(\alpha \gamma W_0 \gamma^{-1})} 
\end{aligned}
\label{npt Wick 2hL extended}
\end{equation}
where $|{\rm Aut} \, W_0 | = | S_{\hL} [\bb{Z}_2] | = L! \, 2^L$.
The sum over $\{ \ell_{ij} \}$ has been successfully removed.


\subsection{$S_{\hL}$ (or edge-based) formula}\label{sec:SL}

In the previous section, we labeled fields $\Phi^{A_P}$ by $P=1,2,\dots, 2\hL$ and derived formulae as a sum over $S_{2\hL}$\,. Here we label Wick-contractions by $1,2, \dots, \hL$, and give another formula in terms of $S_{\hL}$\,.

First, we extend the operator $\cO_i$ by adding identity fields,
\begin{equation}
\hat \cO_i \equiv \cO_{\alpha_i} \times \tr ({\bf 1})^{\hL - L_i} 
\equiv \prod_{p=1}^{\hL}  (\Phi^{\hat A_p^{(i)}})^{a_p}_{a_{\hat \alpha_i(p)}} \,, \qquad
\hat \alpha_i \in S_{\hL} \,.
\label{def:Ohat}
\end{equation}
When $\Phi^{\hat A_p^{(i)}} = {\bf 1}$, the permutation $\hat \alpha_i$ acts as the identity on $p$.

Second, we introduce $n$-tuple Wick-contraction by
\begin{equation}
\contraction{(}{\Phi^{\hat A_1}}{)_{a_1}^{b_1} \, (}{\Phi^{\hat A_2}}
\contraction{(\Phi^{\hat A_1})_{a_1}^{b_1} \, (}{\Phi^{\hat A_2}}{)_{a_2}^{b_2} \, (}{\Phi^{\hat A_3}}
\contraction{(\Phi^{\hat A_1})_{a_1}^{b_1} \, (\Phi^{\hat A_2})_{a_2}^{b_2} \, (}{\Phi^{\hat A_3}}{)_{a_3}^{b_3} \ \dots \ (}{\Phi^{\hat A_n}}
(\Phi^{\hat A_1})_{a_1}^{b_1} \, (\Phi^{\hat A_2})_{a_2}^{b_2} \, (\Phi^{\hat A_3})_{a_3}^{b_3} \ \dots \ (\Phi^{\hat A_n})_{a_n}^{b_n}
= h^{\hat A_1 \hat A_2 \dots \hat A_n} \ 
\delta^{b_2}_{a_1} \, \delta^{b_3}_{a_2} \ \dots \ \delta^{b_n}_{a_1} \,.
\label{UNc Wick n-tuple}
\end{equation}
We demand that $h^{\hat A_1 \hat A_2 \dots \hat A_n}$ is equal to the two-point inner-product $g^{AB}$, if $(n-2)$ of the flavor indices $\{ \hat A_1 \,, \hat A_2 \,, \dots \,, \hat A_n \}$ are the identity field ${\bf 1}$, and otherwise $h^{\hat A_1 \hat A_2 \dots \hat A_n} = 0$.
For example, triple contraction is given by,\footnote{We use the same symbol $h^{AB}$ as in \eqref{def:hAB}, because both $h^{AB}$'s remove the self-contractions.}
\begin{equation}
h^{ABC} = 
h^{AB} \, \delta^C_{\bf 1} 
+ h^{BC} \, \delta^A_{\bf 1} + h^{CA} \, \delta^B_{\bf 1} \,, \qquad
h^{AB} = \begin{cases}
g^{AB} &\quad (A \neq {\bf 1}, B \neq {\bf 1}) \\
0 &\quad ({\rm otherwise})
\end{cases}
\label{def:hABC}
\end{equation}
and in general
\begin{equation}
h^{A_1 A_2 \dots A_n} = \frac{1}{2 (n-2)!} \, \sum_{\pi \in S_n} h^{A_{\pi(1)} A_{\pi(2)}} \prod_{j=3}^n \delta^{A_{\pi(j)}}_{\bf 1} \,.
\label{def:hA1An}
\end{equation}
With this definition of $h^{A_1 A_2 \dots A_n}$, one can see that the $n$-tuple Wick-contraction \eqref{UNc Wick n-tuple} is equivalent to the pairwise Wick-contraction (see e.g. \eqref{npt Wick original}), by recalling that $\hat \alpha_i (p) = p$ if $\Phi^{\hat A_p^{(i)}} = {\bf 1}$.

Third, we evaluate the $n$-point functions of $\hat \cO_i$ by taking all possible $n$-tuple Wick-contractions. 
One $n$-tuple Wick-contraction from $\vev{ \prod_i \hat \cO_i }$ can be specified by
\begin{equation}
\( W_{12} \,, W_{23} \,, \dots \,, W_{n1} \) \in S_{\hL}^{\otimes n}, \qquad
W_{12} \, W_{23} \, \dots \, W_{n1} = 1
\end{equation}
where $W_{ij} \in S_{\hL}$ tells which fields of $\hat \cO_i$ and $\hat \cO_j$ are paired together inside the $n$-tuple contraction. 
These permutations $\{ W_{ij} \}$ satisfy $W_{ij} W_{jk} = W_{ik}$\,, and have the trivial monodromy $W_{ii}=1$.
For example, three-point functions are given by
\begin{align}
\Vev{ \prod_{i=1}^3 \hat \cO_i { (x_i) } }
&\equiv \sum_{W_{12} \,, W_{23} \,, W_{31}} 
\prod_{p,q,r=1}^{\hL} \, 
h^{\hat A_p \hat B_q \hat C_r} \, 
\delta^{b_q}_{a_{\hat \alpha_1(p)}} \, 
\delta^{c_r}_{b_{\hat \alpha_2 (q)}} \, 
\delta^{a_p}_{c_{\hat \alpha_3 (r)}} \ \Big|_{p=W_{12} (q), \ q = W_{23} (r), \ r = W_{31} (p)} 
\notag \\
&= \sum_{W_{12} \,, W_{23} \,, W_{31}} 
\hspace{-7mm} {}' \hspace{7mm}
\prod_{p}^{\hL} \, 
h^{\hat A_p \hat B_{W_{12}^{-1} (p)} \hat C_{W_{31} (p)}} \, 
\delta^{a_p}_{a_{\hat \alpha_1 W_{12} \hat \alpha_2 W_{23} \hat \alpha_3 W_{31} (p)}} 
\notag \\
&= \sum_{W_{12} \,, W_{23} \,, W_{31}}
\hspace{-7mm} {}' \hspace{7mm}
\( \prod_{p}^{\hL} \, h^{\hat A_p \hat B_{W_{12}^{-1} (p)} \hat C_{W_{31} (p)}} \)
\delta_{\hL} \( \Omega \, \hat \alpha_1 \, W_{12} \, \hat \alpha_2 \, W_{23} \, \hat \alpha_3 \, W_{31} \) 
\label{SL 3pt comp}
\end{align}
where $\delta_L$ and $\Omega$ are defined in \eqref{def:deltaL} and \eqref{def:Omega}, respectively.
The symbol $\sum'$ means that the sums over $W_{12} \,, W_{23} \,, W_{31}$ are constrained by
\begin{equation}
W_{12} \, W_{23} \, W_{31} = 1 \ \in \ S_{\hL} \,.
\label{monodromy constraint}
\end{equation}
We can solve the monodromy constraint formally by introducing
\begin{equation}
W_{ij} = U_i \, U_j^{-1} \,, \qquad U_k \in S_{\hL} \,.
\label{def:Wij to Ui}
\end{equation}
Recall that $W_{ij}$ is a pairing function between $\cO_i$ and $\cO_j$\,. The new permutation $U_i$ acts only on $\hat \alpha_i$\,.
The equation \eqref{SL 3pt comp} becomes
\begin{equation}
\Vev{ \prod_{i=1}^3 \hat \cO_i { (x_i) } }
= \frac{1}{|S_L|} \, \sum_{U_1 \,, U_2 \,, U_3 \in S_{\hL}}
\( \prod_{q=1}^{\hL} \, h^{\hat A_{U_1(q)} \hat B_{U_2 (q)} \hat C_{U_3 (q)}} \)
\delta_{\hL} \( \Omega \, U_1^{-1} \, \hat \alpha_1 \, U_1 U_2^{-1} \, \hat \alpha_2 \, U_2 \, U_3^{-1} \, \hat \alpha_3 \, U_3 \) .
\label{SL 3pt comp2}
\end{equation}
We introduced the factor $|S_L| = L!$ because the simultaneous transformation $U_k \ \to U_k \, \gamma$ with $\forall \gamma \in S_{\hL}$ does not change $W_{ij}$ in \eqref{def:Wij to Ui}.

The formula for general $n$-point functions of $\hat \cO_i$ is
\begin{equation}
\begin{gathered}
\Vev{ \prod_{i=1}^n \hat \cO_i { (x_i) } }
= \frac{1}{\hL!} \ \sum_{\{U_i\} \in S_L^{\otimes n} }
\( \prod_{p=1}^{\hL} \, h^{\chA^{(1)}_{p} \chA^{(2)}_{p} \dots \chA^{(n)}_{p} } \) 
\delta_{\hL} \( \Omega \, 
\chalpha_1 \, \chalpha_2 \, \dots \, \chalpha_n \)
\\[1mm]
\chA^{(k)}_p \equiv \hat A^{(k)}_{U_k(p)} \,,\qquad
\check \alpha_k \equiv U_k^{-1} \, \hat \alpha_k \, U_k \,.
\end{gathered}
\label{SL tree n-pt} 
\end{equation}
This quantity is proportional to the original correlator of $\{ \cO_i \}$. 
The proportionality constant comes from the permutations of the identity fields in \eqref{def:Ohat}. 
Since $\hat \cO_i$ has $(\hL - L_i)$ identity fields, we find
\begin{equation}
\Vev{ \prod_{i=1}^n \cO_i (x_i) } = \frac{1}{\ds \prod_{i=1}^n (\hL - L_i)!} \ 
\Vev{ \prod_{i=1}^n \hat \cO_i { (x_i) } } .
\label{normalize hat corr}
\end{equation}

\subsubsection{Symmetry of the $S_{\hL}$ formula}\label{sec:symmetry SL}

We summarize the symmetry of the formula \eqref{SL tree n-pt}.

First, the $n$-point formula is invariant under the simultaneous redefinition $\{ U'_i \} = \{ U_i \hat \gamma \}$. This corresponds to relabeling the Wick-contractions $p \to \hat \gamma (p)$ with $\hat \gamma \in S_{\hL}$\,. We can use this redundancy to set e.g. $U_1=1$.

Second, the extended operator \eqref{def:Ohat} has the gauge symmetry
\begin{equation}
\cO^{\hat A_1 \hat A_2 \dots \hat A_{\hL}}_{\hat \alpha_i} 
= \cO^{\hat A_{\hat \gamma(1)} \hat A_{\hat \gamma(2)} \dots \hat A_{\hat \gamma(\hL)}}_{\hat \gamma^{-1} \hat \alpha_i \hat \gamma} \,, \qquad
\hat \gamma \in S_{L_i} \times S_{\hL - L_i} \subset S_{\hL} 
\label{def:hat gauge symm}
\end{equation}
generalizing \eqref{def:gauge symm}. 
Our $n$-point formula is also invariant under the gauge transformations \eqref{def:hat gauge symm} for each fixed $i$, because the change $( \hat \alpha_i \,, \hat A^{(i)}_p ) \to ( \gamma_i^{-1} \hat \alpha_i \gamma_i \,, \hat A^{(i)}_{\gamma_i (p)} )$ can be absorbed by the redefinition $U'_i = \gamma_i \, U_i$\,.

Third, the $n$-pt formula is invariant under any $S_n$ permutations,
\begin{equation}
\vev{ \cO_1 \cO_2 \dots \cO_n } = \vev{ \cO_{\pi(1)} \cO_{\pi(2)} \dots \cO_{\pi(n)} }, \qquad
\forall \pi \in S_n \,.
\label{npt perm symm}
\end{equation}
This follows from $S_n$ invariance of the $n$-tuple metric $h^{A_1 A_2 \dots A_n}$ in \eqref{def:hA1An}, and the redefinition
\begin{equation}
U_k^{-1} \, \hat \alpha_k \, U_k \, U_{k+1}^{-1} \, \hat \alpha_{k+1} \, U_{k+1}
=
V_{k+1}^{-1} \, \hat \alpha_{k+1} \, V_{k+1} \, U_k^{-1} \, \hat \alpha_k \, U_k \,, \qquad
\( V_{k+1}^{-1} = U_k^{-1} \, \hat \alpha_k \, U_k \, U_{k+1}^{-1} \).
\end{equation}
In other words, two conjugacy classes commute.
Recall that we introduced an artificial cyclic ordering $(12 \dots n)$ in \eqref{UNc Wick n-tuple}, apparently breaking $S_n$ to $\bb{Z}_n$. We could use the $S_n$\,-invariant Wick-contraction rule,
\begin{equation}
\delta^{j_2}_{i_1} \, \delta^{j_3}_{i_2} \ \dots \ \delta^{j_n}_{i_1}
\quad \to \quad 
\sum_{\pi \in {\rm Der}_n}
\frac{1}{|{\rm Der}_n|} \, \delta^{j_1}_{\pi(i_1)} \, \delta^{j_2}_{\pi(i_2)} \ \dots \ \delta^{j_n}_{\pi(i_n)} \,,
\label{UNc Wick n-tuple2}
\end{equation}
where ${\rm Der}_n$ is the derangement of $n$ elements (permutations without fixed points). 
The two definitions give the same result, because $h^{A_1 A_2 \dots A_n}$ is locally a two-point inner-product, and the order of the color indices does not matter.

\subsubsection{Cayley graph}\label{sec:SL Cayley}

\begin{figure}[t]
\begin{center}
\includegraphics[scale=.7]{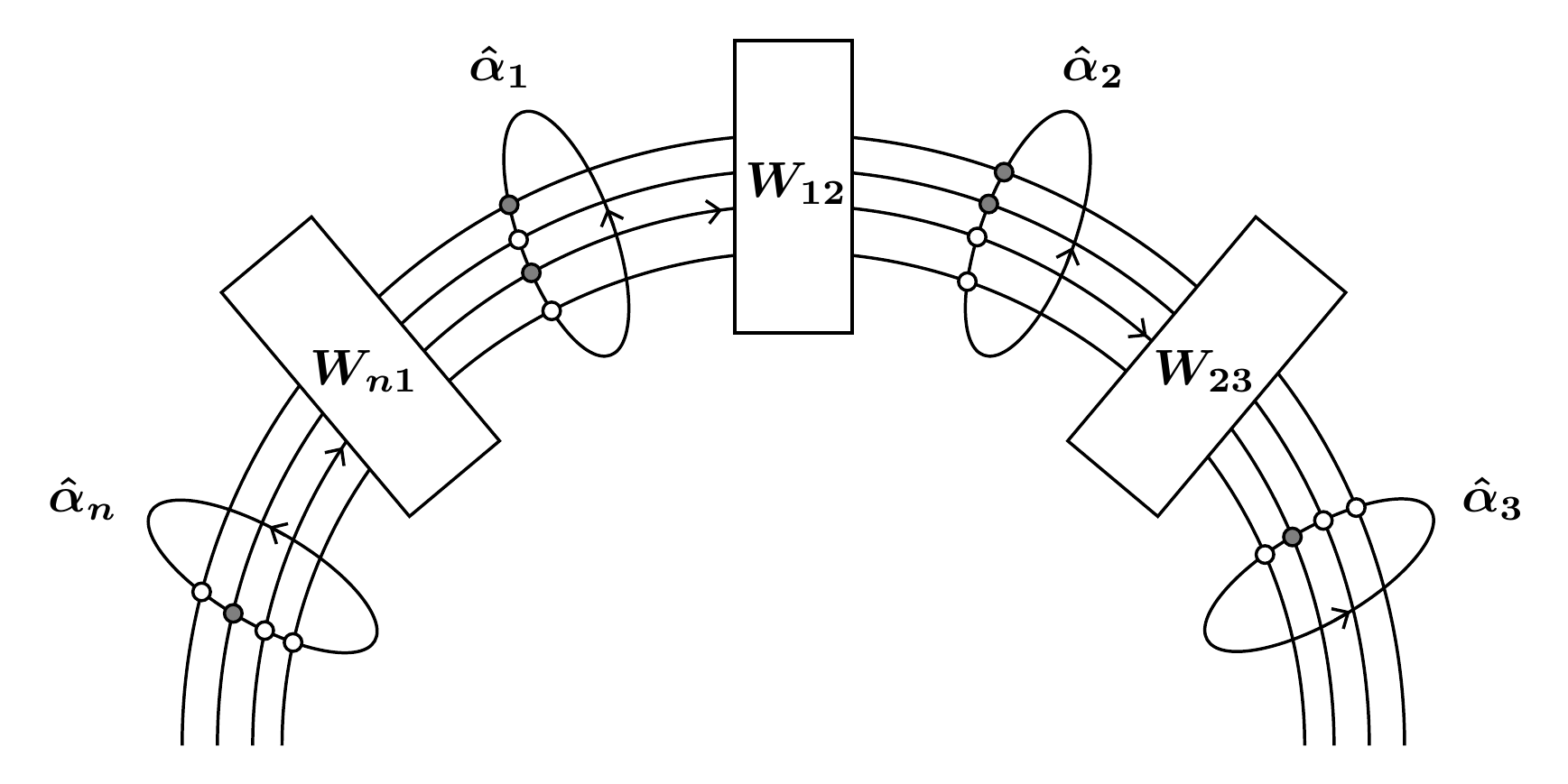}
\caption{The Cayley graph in $S_{\hL}$ formula, generated by $\{ \hat\alpha_i \}$ and $\{ W_{ij} \}$\,. Gray circles represent the scalar fields, and white circles the identity fields.}
\label{fig:SL_Cayley}
\end{center}
\end{figure}

We discuss how to draw a Cayley graph based on the $S_{\hL}$ formula.
The main idea is to define an $n \times \hL$ Wick-contraction matrix
\begin{equation}
\hmu = \begin{pmatrix}
1 & \halpha_1 (1) & \halpha_1^2 (1) & \dots & \halpha_1^{\hL-1} (1) \\[1mm]
W_{21} (1) & W_{21} \, \halpha_1 (1) & W_{21} \, \halpha_1^2 (1) & \dots & W_{21} \, \halpha_1^{\hL-1} (1) \\
\vdots & & & & \vdots \\
W_{n1} (1) & W_{n1} \, \halpha_1 (1) & W_{n1} \, \halpha_1^2 (1) & \dots & W_{n1} \, \halpha_1^{\hL-1} (1)
\end{pmatrix}, \qquad
W_{k1} = U_k \, U_1^{-1} \,.
\label{def:hmu mat}
\end{equation}
In general $\halpha_k$ permutes the $k$-th row of $\hmu$ in a non-trivial way. 
We wrote the matrix $\hmu$ in the form where $\halpha_1$ acts trivially.
The $p$-th column of $\hmu$ corresponds to the $p$-th flavor factor
\begin{multline}
\( \halpha_1^{p-1} (1) , W_{21} \, \halpha_1^{p-1} (1) , \dots W_{n1} \, \halpha_1^{p-1} (1) \)
\equiv \( U_1(p) , U_2(p) , \dots , U_n (p) \)
\\
\quad \leftrightarrow \quad
h^{\hat A^{(1)}_{U_1 (p)} \hat A^{(2)}_{U_2 (p)} \dots \hat A^{(n)}_{U_n (p)} } \,.
\end{multline}
Thanks to the monodromy constraint $W_{11}=1$, we can think of $\hmu$ as an element of $\bb{Z}_n^{\otimes \hL}$ consisting of $\hL$ columns.\footnote{The matrix $\hmu$ has the symmetry $S_n^{\otimes \hL}$ if we sum over $\{U_k\}$. Here we consider a gauge-fixed version.} 
The two groups
\begin{equation}
\halpha = \otimes_{k=1}^n \halpha_k \subset S_{\hL}^{\otimes n}  
\quad {\rm and } \quad
\hmu \in \bb{Z}_n^{\otimes \hL}
\end{equation}
can be used to draw a Cayley graph of the $n$-point function the $S_{\hL}$ formula, as in Figure \ref{fig:SL_Cayley}.
This graph looks different from the Cayley graph in the $S_{2\hL}$ formula. In particular, planar graphs in the $S_{2\hL}$ formula may not be realized as a planar graph in the $S_{\hL}$ formula, even though the number of color cycles is the same for both.

Intuitively, the $n$-tuple Wick-contraction can be regarded as a hoop with two hooks. There are $\hL$ hoops in total. We hook the hoop at a pair of operators $( \cO_i \,, \cO_j )$, where each operator $\cO_i$ has $L_i$ holes where one can hook the hoop. 
The position of the holes is permuted by $U$.

By knowing the position of the identity fields in $\hat \cO_i$ from \eqref{def:Ohat}, we can reconstruct the Wick-contraction structure immediately from $\hmu$. To see it, consider the following example. One possible Wick-contraction of $\Vev{ \cO_1 \, \cO_2 \, \cO_3 }$ with $(L_1,L_2,L_3)=(3,4,3)$ can be represented as
\begin{equation}
\contraction{\Big\langle \tr (}{\Phi^{A_1}}{ \Phi^{A^2} \Phi_{A_3} ) \, \tr (}{\Phi^{B_1}}
\contraction[2ex]{\Big\langle \tr (\Phi^{A_1} }{\Phi^{A_2}}{ \Phi^{A_3} ) \, \tr (\Phi^{B_1} }{\Phi^{B_2}}
\contraction[3ex]{\Big\langle \tr (\Phi^{A_1} \Phi^{A_2} }{\Phi^{A_3}}{ ) \, \tr (\Phi^{B_1} \Phi^{B_2} \Phi^{B_3} \Phi^{B_4} ) \, \tr (}{\Phi^{C_1}}%
\contraction{\Big\langle \tr (\Phi^{A_1} \Phi^{A_2} \Phi^{A_3} ) \, \tr (\Phi^{B_1} \Phi^{B_2} }{\Phi^{B_3}}{  \Phi^{B_4} ) \, \tr (\Phi^{C_1} \Phi^{C_2} }{\Phi^{C_3}}%
\contraction[2ex]{\Big\langle \tr (\Phi^{A_1} \Phi^{A_2} \Phi^{A_3} ) \, \tr (\Phi^{B_1} \Phi^{B_2} \Phi^{B_3} }{\Phi^{B_4}}{ ) \, \tr (\Phi^{C_1} }{\Phi^{C_2} }%
\Big\langle \tr (\Phi^{A_1} \Phi^{A_2} \Phi^{A_3} ) \, \tr (\Phi^{B_1} \Phi^{B_2} \Phi^{B_3} \Phi^{B_4} ) \, \tr (\Phi^{C_1} \Phi^{C_2} \Phi^{C_3} ) \Big\rangle
\quad \leftrightarrow \quad 
\begin{pmatrix}
A_1 & A_2 & A_3 & & \\
B_1 & B_2 & & B_3 & B_4 \\
& & C_1 & C_3 & C_2
\end{pmatrix} .
\label{hatL embedding}
\end{equation}
We can slightly generalize the same matrix so that it has double columns,\footnote{Since the flavor inner-product $g^{AB}$ is symmetric, we do not distinguish two ways to double the columns, i.e.
$$\begin{pmatrix} A \\ B \end{pmatrix}
\sim \begin{pmatrix} A & \\ & B \end{pmatrix}
\sim \begin{pmatrix} & A \\ B & \end{pmatrix}.$$}
\begin{equation}
\begin{pmatrix}
A_1 & A_2 & A_3 & & \\
B_1 & B_2 & & B_3 & B_4 \\
& & C_1 & C_3 & C_2
\end{pmatrix} 
\quad \Leftarrow \quad 
\begin{pmatrix}
A_1 & & A_2 & & A_3 & & & & & \\
& B_{1'} & & B_{2'} & & & B_3 & & B_4 & \\
& & & & & C_{1'} & & C_{3'} & & C_{2'}
\end{pmatrix} .
\label{hatL embedding double}
\end{equation}
This notation will be used later, e.g. in \eqref{def:hmu}.
To the empty entries, let us assign dummy indices $\{ \hat A_4 \,, \hat A_5 \,, \hat B_5 \,, \hat C_4 \,, \hat C_5 \}$ corresponding to ${\bf 1}$ as
\begin{equation}
\mu = \begin{pmatrix}
A_1 & A_2 & A_3 & & \\
B_1 & B_2 & & B_3 & B_4 \\
& & C_1 & C_3 & C_2
\end{pmatrix}
\ \ \to \ \ 
\begin{pmatrix}
A_1 & A_2 & A_3 & \hat A_4 & \hat A_5 \\
B_1 & B_2 & \hat B_5 & B_3 & B_4 \\
\hat C_4 & \hat C_5 & C_1 & C_3 & C_2
\end{pmatrix}
= \hmu.
\label{hatL embedding2}
\end{equation}
We reconstruct the original Wick-contraction \eqref{hatL embedding} by remembering the label of the dummy indices.
The Wick contraction matrix plays an important r\^ole in the rest of the paper.

\subsubsection{Extremal case}

We show that the $n$-point function \eqref{SL tree n-pt} in the extremal case $\sum_{i=1}^{n-1} L_i = L_n$ reduces to the two-point function \eqref{two-point MT correlator}.
In the extremal correlator, all Wick contractions should be taken between $\cO_i$ and $\cO_n$ for $i=1,2, \dots, n-1$. Thus
\begin{equation}
\hL = \sum_{i=1}^{n-1} \ell_{in} = \sum_{i=1}^{n-1} L_i = L_n
\end{equation}
Using the gauge symmetry, we may fix the position of the scalar and identity fields of $\hat \cO_i$ as
\begin{equation}
\Phi^{\hat A_p^{(i)}} = 1 \quad \text{unless} \quad L'_i \le p < L'_{i+1} \,, \qquad
L'_i = \sum_{k=1}^{i-1} \ell_{kn} \,, \quad L'_1=0,
\end{equation}
and assume that $\alpha_i \in S_{L_i}$ acts on the range $L'_i \le p < L'_{i+1}$\,.
We can restrict the sum over $\{U_i\} \in S_L^{\otimes n}$ in \eqref{SL tree n-pt} as
\begin{equation}
\( U_1 \,, U_2 \,, \dots U_{n-1} \,, U_n \) \quad \to \quad
\( u_1 \,, u_2 \,, \dots u_{n-1} \,, U_n \) \in \( S_{L_1} \,, S_{L_2} \,, \dots \,, S_{L_{n-1}} \,, S_{\hL} \).
\end{equation}
Then the flavor factor simplifies as
\begin{equation}
\prod_{p=1}^{\hL} \, h^{\chA^{(1)}_{p} \chA^{(2)}_{p} \dots \chA^{(n)}_{p} }
=
\( \prod_{p=L'_i+1}^{L'_{i+1}} \, \frac{g^{A^{(i)}_{u_i(p)} A^{(n)}_{U_n(p)} }}{x_{in}^2} \)
\equiv 
 \prod_{q=1}^{\hL} \, \tilde g^{B_{V U_n^{-1} (q)} A^{(n)}_{q} } 
\end{equation}
with $p = U_n (q)$, and the product of $\alpha_i$ can be unified as
\begin{equation}
\begin{gathered}
u_1^{-1} \alpha_1 u_1 \, u_2^{-1} \alpha_2 u_2 \dots u_{n-1}^{-1} \alpha_{n-1} u_{n-1}
= V^{-1} \beta V \in S_{\hL} \,,
\\[2mm]
V = u_1 \circ u_2 \circ \dots \circ u_{n-1}, \qquad
\beta = \alpha_1 \circ \alpha_2 \circ \dots \circ \alpha_{n-1} \in \prod_{i=1}^{n-1} S_{L_i} \subset S_{\hL} \,.
\end{gathered}
\end{equation}
By writing $V_n \equiv V \, U_n^{-1}$\,, we find that one of the sums over $V$ and $U_n$ is trivial.
The $n$-point function becomes
\begin{equation}
\Vev{ \prod_{i=1}^n \hat \cO_i { (x_i) } }
=
\sum_{V_n \in S_{\hL}}
\( \prod_{p=1}^{\hL} \, \tilde g^{B_{V_n^{-1} (p)} A^{(n)}_{p} } \)
\delta_{\hL} \( \Omega \,  \beta V_n \, \alpha_n \, V_n^{-1}\)
\end{equation}
which agrees with the two-point function \eqref{two-point MT correlator}.

\subsection{Face-based formulae}\label{sec:face-based formula}

From a Cayley graph generated by $(\alpha , W)$, we define the dual graph by
\begin{equation}
{\rm Dual} \,: \, (\alpha , W) \ \mapsto \ (\omega, W), \qquad
\omega \equiv \alpha W .
\label{def:dual omega}
\end{equation}
We want to express the $n$-point functions as a sum over the face permutation $\omega$, instead of a sum over $W$,
\begin{equation}
G_n = \sum_{\omega} \ \bb{F} ( \omega )
\label{n-pt S*2hL-1}
\end{equation}
where $\bb{F}$ computes the color and flavor factors associated to $\omega$.

In order to define $\bb{F}$, we need to parametrize the faces of the graph. 
The Feynman graphs in the double-line notation have two types of edges, those from Wick-contractions and the others from external operators. 
The red lines in Figure \ref{fig:Cayley644} signify the former, and the blue ones signify the latter.
We call the edge from Wick-contractions the Wick-edge, and the edge from external operators the {\it rim}. Each Wick-edge is connected to a pair of rims, and each rim is oriented by $\alpha = \prod_{i=1}^n \alpha_i \in S_{2\hL}$\,.

\begin{figure}[t]
\begin{center}
\includegraphics[scale=.85]{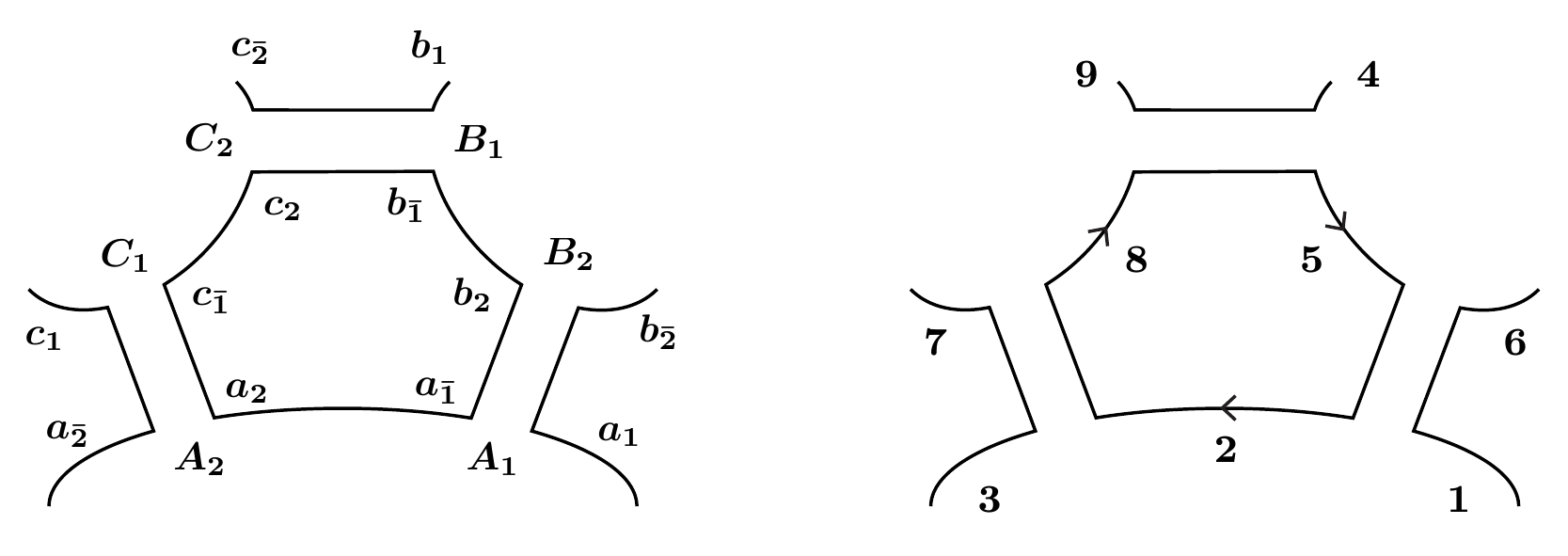}
\caption{(Left) Wick-contraction in the half-edge notation, see Appendix \ref{app:double line}. This is the Cayley graph generated by $\iota_\alpha = (a_{1'} a_2)(b_{1'} b_2) (c_{1'} c_2) \dots$ and $W = (A_1 B_2)(B_1 C_2)(C_1 A_1) \dots $. (Right) The same graph whose rim between $a_{\bar P}$ and $a_{\alpha(P)}$ is labeled as $P$.}
\label{fig:half-edge}

\bigskip
\includegraphics[scale=.8]{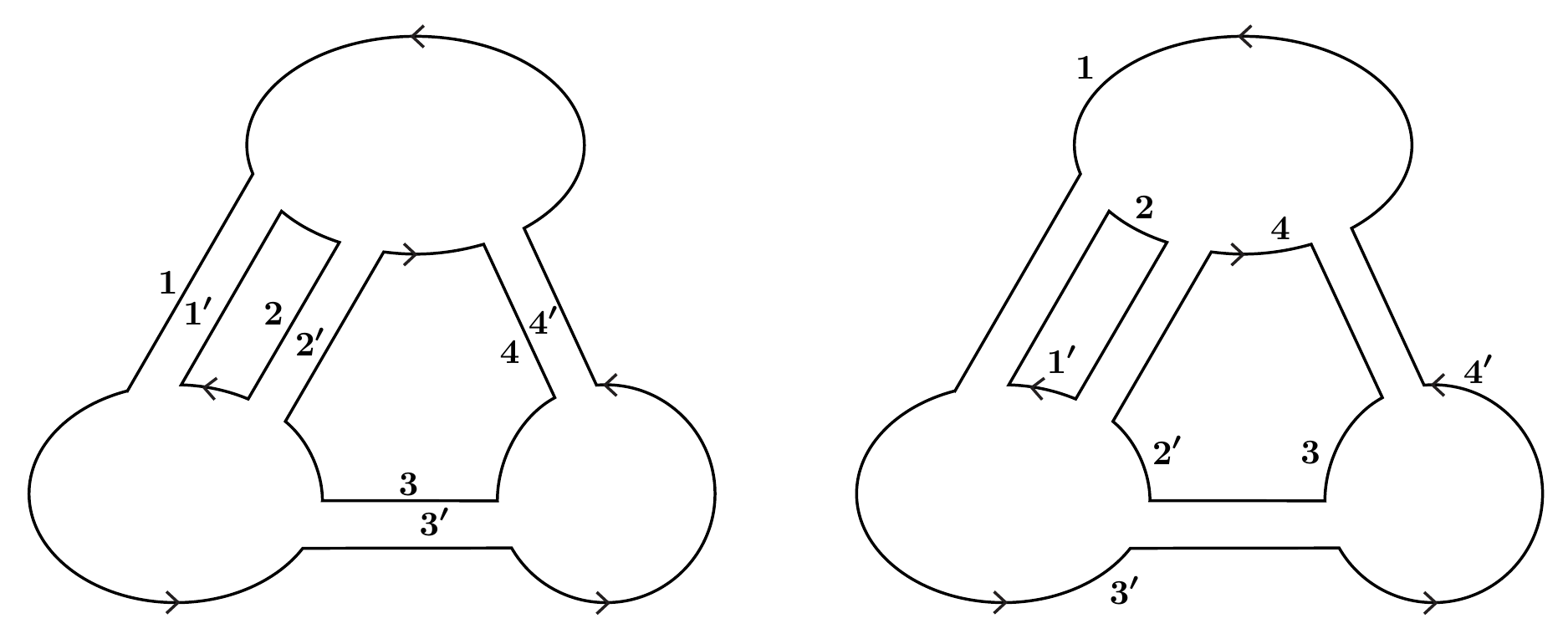}
\caption{(Left). Wick-contraction in the double-line notation with the numbers $\{ 1, 1', 2, 2', \dots \}$ labeling Wick-edges. (Right). The same graph whose rims are labeled. Both figures correspond to $\omega = (1' 2)(2' 4 3)(1 3' 4')$.}
\label{fig:half-edge2}
\end{center}
\end{figure}

There are two ways to label rims, depending on whether one simplifies $\alpha$ or $W$.
In the first method, we label the rims by $\{ 1, 2, \dots, 2\hL \}$ as in Figure \ref{fig:half-edge}. These numbers are ordered in the same way as $\alpha$ explained in \eqref{parametrize alpha_k}.\footnote{For example, $\alpha_2$ contains $L_2$ rims labeled by $L_1+1, L_1+2, \dots , L_1+L_2$.}
In the second method, we label the Wick-edges by $\{ 1, 1', 2, 2', \dots, \hL, \hL' \}$. The Wick-contraction is given by $W = \prod_p (p \, p')$. Since the whole graph is oriented, each rim starts from a Wick-edge and ends at another Wick-edge. We label the rim by using the ending Wick-edge as in Figure \ref{fig:half-edge2}.

\subsubsection*{Rim-labels by $W$}

We consider $n$-point functions of single-trace operators by using the rims labeled by $W$. 
The rim-labels by $\alpha$ will be studied in Appendix \ref{sec:face rim}.
For simplicity, we assume that all external operators are single-traces.

To begin with, we show the equivalence between the set of unlabeled graphs and face permutations.
The face permutation should not have one-cycles, because the one-cycles come from self-contractions. 
We define $S_{2 \hL}^\times \subset S_{2 \hL}$ by
\begin{equation}
\omega \in S_{2 \hL}^\times \qquad \Leftrightarrow \qquad
p \neq \omega (p) \quad (\forall p).
\label{def:S2L times}
\end{equation}
It follows that
\begin{equation}
\pare{ \text{Feynman graphs with $2\hL$ unlabeled vertices} } \quad \leftrightarrow \quad 
\omega \in S_{2 \hL}^\times /S_{\hL} [\bb{Z}_2] .
\label{graph coset equivalence}
\end{equation}
In RHS, we define $\omega$ as the permutation of $\{ 1, 1', 2, 2', \dots, \hL, \hL' \}$ modulo the action of $S_{\hL} [\bb{Z}_2]$ explained in \eqref{app:SLZ2 action}.
Owing to the redundancy of relabeling, $\omega$ is in one-to-one correspondence with the coset $S_{2 \hL}^\times /S_{\hL} [\bb{Z}_2]$, where the denominator has the adjoint action like \eqref{def:moduli cW}.
We have already explained how to construct a face permutation from a given graph in the double-line notation; label the Wick edges using $\{ 1, 1', 2, 2', \dots, \hL, \hL' \}$ and export them to rims.
Conversely, we can construct a graph from the face permutation as follows.
Consider the permutation $\alpha_\bullet \equiv \omega \, W_\bullet \in S_{2 \hL}$ with
\begin{equation}
W_\bullet \equiv \prod_{p=1}^{\hL} \( p \, p' \) .
\label{canonical W}
\end{equation}
If we regard $\alpha_\bullet $ as the permutation for the external single-trace operators, we can make all rims oriented. 
By connecting the end-points of the rim pairs $p, p'$, we obtain a Feynman diagram in the double-line notation. Hence the relation \eqref{graph coset equivalence} is proven.

We make two more comments.
First, the relation \eqref{graph coset equivalence} tells nothing about the cycle type of $\alpha_\bullet$\,. 
To compute $G_n$ we need to restrict the sum over $\omega$ further.
Second, in this construction $\omega$ produces an unlabeled graph as exemplified in Figure \ref{fig:half-edge2}. 
In other words, the rim labels $\{ 1, 1', 2, 2', \dots, \hL, \hL' \}$ are not related to the flavor indices of the external operators, and the identification is not unique.
Generally, an unlabeled graph is an equivalence class of the relabeling group acting on labeled graphs.
Thus, the function $\bb{F}$ in \eqref{n-pt S*2hL-1} is a map from a {\it set of graphs} to a polynomial of $(N_c, g^{AB})$, where the graph set is a certain group orbit.

\bigskip
We define $\bb{F}$ in the following way. 
To begin with, we interpret a permutation as a set of ordered lists. For example, a permutation of the cycle type $[2^2]$ can be interpreted as eight ordered lists,
\begin{multline}
S_4 \ \ni \ (12)(1'2') \ \stackrel{\bf CL}{\longleftrightarrow}\ 
\Bigl\{ (1,2)(1',2'), (2,1)(1',2'),(1,2)(2',1'), (2,1)(2',1'),
\\
(1',2')(1,2), (1',2')(2,1), (2',1')(1,2), (2',1')(2,1) \Bigr\}
\label{example:cycle to list}
\end{multline}
where {\bf CL} stands for the map between cycles to lists.
Generally, a permutation of the cycle type $\lambda = [1^{\lambda_1} 2^{\lambda_2} \dots ]$ can be interpreted as $| {\bf CL} (\lambda) |$ ordered lists, where
\begin{equation}
\abs{ {\bf CL} (\lambda) } = \abs{ \prod_k S_{\lambda_k} [\bb{Z}_k] } 
= \prod_k \( \lambda_k! \ k^{\lambda_k} \).
\label{abs CL lambda}
\end{equation}
The multiple interpretations come from the translation of each cycle and permuting the cycles of the same length.
In other words, ${\bf CL} (\alpha_\bullet)$ is isomorphic to the stabilizer subgroup,\footnote{In the literature, ${\rm Stab} (\alpha_\bullet)$ is also called the automorphism group ${\rm Aut} (\alpha_\bullet)$.}
\begin{equation}
{\rm Stab} (\alpha_\bullet) \equiv 
\Bigl\{ \gamma \in S_{2 \hL} \mid \gamma \alpha_\bullet \gamma^{-1} = \alpha_{\bullet} \Bigr\} .
\end{equation}
Indeed, if $\alpha_\bullet$ has the cycle type $\lambda = [1^{\lambda_1} 2^{\lambda_2} \dots ]$, the order of ${\rm Stab} (\alpha_\bullet)$ is\footnote{Since all operators are assumed to be single-traces, $\lambda_k$ is the number of single-trace operators of length $k$.}
\begin{equation}
\abs{{\rm Stab} (\alpha_\bullet) } = \prod_k S_{\lambda_k} [\bb{Z}_k] = | {\bf CL} (\lambda) | .
\label{CL as Stab alpha}
\end{equation}

Given the permutation $\alpha_\bullet$\,, we pick up one element from ${\bf CL} (\alpha_\bullet)$ and write it as $\beta \equiv (\beta_1\,, \dots \,, \beta_n)$.\footnote{At this point $n$ is a free parameter. We will equate $n$ with the number of external single-trace operators in \eqref{n-pt S*2hL cycle type}.}
In the example of Figure \ref{fig:half-edge2}, we find $\alpha = (124)(1'3'2')(34')$. 
Let us choose an ordered list as
\begin{equation}
\beta = (1',3',2')(2,4,1)(3,4') \ \in \ {\bf CL} (\alpha) .
\end{equation}
The choice of $\beta$ picks up a Feynman graph, i.e. one set of Wick-contractions appearing in
\begin{equation}
\Vev{ \cO_1 \, \cO_2 \, \cO_3 } = \Big\langle
\tr( \Phi^{A_1} \, \Phi^{A_2} \, \Phi^{A_{3}}) \,
\tr( \Phi^{A_4} \, \Phi^{A_5} \, \Phi^{A_{6}}) \, 
\tr ( \Phi^{A_{7}} \, \Phi^{A_{8}} ) \Big\rangle.
\end{equation}
Let us make this point clearer. The ordered list $\beta \in {\bf CL} (\alpha_\bullet)$ relates the rim labels to the flavor indices as
\begin{equation}
\beta_i : \{1,1', 2, 2', \dots \} \ \to \ \{ 0,1,2,\dots, L_i \} 
\label{def:beta_i}
\end{equation}
where $\beta_i^{-1} (x)$ is uniquely defined if and only if $x \neq 0$.\footnote{Here $x=0$ implies that the corresponding rim label is not assigned to $\cO_i$\,.}
We parameterize $\alpha$ as
\begin{equation}
\alpha 
= \begin{pmatrix} \alpha_1 \\ \alpha_2 \\ \vdots \\ \alpha_n \end{pmatrix} 
= \begin{pmatrix}
\( \beta_1^{-1} (1) \, \beta_1^{-1} (2) \, \dots \, \beta_1^{-1} (L_1) \) \\
\( \beta_2^{-1} (1) \, \beta_2^{-1} (2) \, \dots \, \beta_2^{-1} (L_2) \) \\
\vdots \\ 
\( \beta_n^{-1} (1) \, \beta_n^{-1} (2) \, \dots \, \beta_n^{-1} (L_n) \) 
\end{pmatrix} 
\end{equation}
and construct the Wick-contraction matrix (see \eqref{hatL embedding double}),
\begin{equation}
\hmu_\beta = \begin{pmatrix}
\mu_{1,1} & \mu_{1, 1'} & \mu_{1,2} & \mu_{1, 2'} & \dots & \mu_{1, \hL'} \\ 
\mu_{2,1} & \mu_{2,1'} & \mu_{2,2} & \mu_{2, 2'} & \dots & \mu_{2, \hL'} \\ 
\vdots \\ 
\mu_{n,1} & \mu_{n,1'} & \mu_{n,2} & \mu_{n, 2'} & \dots & \mu_{n, \hL'} 
\end{pmatrix}, \qquad
\mu_{i,s} = 
\begin{cases}
A^{(i)}_{\beta_i (s)} &\quad (\beta_i (s) \neq 0) \\
{\bf 1} &\quad (\beta_i (s) = 0).
\end{cases}
\label{def:hmu}
\end{equation}
The matrix $\hmu_\beta$ is defined uniquely for each $\beta \in {\bf CL} (\alpha_\bullet)$.
We associate a flavor factor to $\hmu_\beta$ as
\begin{equation}
{\bf F} (\hmu_\beta) = \prod_{s=1}^{\hL} h^{\mu_{1,s} \, \mu_{1, s'} \, \mu_{2,s} \, \mu_{2,s'} \, \dots \, \mu_{n,s} \, \mu_{n,s'}} ,
\label{def:bfF hmub}
\end{equation}
where $h^{\mu_{1,s} \, \mu_{1,s'} \, \dots}$ is defined by 
\begin{equation}
h^{A_1 A_{1'} \dots A_{n} A_{n'}} = \frac{1}{2 (2n-2)!} \, \sum_{\pi \in S_{2n}} h^{A_{\pi(1)} A_{\pi(1')}} \prod_{j=2}^{n} \delta^{A_{\pi(j)}}_{\bf 1} \, \delta^{A_{\pi(j')}}_{\bf 1} 
\label{def:hA1A2n}
\end{equation}
like \eqref{def:hA1An}, and satisfies
\begin{equation}
h^{\mu_{1,s} \, \mu_{1,s'} \, \mu_{2,s} \, \mu_{2,s'} \, \dots \, \mu_{n,s} \, \mu_{n,s'}}
= h^{A^{(i_s)}_s \, A^{(j_s)}_{s'}} 
\qquad \text{for some $( i_s \,,j_s )$ from $1,2,\dots, n$}.
\label{hA1-2n reduce}
\end{equation}
The permutation $S_{2n}$ in \eqref{def:hA1A2n} runs over two columns of $\hmu_\beta$, where two out of $2n$ entries should not be ${\bf 1}$.
From \eqref{hA1-2n reduce} we obtain the set of $\hL$ pairs
\begin{equation}
\pare{ (A^{(i_1)}_1 , A^{(j_1)}_{1'}) \, (A^{(i_2)}_2 , A^{(j_2)}_{2'}) \, \dots \, 
(A^{(i_{\hL})}_{\hL} , A^{(j_{\hL})}_{\hL'}) }
\end{equation}
which is essentially $W \in \bb{Z}_2^{\otimes \hL}$ in Section \ref{sec:S2L formalsim}.
The self-contractions are removed by the property of $h^{AB}$ in \eqref{def:hAB}.

We should sum over the choices $\beta \in {\bf CL} (\alpha_\bullet)$ entering in $\hmu_\beta$\,.
This procedure corresponds to labeling unlabeled graphs.
Then we define the function on faces by
\begin{equation}
\bb{F} (\omega) = N_c^{C(\omega)} \sum_{\beta \, \in \, {\rm Stab} (\alpha_\bullet)}
{\bf F} (\hmu_\beta ) \,,
\label{gen: construct F}
\end{equation}
with ${\bf F} (\hmu_\beta )$ in \eqref{def:bfF hmub}.
Using the correspondence \eqref{graph coset equivalence}, we can write the $n$-point function of single-trace operators as
\begin{equation}
G_n = \frac{1}{|S_{\hL} [\bb{Z}_2]|} \, \sum_{\omega\in S_{2 \hL}^\times } \ \bb{F} ( \omega ) \, 
\delta_{\text{cycle-type}} ( \alpha_{\rm ex}^{-1} \,, \omega W_\bullet ) .
\label{n-pt S*2hL cycle type}
\end{equation}
We inserted the $\delta$-function in order to guarantee that $\alpha_\bullet = \omega W_\bullet$ has the same cycle type of $\alpha_{\rm ex}$ specified by the external operators.
We rewrite \eqref{n-pt S*2hL cycle type} in terms of the conjugacy class \eqref{def:conj class} of $S_{2\hL}$ as
\begin{equation}
G_n = \frac{1}{|S_{\hL} [\bb{Z}_2]|} \, \frac{(2 \hL)!}{| {\rm Stab} (\alpha_{\rm ex}) |} \,
\sum_{\omega\in S_{2 \hL}^\times } \ \bb{F} ( \omega ) \, 
\delta_{2 \hL} ( [ \alpha_{\rm ex}^{-1}] \, \omega W_\bullet ) .
\label{n-pt S*2hL-3}
\end{equation}
We checked this formula by {\tt Mathematica} for simple cases.


\section{Pants decomposition}\label{sec:PF pants}

Let us discuss the geometric interpretation of the $n$-point function in the $S_{\hL}$ formalism.
We rewrite the formula \eqref{SL tree n-pt}  as a product of three-point functions, or pairs of pants, schematically as
\begin{equation}
\begin{aligned}
\cX [ \Sigma_{0,n} ] 
&\equiv \frac{1}{\hL!} \ \sum_{U}
\( \prod_{p=1}^{\hL} \, h^{\chA^{(1)}_{p} \chA^{(2)}_{p} \dots \chA^{(n)}_{p} } \) 
\delta_{\hL} \( \Omega \, 
\chalpha_1 \, \chalpha_2 \, \dots \, \chalpha_n \)
\\[1mm]
&\sim {\rm Glue} \( \cX [ \cS_1 ] , \cX [ \cS_2 ] , \dots , \cX [ \cS_{n-2} ] \) 
\end{aligned}
\label{correlator as TFT}
\end{equation}
where $\{ \cS_f \}$ defines a pants decomposition of $\bb{CP}_1$ with $n$ punctures,
\begin{equation}
\Sigma_{0,n} = \cS_1 \amalg \cS_2 \amalg \dots \amalg \cS_{n-2} \,.
\label{def:decomp Sigma}
\end{equation}
The last equation of \eqref{correlator as TFT} will be made precise in \eqref{pants decomp npt}.
This expression implies that the tree-level $n$-point function is a certain topological partition function on $\bb{CP}_1$ with $n$ boundaries and defects. 
We call it topological, because it is invariant under different pants decompositions, and does not depend on the complex structure of $\Sigma_{0,n}$\,.\footnote{We can also think of $\Sigma_{0,n}$ as $\bb{CP}_1$ with $n$ boundaries, because the topological partition function does not depend on the length of boundary circles.}
The punctures represent external operators $\{ \hat \cO_i \}$, and each factor $\cX [ \cS_f ]$ represents a local three-point interaction.
The defects carry all the information about $N_c$\,. This is why higher-genus Riemann surfaces do not show up.
The sum over $U$'s in \eqref{correlator as TFT} can be thought of as a sum over different twisted sectors, or the partition function of a topological gauge theory with finite gauge groups \cite{Dijkgraaf:1989pz,Fukuma:1993hy,Chung:1993xr,Lazaroiu:2000rk,MooreLec01,Moore:2006dw}.\footnote{Beware that \eqref{correlator as TFT} is {\it not} the topological string partition function on $\bb{P}^1$ mentioned in Introduction.}

\bigskip
To derive \eqref{correlator as TFT}, consider the color and flavor factors separately.

\begin{figure}[t]
\begin{center}
\includegraphics[scale=0.8]{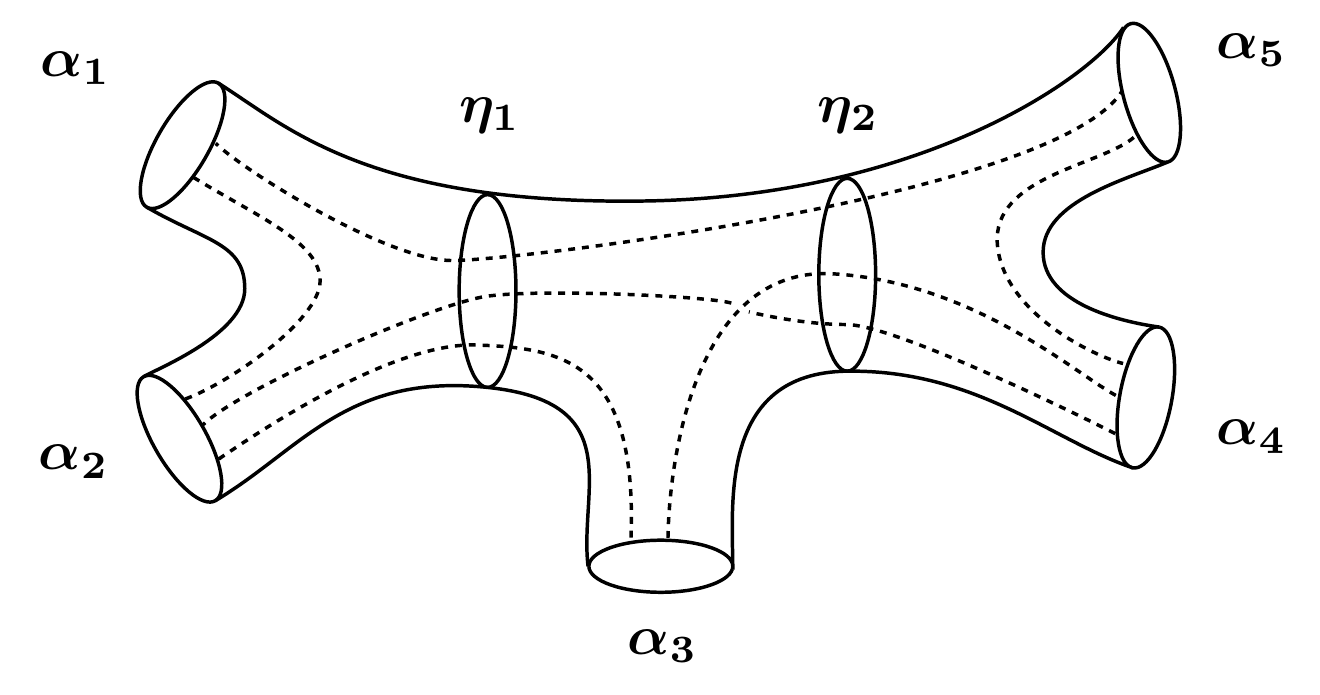}
\caption{Pants decomposition of the five-point function $\vev{\cO_1 \cO_2 \cO_3 \cO_4 \cO_5}$, with intermediate states denoted by $\cO_{\eta_1}$ and $\cO_{\eta_2}$\,. Although this figure has a fixed number of bridge lengths, the sum over $\{ \ell_{ij} \}$ has already been performed in our formula.}
\label{fig:TFT recursion}
\end{center}
\end{figure}

The color factor decomposes as follows. Thanks to the permutation symmetry \eqref{npt perm symm}, without loss of generality we may assume $\cS_1$ contains $\{ \alpha_1 \,, \alpha_2 \}$, $\cS_k$ contains $\alpha_{k+1}$ for $2 \le k \le n-3$, and $\cS_{n-2}$ has $\{ \alpha_{n-1} \,, \alpha_n \}$. 
Then we insert the resolution of identity on the permutation basis\footnote{This relation is not the same as the resolution of identity in the Hilbert space of gauge theory. In general, the inverse of tree-level two-point functions is much more complicated than the inverse of permutations.}
\begin{equation}
\delta_{\hL} \( \alpha \beta \) = \sum_{\eta \in S_{\hL}} \delta_{\hL} \( \alpha \eta \) \, \delta_{\hL} \( \eta^{-1} \beta \).
\label{resolution of identity}
\end{equation}
Now we rewrite the $\delta$-function as
\begin{multline}
\delta_{\hL} \( \Omega \, \chalpha_1 \, \chalpha_2 \, \dots \, \chalpha_n \)
=
\sum_{\homega_1 \,, \homega_2 \,, \dots \,, \homega_{n-2} \in S_{\hL}} \ 
\sum_{\eta_1 \,, \eta_2 \,, \dots \,, \eta_{n-3} \in S_{\hL}}
\delta_{\hL} \( \Omega \, [\homega_1] \, [\homega_2] \, \dots \, [\homega_{n-2}] \) \ \times
\\
\delta_{\hL} \( \homega_1^{-1} \, \chalpha_1 \, \chalpha_2 \, \eta_1 \) \ 
\prod_{k=2}^{n-3} \delta_{\hL} \( \homega_k^{-1} \, \eta_{k-1}^{-1} \, \chalpha_{k+1} \, \eta_k \) \ 
\delta_{\hL} \( \homega_{n-2}^{-1} \, \eta_{n-3}^{-1} \, \chalpha_{n-1} \, \chalpha_n \) 
\label{pants decomp color}
\end{multline}
where $[ \omega_k ]$ is the conjugacy class defined in \eqref{def:conj class}. Since two conjugacy classes commute, the first line is manifestly invariant under the permutation of different $\omega$'s.
This decomposition is similar to the following set of OPE's,\footnote{Strictly speaking, this operation is not OPE. Generally, an OPE produces several terms, while inserting the resolution of identity produces one term.}
\begin{equation}
\cO_1 \times \cO_2 \to \cO_{\eta_1} \,, \qquad
\cO_{\eta_{k-1}} \times \cO_{k+1} \to \cO_{\eta_k} \quad (2 \le k \le n-3), \qquad
\cO_{n-3} \times \cO_{n-1} \to \cO_n \,.
\end{equation}
We identify the $\delta$-function $\delta_{\hL} \( \homega_k^{-1} \, \eta_{k-1}^{-1} \, \chalpha_{k+1} \, \eta_k \)$ as the color part of $\cX [ \cS_k ]$.
This structure is also depicted in Figure \ref{fig:TFT recursion}.

In order to split the flavor part of the formula \eqref{correlator as TFT}, we introduce the functions
\begin{equation}
H^{A_1 A_2 A_3} (\phi) = \phi^{-1} \, h^{A_1 A_2 A_3} + \prod_{j=1}^3 \delta^{A_j}_{\bf 1} \,, \qquad
\zeta_{AB} (\phi) = \phi \, g_{AB} + \delta_A^{\bf 1} \, \delta_B^{\bf 1}
\label{def:H-phi}
\end{equation}
where $\phi$ is an extra parameter to count the scalars (i.e. non-identity fields), and $g_{AB}$ is the inverse metric
\begin{equation}
g^{AB} \, g_{BC} \, g^{CD} = g^{AD} \,, \qquad
g^{\bf 1 \bf 1} = g_{\bf 1 \bf 1} = 0.
\label{def:inverse g}
\end{equation}
According to \eqref{def:hABC}, $h^{A_1 A_2 \dots A_n} \neq 0$ when it has precisely two scalars.
Thus $H^{A_1 A_2 A_3} \neq 0$ when it has zero or two scalar indices.

By using \eqref{def:H-phi} and \eqref{def:inverse g}, we find the identity
\begin{multline}
\prod_{p=1}^{\hL} h^{\chA^{(1)}_{p} \chA^{(2)}_{p} \dots \chA^{(n)}_{p} } 
=
\prod_{p=1}^{\hL} \ \Res_{\phi_p =0} \ \Biggl[
\prod_{k=2}^{n-2} \zeta_{B^{(k-1)}_p C_p^{(k-1)}} (\phi_{p}) 
\ \times
\\
H^{\chA^{(1)}_p \chA^{(2)}_p B_p^{(1)}} (\phi_{p}) \ 
\( \prod_{k=3}^{n-2} 
H^{C_p^{(k-2)} \chA^{(k)}_p B_p^{(k-1)}} (\phi_{p}) \) \ 
H^{C_p^{(n-3)} \chA^{(n-1)}_p \chA^{(n)}_p } (\phi_{p}) 
\Biggr] .
\label{pants decomp flavor}
\end{multline}
To prove it, recall that precisely two indices in $\{ \chA^{(1)}_p \,, \chA^{(2)}_p \,, \dots \,, \chA^{(n)}_p \}$ on LHS are scalars for each $p$.
On RHS, there are $(3n-6)$ flavor indices for each $p$
\begin{equation}
\Bigl\{ \chA^{(1)}_p \,, \chA^{(2)}_p \,, B_p^{(1)} \mid
C_p^{(1)} \,, \chA^{(3)}_p \,, B_p^{(2)} \mid
\dots \mid
C_p^{(n-4)} \,, \chA^{(n-2)}_p \,, B_p^{(n-3)} \mid
C_p^{(n-3)} \,, \chA^{(n-1)}_p \,, \chA^{(n)}_p \Bigr\} 
\end{equation}
and the factors $H$ or $\zeta$ are trivial if they consist of the identity fields only.
We look for the term of $O(\phi_p^{-1})$. This comes from the case
\begin{equation}
\( \text{ Number of $H$'s carrying scalars } \)
- \( \text{ Number of $\zeta$'s carrying scalars } \) = 1.
\end{equation}
Since $g_{BC}$ is a diagonal metric, this condition is satisfied precisely when two of $\{ \chA^{(1)}_p \,, \chA^{(2)}_p \,, \dots \,, \chA^{(n)}_p \}$ are scalars. In particular, all the $B$'s and $C$'s between a pair of $A$'s should be scalars,
\begin{equation}
\chA^{(k)}_p \neq {\bf 1} \ \ {\rm and} \ \ \chA^{(\ell)}_p \neq {\bf 1} \quad \Rightarrow \quad
B_p^{(m)} \neq {\bf 1} \ \ {\rm and} \ \ C_p^{(m)} \neq {\bf 1} \quad (k-1 \le m \le \ell-2).
\end{equation}
This completes the proof.

\bigskip
Let us combine the pants decomposition of the color and flavor factors \eqref{pants decomp color}, \eqref{pants decomp flavor}, and substitute them into the $n$-point formula \eqref{correlator as TFT}.
The result is
\begin{equation}
\cX [ \Sigma_{0,n} ] = \frac{1}{\hL!} \ \sum_{ \homega, \eta }
\delta_{\hL} \( \Omega \, [\homega_1] \, [\homega_2] \, \dots \, [\homega_{n-2}] \) \cdot \ 
\prod_{p=1}^{\hL} \ \Res_{\phi_p =0} \ \Bigl[ \prod_{k=1}^{n-3} \zeta_{B^{(k)}_p C_p^{(k)}} (\phi_p)  \ 
\prod_{f=1}^{n-2} \cX [\cS_f ] \Bigr]
\label{pants decomp npt}
\end{equation}
where
\begin{equation}
\begin{aligned}
\cX [\cS_1 ] &= \sum_{U_1 \,, U_2 \in S_{\hL} } \delta_{\hL} \( \homega_1^{-1} \, \chalpha_1 \, \chalpha_2 \, \eta_1 \)
\prod_{p} \, H^{\chA^{(1)}_p \chA^{(2)}_p B_p^{(1)}} (\phi_{p}) ,
\\
\cX [\cS_k ] &= \sum_{U_{k+1} \in S_{\hL} } \delta_{\hL} \( \homega_k^{-1} \, \eta_{k-1}^{-1} \, \chalpha_{k+1} \, \eta_k \)
\prod_{p} \, H^{C_p^{(k-1)} \chA^{(k+1)}_s B_p^{(k)}} (\phi_{p}),
\qquad (2 \le k \le n-3)
\\
\cX [\cS_{n-2}] &= \sum_{U_{n-1} \,, U_n \in S_{\hL}} \delta_{\hL} \( \homega_{n-2}^{-1} \, \eta_{n-3}^{-1} \, \chalpha_{n-1} \, \chalpha_n \) 
\prod_{p} \, H^{C_p^{(n-3)} \chA^{(n-1)}_p \chA^{(n)}_p } (\phi_{p}) .
\end{aligned}
\label{def:cX patch}
\end{equation}
The equation \eqref{pants decomp npt} gives the pants decomposition of the $n$-point function addressed in \eqref{correlator as TFT}. The gluing operation consists of the sum over the defect factors $\{ \omega_f \}$, the residue operation and inverse metrics $\{ \zeta_{BC} \}$.

\bigskip
Note that the above discussion is different from TFT interpretation discussed in \cite{Kimura:2016bzo}.
There we required the invariance of the topological partition function under {\it any} cell decompositions of the two-dimensional surface, where each 2-cell is a strip.
In other words, the $n$-point functions should decompose into a collection of strips, where each strip collects the local Wick-contractions between $\cO_i$ and $\cO_j$\,. However, it turns out that $G_n$ in the $S_{\hL}$ formalism is compatible with such rewriting only if $n \le 4$, or equivalently as long as the complete graph $K_n$ is planar.
For $n \ge 5$, we find obstruction for rewriting, unless we assume certain commutation relation between the permutations for external operators.

\section{Skeleton reduction}\label{sec:reduction}

We will perform a skeleton reduction of Feynman graphs.

\begin{figure}[t]
\begin{center}
\includegraphics[scale=1.45]{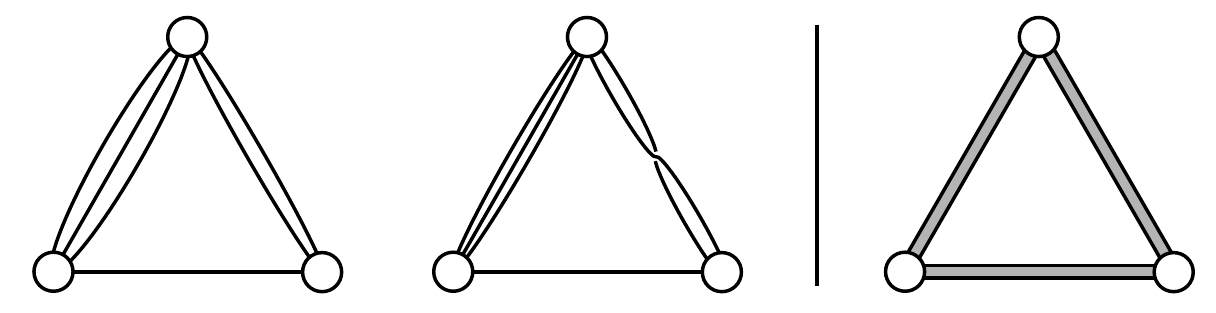}
\caption{Skeleton reduction of a Feynman graph. Planar Wick-contractions (left) and non-planar consecutive Wick-contractions (center) are reduced to the same skeleton graph (Right).}
\label{fig:skeleton-red}
\end{center}
\end{figure}

Let us first clarify what we mean by skeleton reduction.
One possible definition of the skeleton reduction is to make a new graph by assembling all edges connecting the same pair of vertices.
However, this definition is ambiguous for us, because the Feynman graphs in the double-line notation have the color structure coming from the cyclic ordering of edges around each vertex.
We define {\it skeleton} as a type of graphs which does not have faces made of two or fewer edges.
This definition is convenient because we will eventually relate the skeleton-reduced graph with the complex structure of a Riemann surface.
There are a few ways to reduce the original Feynman graphs satisfying this condition.
The skeleton reduction used in \cite{Gopakumar:2005fx,Bargheer:2017nne,Bargheer:2018jvq} assembles the planar, ladder-type internal Wick-contractions, which we call ladder-skeleton.
Our skeleton reduction includes non-planar internal Wick-contractions, which simplifies the topology of the reduced graphs.
Figure \ref{fig:skeleton-red} shows a simple example of skeleton reduction.

For simplicity, we consider only the $n$-point functions of single-traces in this section.
Our notation is summarized in Appendix \ref{app:skeleton}, and the implementation of the reduction procedure is explained in Appendix \ref{app:details skeleton}.

\subsection{Open two-point functions}\label{sec:open 2pt}

\begin{figure}[t]
\begin{center}
\includegraphics[scale=0.75]{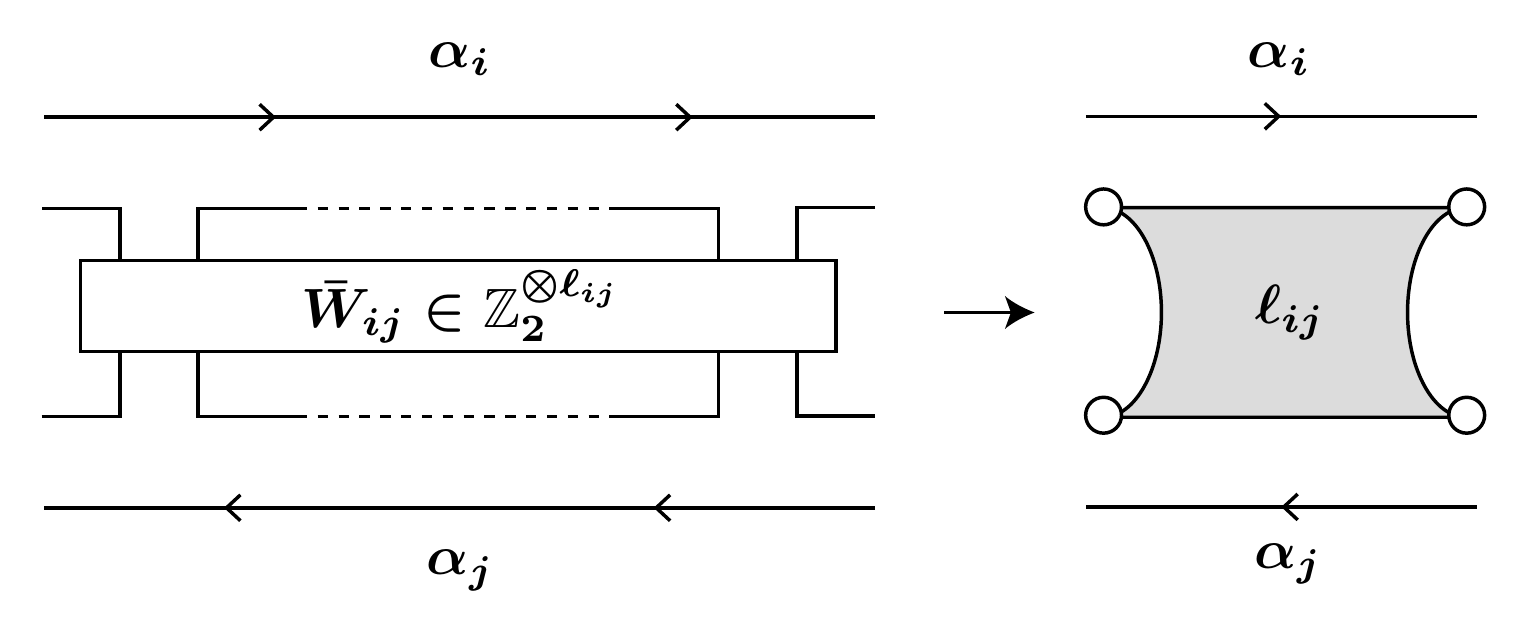}
\caption{Computing an open two-point function between $\cO_i$ and $\cO_j$ by taking the Wick-contractions of consecutive $\ell_{ij}$ fields. White circles represent the color indices at the end-points. Both $\alpha_i$ and $\alpha_j$ preserve the counterclockwise order.}
\label{fig:Open2pt}
\end{center}
\end{figure}

Let us denote a sequence of $\ell$ numbers by
\begin{equation}
\( p+1 \ \sim \ p+\ell \) \ \equiv \ \pare{ p+1, p+2, \dots, p+\ell } 
\label{def:sequence numbers}
\end{equation}
and a sequence of consecutive fields by
\begin{equation}
( {\bf \Phi}^{\bsA_{1 \sim \ell}} )^{a_1}_{a_{\ell+1}} = (\Phi^{A_1})^{a_1}_{a_2} \, (\Phi^{A_2})^{a_2}_{a_3} \, \dots (\Phi^{A_\ell})^{a_\ell}_{a_{\ell+1}} \,.
\label{def:sequence fields}
\end{equation}
We define open two-point functions by taking all possible Wick-contractions between a pair of sequences,
\begin{equation}
\contraction[1.5ex]{( }{{\bf \Phi}}{^{{\bsA}^{(i)}_{p +1 \sim p+\ell}} )^{a_{p+1}}_{a_{p+\ell+1}} \, 
( }{{\bf \Phi}}
( {\bf \Phi}^{{\bsA}^{(i)}_{p +1 \sim p+\ell}} )^{a_{p+1}}_{a_{p+\ell+1}} \, 
( {\bf \Phi}^{{\bsA}^{(j)}_{q +1 \sim q + \ell'}} )^{b_{q+1}}_{b_{q+\ell'+1}} 
= \delta_{\ell, \ell'} \, \sum_{\bW} \ \prod_{k=1}^\ell \, 
g^{A_{p+k} B_{q+k'}} \, 
\delta^{a_{p+k}}_{b_{q+k'+1}} \, \delta^{b_{q+k'}}_{a_{p+k+1}} \ 
\Big|_{q + k' = \bW^{-1} (p + k)} 
\label{def:open Wick 2pt}
\end{equation}
where $k' = \tau(k)$ for some $\tau \in S_\ell$\,. 
We sum $\bW$ over the space of ``local'' Wick-contractions
\begin{equation}
\bW \in \pare{ \prod_{k=1}^{\ell} \Bigl( p+k , q+ \tau(\ell-k) \Bigr) \, \Big| \, \tau \in S_\ell }, \qquad
\bW \in \bb{Z}_2^{\otimes \ell}, \quad
\bW^2 = 1.
\label{def:bW tau}
\end{equation}
We inverted the argument of $\tau$ by $k \mapsto \ell-k$ to keep track of the orientation of the whole graph. We assign the counterclockwise ordering to all external operators. 
This means that $\alpha_i$ and $\alpha_j$ rotate the color indices in the opposite directions, as explained graphically in Figure \ref{fig:Open2pt}.

Let us simplify the color structure of open two-point functions.
In \eqref{def:open Wick 2pt}, RHS contains $2 \ell$ color indices, while LHS has the color structure which we express by a matrix $R^{a_{p+1} \, b_{q+1}}_{a_{p+\ell+1} \, b_{q+\ell+1} }$\,. The remaining $2 ( \ell-1 )$ color indices should be contracted. Let us write
\begin{equation}
\begin{aligned}
\prod_{k=1}^\ell \, \delta^{a_{p+k}}_{b_{q+k'+1}} \, \delta^{b_{q+k'}}_{a_{p+k+1}} \ \Big|_{q + k' = \bW^{-1} (p + k)}
&= \prod_{k=1}^{\ell} \, \delta^{a_{p+k}}_{b_{\alpha_j \bW^{-1} (p+k)}} \, \delta^{b_{\bW^{-1}(p+k)}}_{a_{\alpha_i (p+k)}} 
\\
&\equiv N_c^{Z (\alpha_i \bW \alpha_j \bW^{-1} | \, a_{p+1} \,, b_{q+1} )} \,
R^{a_{p+1} \, b_{q+1}}_{a_{p+\ell+1} \, b_{q+\ell+1} } \,.
\end{aligned}
\label{def:R2pt}
\end{equation}
Here $Z ( \tilde\omega | a , b )$ counts the number of cycles in $\tilde\omega$ without $a$ or $b$\,.
To understand $Z ( \tilde\omega | a , b )$, recall that the permutation $\tilde\omega = \alpha_i \bW \alpha_j \bW^{-1}$ represents the faces of the part of the original Feynman graph.\footnote{Since $\bW$ is a set of local Wick-contractions as in \eqref{def:bW tau}, $\tilde \omega$ knows only the faces inside $\cO_i$ and $\cO_j$\,.}
Since all Wick-contractions are locally taken between $\cO_i$ and $\cO_j$\,, the repeated applications of $\omega$ send the upper indices $(a_{p+1}, b_{q+1})$ to the lower indices $(a_{p+\ell+1} , b_{q+\ell+1})$. 
In other words, we cannot go outside the interval between $(a_{p+1}, b_{q+1})$ and $(a_{p+\ell+1} , b_{q+\ell+1})$, because all edges are consecutive.
If we write
\begin{multline}
\tilde \omega = \Big( a_{p+1} \,, \omega (a_{p+1}) \,, \omega^2 (a_{p+1}) \,, \dots \,, c \,, \dots \Big) \ \times
\notag \\[1mm]
\Big( b_{q+1} \,, \omega (b_{q+1}) \,, \omega^2 (b_{q+1}) \,, \dots \,, c' \,, \dots \Big) \times \ 
(\text{the other internal cycles} )
\end{multline}
where $(c,c')$ is equal to $(a_{p+\ell+1} \,, b_{q+\ell+1})$ or $(b_{q+\ell+1} \,, a_{p+\ell+1})$, we find
\begin{equation}
Z (\tilde \omega | \, a_{p+1} \,, b_{q+1} ) \equiv ( \text{the number of the other internal cycles} ) .
\label{def:Z(wab)}
\end{equation}

By substituting \eqref{def:R2pt} into \eqref{def:open Wick 2pt}, we obtain
\begin{multline}
\contraction[1.5ex]{( }{{\bf \Phi}}{^{\bsA_{p +1 \sim p+\ell}} )^{a_{p+1}}_{a_{p+\ell+1}} \, 
( }{{\bf \Phi}} 
( {\bf \Phi}^{\bsA_{p +1 \sim p+\ell}} )^{a_{p+1}}_{a_{p+\ell+1}} \, 
( {\bf \Phi}^{\bsB_{q +1 \sim q + \ell'}} )^{b_{q+1}}_{b_{q+\ell'+1}} 
\\
= \delta_{\ell, \ell'} \, \sum_{\bW} \ \( \prod_{k=1}^{\ell} \, g^{A_{p+k} B_{\bW^{-1}(p+k)}} \)
N_c^{Z (\alpha_i \bW \alpha_j \bW^{-1} | \, a_{p+1} \,, b_{q+1} )} \,
R^{a_{p+1} \, b_{q+1}}_{a_{p+\ell+1} \, b_{q+\ell+1} } \,.
\label{open Wick 2pt-R}
\end{multline}

\subsubsection*{Powers of $N_c$ and the color $R$-matrix}

As a linear operator, the color matrix $R^{ab}_{a'b'}$ is just a product of two $\delta$-functions.
Its index structure is correlated with the powers of $N_c$ as
\begin{equation}
N_c^{Z (\alpha_i \bW \alpha_j \bW^{-1} | \, a_{p+1} \,, b_{q+1} )} \ R^{a_{p+1} \, b_{q+1}}_{a_{p+\ell+1} \, b_{q+\ell+1} }
= \begin{cases}
N_c^{\ell - 2n + 1} \ \delta^{a_{p+1}}_{b_{q+\ell+1}} \, \delta^{b_{q+1}}_{a_{p+\ell+1}} 
\\[1mm]
N_c^{\ell - 2n} \ \delta^{a_{p+1}}_{a_{p+\ell+1}} \, \delta^{b_{q+1}}_{b_{q+\ell+1}} 
\end{cases}
\quad (\text{for some } n \in \bb{Z}_{\ge 1}) .
\label{Rab selection rule}
\end{equation}
We can prove this selection rule as follows. If we multiply $\delta_{a_{p+1}}^{a_{p+\ell+1}} \, \delta_{b_{q+1}}^{b_{q+\ell+1}}$\,, the equation \eqref{Rab selection rule} becomes part of the two-point function of single-traces, written as
\begin{equation}
\delta_{a_{p+1}}^{a_{p+\ell+1}} \, \delta_{b_{q+1}}^{b_{q+\ell+1}}
\prod_{k=1}^{\ell} \, \delta^{a_{p+k}}_{b_{\alpha_j \bW^{-1} (p+k)}} \, \delta^{b_{\bW^{-1}(p+k)}}_{a_{\alpha_i (p+k)}} 
= N_c^{C (\alpha'_i \bW \alpha'_j \bW^{-1}) }, \qquad 
(\alpha'_i \,, \alpha'_j \in \bb{Z}_{\ell+1} ).
\end{equation}
Since $\alpha'_i$ and $\alpha'_j$ have the same cycle type and belong to the same conjugacy class, we may redefine $\bar W$ to rewrite
\begin{equation}
\alpha'_i \bW \alpha'_j \bW^{-1} = \alpha'_i \, \bW' \, \alpha'_i{}^{-1} \, \bW'{}^{-1} .
\end{equation}
This is an element of the commutator subgroup of $S_{\ell+1}$\,, namely $A_{\ell+1}$\,, consisting of an even number of transpositions. Thus the power of $N_c$ changes only by an even number when $\bW'$ changes, which translates into \eqref{Rab selection rule}.

\begin{figure}[t]
\begin{center}
\includegraphics[scale=0.65]{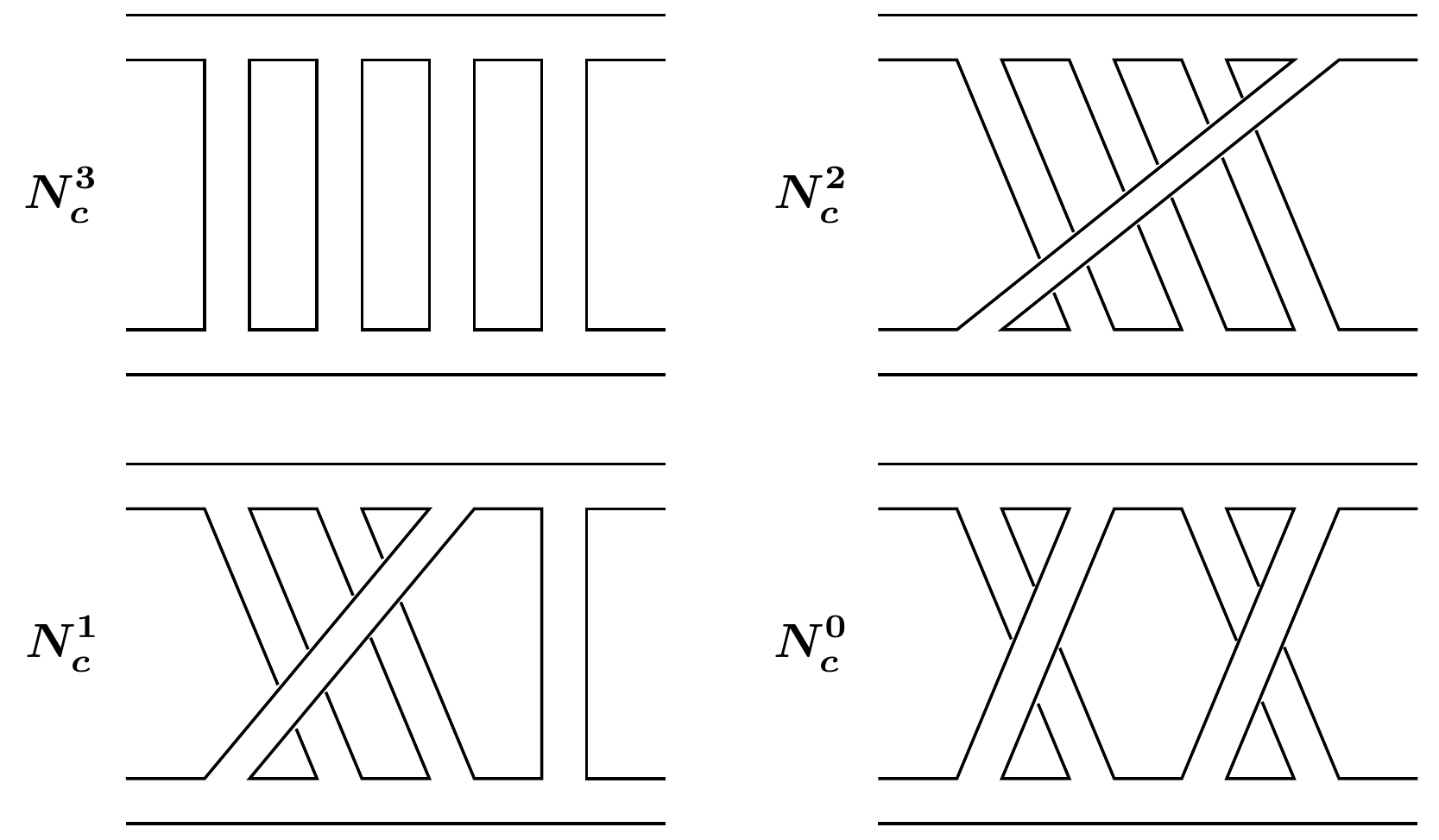}
\caption{Open two-points at $\ell=4$. We find $\tau = (1)(2)(3)(4)$ and $(1234)$ (top left and right), $(123)(4)$ and $(12)(34)$ (bottom left and right); 
$R^{ab}_{a'b'}=\delta^a_{b'} \, \delta^b_{a'}$ and $\delta^a_{a'} \, \delta^b_{b'}$ for left and right figures.}
\label{fig:openZR}
\end{center}
\end{figure}

The powers of $N_c$ are also determined by $\tau \in S_\ell$ in \eqref{def:bW tau}, and we can rewrite
\begin{equation}
N_c^{\bZ (\tau) } \equiv
N_c^{Z (\alpha_i \bW \alpha_j \bW^{-1} | \, a_{p+1} \,, b_{q+1} )} .
\label{def:bZ tau}
\end{equation}
Some examples are shown in Figure \ref{fig:openZR}. $\bZ$ is maximal if $\tau$ is identity,
\begin{equation}
\tau = {\bf 1} \in S_\ell \qquad \Leftrightarrow \qquad
\bZ ( \tau ) = \ell-1
\label{compute Z g=0}
\end{equation}
which corresponds to the planar case. The $O(1/N_c)$ terms come from cyclic permutations,
\begin{equation}
\tau \in \bb{Z}_\ell \,, \ \tau \neq {\bf 1} \qquad \Leftrightarrow \qquad
\bZ ( \tau ) = \ell-2.
\label{compute Z g=1}
\end{equation}
The $O(1/N_c^2)$ terms come from cyclic permutations of lower order with consecutive numbers,
\begin{equation}
\tau = (s , s+1 , \dots,  s+k) \in \bb{Z}_k \ (k < \ell), \ \tau \neq {\bf 1} 
\quad \Leftrightarrow \quad
\bZ ( \tau ) = \ell-3.
\label{compute Z g=2}
\end{equation}

\subsection{Relabeling Wick-contractions}\label{sec:Wick relabel}

The skeleton reduction reorganizes the structure of Wick-contractions.
This structure can be explained in two ways, by using $W_{ij,\rho}$ or equivalently $(V_{ij,\rho} \,, \tau_{ij,\rho})$.

In the first explanation, we introduce $\{ \ell_{ij}^{(\rho)} \}$, where $\rho =1,2, \dots, r_{ij}$ labels non-consecutive sets of Wick-edges between $\cO_i$ and $\cO_j$\,. 
In other words, $r_{ij}$ counts the number of open two-point functions between $\cO_i$ and $\cO_j$\,. 
They satisfy
\begin{equation}
\ell_{ii}^{(\rho)} = 0 , \qquad
\ell_{ij}^{(\rho)} = \ell_{ji}^{(\rho)} , \qquad
\ell_{ij} = \sum_{\rho=1}^{r_{ij}} \ell_{ij}^{(\rho)} , \qquad
\sum_{i<j}^n \sum_{\rho=1}^{r_{ij}} \ell_{ij}^{(\rho)} = \hL ,
\label{sum lij-rho}
\end{equation}
which generalizes \eqref{Wicknum total n-pt}. Note that planar graphs may have $r_{ij}>1$.
In this notation, $G_n$ becomes
\begin{equation}
G_n = \sum_{ \{ \ell_{ij}^{(\rho)} \} }
\prod_{i<j}^n \, \prod_{\rho=1}^{r_{ij}}
\sum_{ W_{ij,\rho} } \, 
\prod_{p=1}^{\ell_{ij}^{(\rho)}} \[ \, g^{A^{(i)}_p A^{(j)}_q } \,
\delta_{b_{\alpha_j (q)}}^{a_p} \,
\delta_{a_{\alpha_i(p)}}^{b_q} \]_{q= W_{ij,\rho}^{-1} (p)} 
\label{npt Wick lij-rho}
\end{equation}
where $\{ W_{ij, \rho} \}$ is completely equivalent to the original $W$,
\begin{equation}
\prod_{i<j}^n \ \prod_{\rho=1}^{r_{ij}} W_{ij, \rho}
= W \, \in \, \bb{Z}_2^{\otimes \hL} \subset S_{2 \hL} \,.
\end{equation}
Each $W_{ij, \rho}$ takes the form
\begin{equation}
W_{ij,\rho} = \prod_{k=1}^{\ell_{ij}^{(\rho)} } \Big(p+k , q+\tau_{ij,\rho} (\ell_{ij}^{(\rho)}-k) \Big) 
\label{def:tau ij-r}
\end{equation}
as in \eqref{def:bW tau}.

The second explanation was given in Introduction.
We split $\cO_i$ into the sequential fields $\{ \bsPhi^{(i)}_r \}$ in \eqref{def:sequence fields}, and call it a partition of $\cO_i$\,,
\begin{equation}
\cO_i = \tr_{L_i} \( \alpha_i \, \Phi^{A_1^{(i)}} \Phi^{A_2^{(i)}} \dots \Phi^{A_L^{(i)}} \)
\quad \stackrel{{\rm partition}}{\longrightarrow} \quad
\olcO_i = \tr_{\bL_i} \( \bar \alpha_i \,  \bsPhi^{(i)}_1 \,  \bsPhi^{(i)}_2 \, \dots \, \bsPhi^{(i)}_{\bL_i} \).
\label{def:partition of cO_i}
\end{equation}
We take open two-point functions as the internal Wick-contractions between $\bsPhi^{(i)}_p$ and $\bsPhi^{(j)}_q$\,.
Now the $n$-point functions contain three sums. 
\begin{itemize}[nosep,leftmargin=16mm]
\item[i)] Sum over the partitions $\cO_i$ into $\{ \olcO_i \}$.
\item[ii)] Sum over the {\it external} Wick-contractions, for all possible pairings $\pare{ (i,p) , (j,q) }$ of the sequential fields $\pare{ \bsPhi^{(i)}_p \,, \bsPhi^{(j)}_q }$.
\item[iii)] Sum over the {\it internal} Wick-contractions, for each $\vev{ \bsPhi^{(i)}_p \, \bsPhi^{(j)}_q }$ as in \eqref{open Wick 2pt-R}.
\end{itemize}
The choice of $\tau_{ij,\rho}$ in \eqref{def:tau ij-r} is equivalent to the internal Wick-contraction.

\begin{figure}
\begin{center}
\includegraphics[scale=0.75]{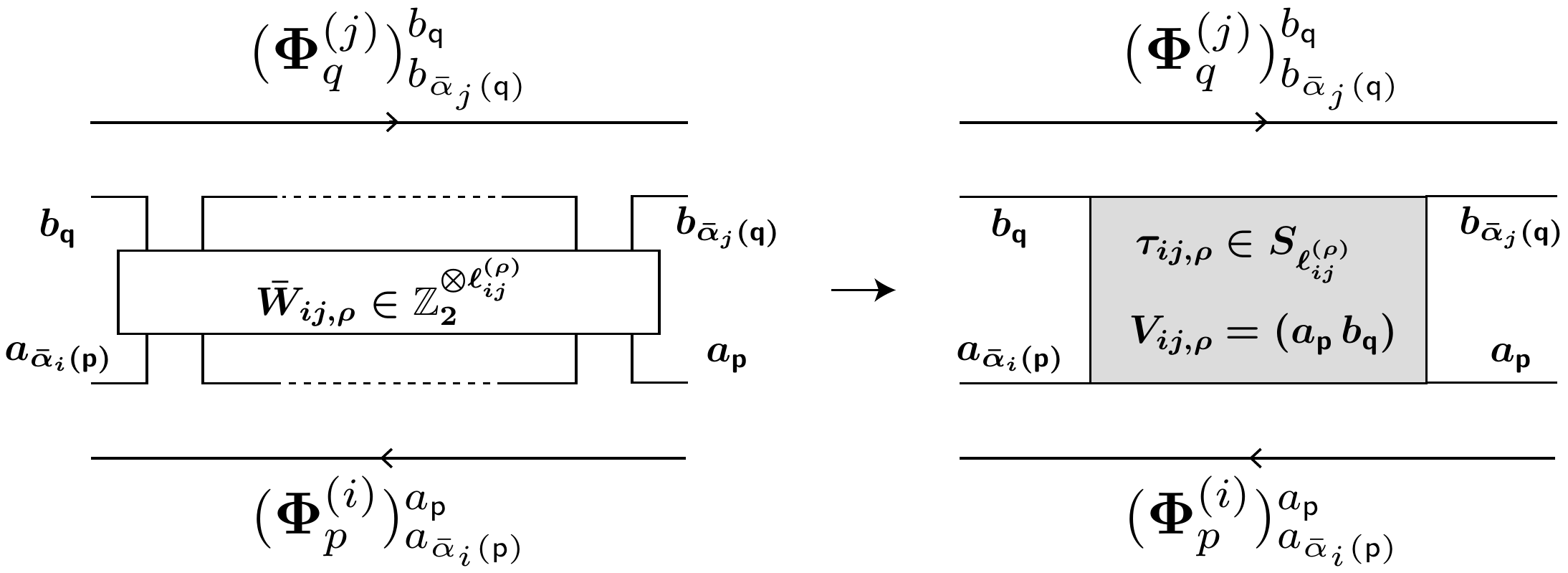}
\caption{Rewriting the open two-point function in Figure \ref{fig:Open2pt} in terms of $(V, \tau)$. 
The color indices at the four end-points are denoted by $\( a_{\sfp} \,, b_{\sfq} \,, a_{\bar \alpha_i (\sfp)} \,, b_{\bar \alpha_j (\sfq)} \)$.}
\label{fig:openVtau}
\end{center}
\end{figure}

\bigskip
For a given partition $\{ \olcO_i \}$, the equivalence of the two explanations can be understood as follows.
We relabel the color indices of $\cO_i$ as
\begin{equation}
p \in \pare{ 1, 2, \dots , L_i } \ = \ 
\Bigl\{ ( \sfp, \bp) \, \Big| \, \bp \in \pare{ 1,2, \dots, \ell_{\sfp} },
\sfp \in \pare{ 1, 2, \dots, \bL_i } \Bigr\}
\label{def:P to sfP}
\end{equation}
where $\sfp$ labels the open end-points and $\bp$ labels the internal indices. 
The equation \eqref{def:partition of cO_i} can be written as
\begin{equation}
\cO_i = \prod_{p=1}^{L_i} \, (\Phi^{A_p})^{a_p}_{a_{\alpha_i (p)}}
\quad \stackrel{{\rm partition}}{\longrightarrow} \quad
\olcO_i 
= \prod_{\sfp=1}^{\bL_i} \, (\bsPhi^{(i)}_p)^{a_{\sfp}}_{a_{\bar \alpha_i (\sfp)}} 
= \prod_{\sfp=1}^{\bL_i} \ 
\prod_{\bp=1}^{\ell_\sfp} (\Phi^{A_{\sfp, \bp}})^{a_{\sfp + \bp -1 }}_{a_{\sfp + \bp} } .
\label{to partition operator}
\end{equation}
Here the reduced color permutation $\bar \alpha_i \in S_{\bL_i}$ shifts the open end-points $\{ \sfp \}$, and thus $a_{\bar \alpha_i (\sfp)} = a_{\sfp + \ell_\sfp}$\,.
We write the Wick-contractions in the new notation as
\begin{equation}
W \, \ni \, (a_p \,, b_q) = (a_{\sfp, \bp} \,, b_{\sfq, \bq} ) \quad \leftrightarrow \quad
(a_\sfp, b_\sfq) \times (\bp, \bq) \, \in \, (V, \tau).
\label{def:W to Vtau}
\end{equation}
The new permutation $V$ is equivalent to the external Wick-contraction. 
The relation between $W_{ij,\rho}$ and $(V_{ij,\rho} \,, \tau_{ij,\rho})$ is also depicted in Figure \ref{fig:openVtau}.

The open two-point function \eqref{open Wick 2pt-R} can be rewritten as
\begin{equation}
\contraction[1.5ex]{( }{\bsPhi^{(i)}_p}{ )^{a_{\sfp}}_{a_{\bar \alpha_i (\sfp)}} \, ( }{\bsPhi^{(j)}_q}
( \bsPhi^{(i)}_p )^{a_{\sfp}}_{a_{\bar \alpha_i (\sfp)}} \, 
( \bsPhi^{(j)}_q )^{b_{\sfq}}_{b_{\bar \alpha_j (\sfq)}} 
= \delta_{\ell_\sfp \,, \ell_\sfq } \, \sum_{\tau \in S_{\ell_\sfp} } \ \( \prod_{k=1}^{\ell_\sfp } \, h^{A_{\sfp, \bk} B_{\sfq, \tau (\bk)} } \)
N_c^{\bZ (\tau)} \ 
R^{a_{\sfp} \, b_{\sfq}}_{a_{\bar \alpha_i (\sfp) } \, b_{\bar \alpha_j (\sfq)}} \ \Big|_{b_\sfq = V(a_\sfp)} 
\label{open Wick 2pt-R relabel}
\end{equation}
where $\bZ (\tau)$ is given in \eqref{def:bZ tau}, and we replaced $g^{AB}$ by $h^{AB}$ in \eqref{def:hAB} for a later purpose.
In this notation, $G_n$ is given by
\begin{equation}
G_n = \sum_{ \{ \text{partition of $\cO_i$} \} } \,
\sum_{V : \, {\rm external}} \, 
\sum_{\tau : \, {\rm internal}} \, 
\prod_{i<j}^n \prod_{\rho=1}^{r_{ij}} \[
\( \prod_{k=1}^{\ell_{ij}^{(\rho)} } \, h^{A^{(i)}_{\sfp, \bk} A^{(j)}_{\sfq, \tau (\bk)} } \)
N_c^{\bZ (\tau_{ij,\rho} )} \ 
R^{a^{(i)}_{\sfp} \, a^{(j)}_{\sfq}}_{a^{(i)}_{\bar \alpha_i (\sfp) } \, a^{(j)}_{\bar \alpha_j (\sfq)}} 
\]
\label{npt Wick lij-rho Vtau}
\end{equation}
where
\begin{equation}
\bb{Z}_2^{\otimes \bL} \ni V = \prod_{i<j}^n \prod_{\rho} V_{ij,\rho} 
= \prod_{i<j}^n \prod_{\rho} (a^{(i)}_{\sfp} \, a^{(j)}_{\sfq}), \qquad
\tau = \prod_{i<j}^n \prod_{\rho} \tau_{ij,\rho} \,.
\end{equation}
In Appendix \ref{app:vertex skeleton}, we simplify \eqref{npt Wick lij-rho Vtau} including the range of summations.

\subsection{Face-based skeleton formula}\label{sec:face skeleton}

We discuss how to reconstruct the $n$-point function of single-trace operators from the graph data, following our discussion in Section \ref{sec:face-based formula}.
In \eqref{npt reduced formula vertex} we fix $\bar\alpha$ and take a sum over $V$. Below we fix $V$ and sum over the face permutation $\nu$ of a skeleton graph.
We also need the set of internal Wick-contractions $\bstau$ to recover the full $n$-point function.

\subsubsection{Sum over the skeleton faces}\label{sec:sum skeleton faces}

To begin with, let us explain how to sum over the faces of skeleton graphs.
We choose a partition
\begin{equation}
\bsl = \(\ell_1 \,, \ell_2 \,, \dots \,, \ell_{\bL} \) \vdash \hL .
\label{fix partition bsl}
\end{equation}
We need $\bL \ge n/2$ for general $n$-point functions, and $\bL \ge n-1$ for connected $n$-point.
We impose the ordering
\begin{equation}
\ell_1 \ge \ell_2 \ge \dots \ge \ell_{\bL} > 0 .
\end{equation}
Since the ordering of $\{ \bsl_i \}$ in \eqref{fix partition bsl} is unimportant, we use $\bsl$ and $[\bsl]$ interchangeably.
Let us denote the $p$-th edge of length $\ell$ by $p_{\ell}$\,, and use $\{ 1_{\ell_1}, 1'_{\ell_1}, \dots, 
\bL_{\ell_{\bL}}, \bL'_{\ell_{\bL}} \}$ to label the edges of a skeleton graph. We set
\begin{equation}
V_\bullet \equiv \prod_{\bp=1}^{\bL} \( p_{\ell_p} \,\, p'_{\ell_p} \) .
\label{canonical V}
\end{equation}
We denote by $\bsr_k$ the number of consecutive Wick-contractions with $k$ fields as in \eqref{def:partition of hL}.
Since there are $\bsr_k$ skeleton edges with length $k$, the symmetry of relabeling is
\begin{equation}
{\rm Aut} \, V = {\rm Aut} \, V (\bsl) \equiv \bigotimes_k S_{\bsr_k} [\bb{Z}_2] \ \subset \ S_{\bL} [\bb{Z}_2] .
\label{def:Aut V}
\end{equation}

There is a bijection between the set of unlabeled ladder-skeleton graphs and face permutations generalizing \eqref{graph coset equivalence},\footnote{The ladder-skeleton graph is made by assembling the planar, ladder-type consecutive Wick-contractions.}
\begin{equation}
\pare{ \text{Ladder-skeleton graphs with $2\bL$ unlabeled vertices} } \quad \leftrightarrow \quad 
\nu \in S_{2 \bL}^{\times \! \times} / {\rm Aut} \, V .
\label{ladder-skeleton graph coset equivalence}
\end{equation}
Here we defined the group $S_{2 \bL}^{\times \! \times} \subset S_{2 \bL}$ without one- or two-cycles,
\begin{equation}
\nu \in S_{2 \bL}^{\times \! \times} \qquad \Leftrightarrow \qquad
p_{\ell} \neq \nu (p_{\ell}) \neq \nu^2 (p_{\ell}) \quad (\forall p_{\ell}).
\label{def:S2bL times}
\end{equation}
In RHS of \eqref{ladder-skeleton graph coset equivalence}, $\nu$ permutes $\{ 1_{\ell_1}, 1'_{\ell_1}, \dots, \bL_{\ell_{\bL}}, \bL'_{\ell_{\bL}} \}$, where $p_{\ell}$ means the $p$-th edge of length $\ell$.
Due to the redundancy of relabeling, $\nu$ is in one-to-one correspondence with the coset $S_{2 \bL}^{\times \! \times} / {\rm Aut} \, V$. 
The correspondence \eqref{ladder-skeleton graph coset equivalence} can be proven as in Section \ref{sec:face-based formula}.
Following Appendix \ref{app:external Wick space}, we define the space $(S_{2 \bL}^{\times \! \times})_{\rm phys}$ as
\begin{equation}
\pare{ \text{Skeleton graphs with $2\bL$ unlabeled vertices} } \quad \leftrightarrow \quad 
\nu \in \( S_{2 \bL}^{\times \! \times} / {\rm Aut} \, V \)_{\rm phys} 
\label{skeleton graph coset equivalence}
\end{equation}
where we exclude the mutually adjacent pairs in RHS.
Unfortunately, we do not find a simple algebraic characterization of $\( S_{2 \bL}^{\times \! \times} / {\rm Aut} \, V \)_{\rm phys}$. 

\begin{figure}[t]
\begin{center}
\includegraphics[scale=0.6]{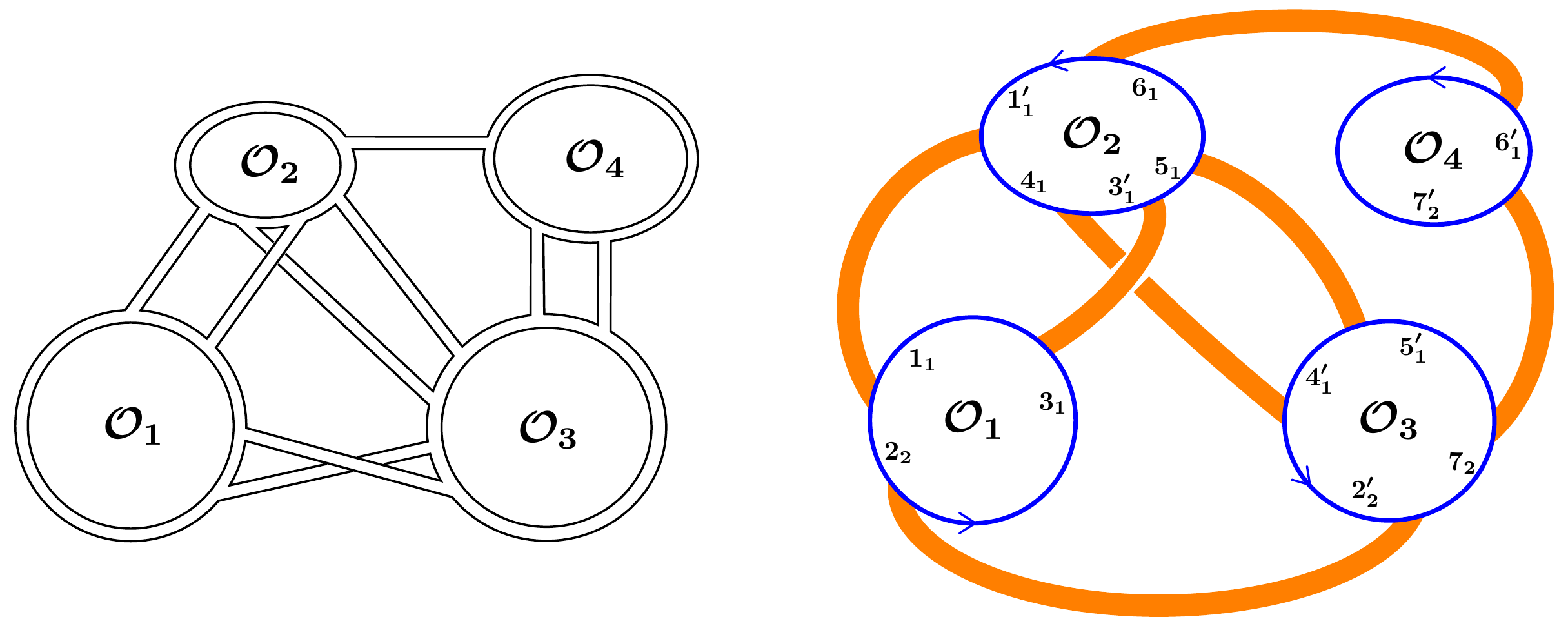}
\caption{Skeleton reduction of a set of the Wick-contractions in $\Vev{ \cO_1 \, \cO_2 \, \cO_3 \, \cO_4 }$ with $(L_1,L_2,L_3,L_4)=(4,5,6,3)$. These graphs are the same as in Figure \ref{fig:ReducedGraph}.}
\label{fig:FaceReducedGraph}
\end{center}
\end{figure}

\subsubsection{Reconstruction of correlators}\label{sec:reconstruction}

Let us explain how to reconstruct correlators from a skeleton graph.
We set $\bar\alpha_\bullet \equiv \nu V_\bullet  \in S_{2 \bL}$ and use $\bar\alpha_\bullet$ to make all skeleton-rims oriented.
We denote the cycle decomposition of $\bar\alpha_\bullet$ by
\begin{equation}
\bar\alpha_\bullet = \( \bar\alpha_1 \,, \bar\alpha_2 \,, \dots \,, \bar\alpha_n \), \qquad
(\bar \alpha_i \in \bb{Z}_{\bL_i} ).
\end{equation}
Parameterize each cycle by using skeleton edges as $\bar\alpha_i = (p^{(i,1)} \, p^{(i,2)} \, \dots \, p^{(i,\bL_i)})$, where $p^{(i,k)}$ is a shorthand of $p_{\ell_{p(i,k)}}$ or $p'_{\ell_{p(i,k)}}$ introduced in \eqref{canonical V}.
Define
\begin{equation}
\Delta (\bar\alpha_i) = \sum_{k=1}^{\bL_i} \ell_{p(i,k)} 
\end{equation}
which should count the number of scalars in $\cO_i$\,.
We impose the conditions that graph data should match those of physical data,
\begin{equation}
n = n_{\rm ex} \,, \quad
\Delta (\bar\alpha_i)  = L_{{\rm ex},i} 
\qquad {\rm for} \quad
\Vev{ \cO_1 \cO_2 \dots \cO_{n_{\rm ex}} }, \quad
( \alpha_i \in S_{L_{{\rm ex},i} } ).
\label{physical constraints}
\end{equation}
These conditions determine the cycle type of $\alpha = \prod_i \alpha_i \in S_{2\hL}$\,. Thus, the equations \eqref{physical constraints} rewrite the constraint $\delta_{\text{cycle-type}} ( \alpha_{\rm ex}^{-1} \,, \omega W_\bullet )$ used in \eqref{n-pt S*2hL cycle type}. 
Figure \ref{fig:FaceReducedGraph} shows an example of the skeleton reduction using the notation introduced here.

Next, we generate a set of ordered lists from the permutation $\alpha_\bullet$ like \eqref{example:cycle to list}. 
In \eqref{abs CL lambda} we defined the map ${\bf CL}$ as the wreath product, namely the composition of the cyclic translations and the permutations of the cycles of the same length. 
Here we consider the cyclic translations and permutations separately.

Let us define the map from a permutation to a set of the ordered reduced lists, denoted by ${\bf CR}$.
This consists of the permutations of the operators with the same dimensions $\Delta (\bar\alpha_i) = \Delta (\bar\alpha_j)$. 
For example, when $\bar\alpha_\bullet = (1_2 \, 2_1)(1'_2 \, 3'_1)(2'_1 \, 3'_1)$ we get
\begin{equation}
{\bf CR} (\bar\alpha_\bullet) = \Bigl\{
(1_2 \,, 2_1)(1'_2 \,, 3'_1)(2'_1 \,, 3'_1), \ 
(1'_2 \,, 3'_1)(1_2 \,, 2_1)(2'_1 \,, 3'_1)
\Bigr\} ,
\label{example CR}
\end{equation}
because $\bar\alpha_1$ and $\bar\alpha_2$ having $\Delta (\bar\alpha_1) = \Delta (\bar\alpha_2) = 3$ can be interchanged.
The map ${\bf CR}$ depends only on the operator dimensions $\Delta (\bar\alpha_i) = \Delta (\bar\alpha_j)$, and they may have $\bL_i \neq \bL_j$\,.
If we denote the cycle type of $\alpha$ by $\lambda = [1^{\lambda_1} 2^{\lambda_2} \dots ] \vdash 2 \hL$, then
\begin{equation}
\abs{ {\bf CR} (\lambda) } = \Big| \prod_k S_{\lambda_k} \Big|
= \prod_k \lambda_k!  \,.
\label{abs CR lambda}
\end{equation}
The cyclic translations will be discussed later.

Choose an element $\bar \beta = (\bar\beta_1\,, \dots \,, \bar\beta_n)$ from ${\bf CR} (\bar\alpha_\bullet)$, and use it to relate the rim labels to the flavor-like indices
\begin{equation}
\bar\beta_i : \{ 1_{\ell_1}, 1'_{\ell_1}, \dots, \bL_{\ell_{\bL}}, \bL'_{\ell_{\bL}} \} \ \to \ \{ 0,1,2,\dots, \bar L_i \} 
\label{def:bar beta_i}
\end{equation}
such that $\bar\beta_i^{-1} (x)$ is uniquely defined if and only if $x \neq 0$.
We parameterize $\bar\alpha$ as
\begin{equation}
\bar\alpha 
= \begin{pmatrix}\bar \alpha_1 \\ \bar\alpha_2 \\ \vdots \\ \bar\alpha_n \end{pmatrix} 
= \begin{pmatrix}
\( \bar\beta_1^{-1} (1) \, \dots \, \bar\beta_1^{-1} (\bL_1) \) \\
\( \bar\beta_2^{-1} (1) \, \dots \, \bar\beta_2^{-1} (\bL_2) \) \\
\vdots \\ 
\( \bar\beta_n^{-1} (1) \, \dots \, \bar\beta_n^{-1} (\bL_n) \) 
\end{pmatrix} .
\end{equation}
We can better understand this notation by introducing the extended reduced operator
\begin{equation}
\hat \olcO_i = \tr \( \bar \alpha_i \,  \bsPhi^{(i)}_1 \,  \bsPhi^{(i)}_2 \, \dots \, \bsPhi^{(i)}_{\bL_i} \)
\times \tr ( {\bf 1} )^{\bL - \bL_i} 
\label{def:extended reduced operator}
\end{equation}
where the length of the sequential fields $\bsPhi^{(i)}_r$ is found from $\bar\beta_i^{-1} (r) \in \{ 1_{\ell_1}, 1'_{\ell_1}, \dots, \bL_{\ell_{\bL}}, \bL'_{\ell_{\bL}} \}$ for $r \neq 0$.
The total number of scalar fields in $\hat \olcO_i$ is given by
\begin{equation}
L_i = \sum_{E \ {\rm from} \ \pare{ \bar\beta_i^{-1} (1), \, \dots \, , \bar\beta_i^{-1} (\bL_i) } } \ell_E \,.
\label{def:Li from beta}
\end{equation}
We construct the reduced Wick-contraction matrix by
\begin{equation}
\bar\mu_{\bar\beta} = \begin{pmatrix}
\bar\mu_{1,1} & \bar\mu_{1, 1'} & \dots & \bar\mu_{1, \bL} & \bar\mu_{1, \bL'} \\ 
\bar\mu_{2,1} & \bar\mu_{2,1'} & \dots & \bar\mu_{2, \bL} & \bar\mu_{2, \bL'} \\ 
\vdots \\ 
\bar\mu_{n,1} & \bar\mu_{n,1'} & \dots & \bar\mu_{n, \bL} & \bar\mu_{n, \bL'} 
\end{pmatrix}, \qquad
\bar\mu_{i,s} = 
\begin{cases}
\bsA^{(i)}_{\bar\beta_i (s)} &\quad (\bar\beta_i (s) \neq 0) \\
{\bf 1} &\quad (\bar\beta_i (s) = 0).
\end{cases}
\end{equation}
The reduced Wick-contraction matrix $\bar\mu_{\bar\beta}$ is in one-to-one correspondence with the external Wick-contraction of the reduced operators. For example, we find
\begin{multline*}
\contraction{\Big\langle \tr (}{\bsPhi_1^{(1)}}{ \bsPhi_2^{(1)} ) \, \tr (}{\bsPhi_1^{(2)}}%
\contraction[2ex]{\Big\langle \tr (\bsPhi_1^{(1)} }{\bsPhi_2^{(1)}}{ ) \, \tr (\bsPhi_1^{(2)} \bsPhi_2^{(2)} ) \, \tr (}{\bsPhi_1^{(3)}}%
\contraction{\Big\langle \tr (\bsPhi_1^{(1)} \bsPhi_2^{(1)} ) \, \tr (\bsPhi_1^{(2)} }{\bsPhi_2^{(2)}}{ ) \, \tr (\bsPhi_1^{(3)} }{\bsPhi_2^{(3)}}%
\Big\langle \tr (\bsPhi_1^{(1)} \bsPhi_2^{(1)} ) \, \tr (\bsPhi_1^{(2)} \bsPhi_2^{(2)} ) \, \tr (\bsPhi_1^{(3)} \bsPhi_2^{(3)} ) \Big\rangle \quad \leftrightarrow \quad 
\\[2mm]
\begin{pmatrix}
\bsA_{1 \sim \bsl_{1|1}} & & \bsA_{\bsl_{1|1}+1 \sim L_1} & & & \\
& \bsB_{1 \sim \bsl_{2|1}} & & & \bsB_{\bsl_{2|1}+1 \sim L_2} & \\
& & & \bsC_{1 \sim \bsl_{3|1}} & & \bsC_{\bsl_{3|1}+1 \sim L_3} 
\end{pmatrix} . 
\label{hatbL embedding}
\end{multline*}
We introduce the flavor factor for the $E$-th edge by
\begin{equation}
h \( \bstau \, \Big| \, \bsA^{(i)}_{\sfp} , \bsA^{(j)}_{\sfq} \) = 
\prod_{\bk=1}^{\ell_E } h^{\bsA^{(i)}_{\sfp, \bk} \, \bsA^{(j)}_{\sfq,\tau_E( \bk)}} \,.
\label{def:hBB cons}
\end{equation}
which agrees with the flavor factor in \eqref{open Wick 2pt-R relabel}.
We generalize $h$ as
\begin{equation}
h \( \bstau \, \Big| \, \bsA_1 \, \bsA_{1'} \dots \bsA_n \, \bsA_{n'}\) = \frac{1}{2 (2n-2)!} \, 
\sum_{\pi \in S_{2n}} h \( \bstau \, \Big| \, \bsA_{\pi(1)} \,, \bsA_{\pi(1')} \) 
\prod_{j=2}^{n} \delta^{\bsA_{(\pi(j))}}_{\bf 1} \, \delta^{\bsA_{(\pi(j'))}}_{\bf 1} \,.
\label{def:hA1A2n W}
\end{equation}
like \eqref{def:hA1A2n}.
We define the flavor function of $\bar\mu_{\bar\beta}$ by
\begin{equation}
{\bf F} \( \bstau |\, \bar\mu_{\bar\beta} \) = \prod_{s=1}^{\bL} \, 
h \( \bstau \, \Big| \, \bar\mu_{1,s} \ \bar\mu_{1, s'} \ \bar\mu_{2,s} \ \bar\mu_{2,s'} \, \dots \, 
\bar\mu_{n,s} \ \bar\mu_{n,s'} \)
\label{def:bfF on bmu}
\end{equation}
where RHS is a generalization of \eqref{def:hA1An}.

Now we consider the cyclic translations $\bb{Z}_\alpha = \prod_i \bb{Z}_{L_i}$\,.
In fact, although the formula \eqref{def:bfF on bmu} computes the flavor factor for a set of reduced operators $\olcO_i$\,, we did not specify the relation between $\olcO_i$ and $\cO_i$\,
Let us rewrite $h ( \bstau \, | \, \bsA^{(i)}_{r} , \bsA^{(j)}_{s} )$ in \eqref{def:hBB cons} as
\begin{equation}
h \( \bstau \, \Big| \, \bsA^{(i)}_{\sfp} , \bsA^{(j)}_{\sfq} \) = 
\prod_{k=1}^{\ell_E } h^{\bsA^{(i)}_{p_0 +k} \, \bsA^{(j)}_{q_0 +\tau_E(\ell_E - k)}} \,,
\quad (\text{for some } \ p_0, q_0)
\end{equation}
using \eqref{def:bW tau}.
The flavor index $p_0  \in \{ 1,2, \dots , L_i \}$ determines the relation between the reduced operator $\olcO_i$ in \eqref{to partition operator} and the original operator $\cO_i$ in \eqref{def:pm basis}.
The flavor factors with different $p_0$ are related by the cyclic translations $\bb{Z}_\alpha$
\begin{equation}
\bb{Z}_\alpha \, \ni \, \bsz = (z_1, z_2, \dots, z_n), \qquad
z_i \, : \, p_0 \mapsto p_0 +1 \quad ({\rm mod} \ L_i).
\label{action of Zalpha on flavor}
\end{equation}
This symmetry was also part of the stabilizer \eqref{CL as Stab alpha}.
We write
\begin{equation}
{\bf F} ( \bstau | \bsz \cdot \bar\mu_{\bar\beta} ) \equiv
\prod_{s=1}^{\bL} \bsz \cdot 
h \( \bstau \, \Big| \, \bar\mu_{1,s} \ \bar\mu_{1, s'} \ \dots \, \bar\mu_{n,s} \ \bar\mu_{n,s'} \),
\qquad
(\bsz \in \bb{Z}_\alpha)
\end{equation}
and take a sum over the entire orbit of $\bb{Z}_\alpha$\,.

\bigskip
Finally we define the function on the skeleton-faces by
\begin{equation}
\bb{F} ( \bstau |\, \nu ) = 
N_c^{\bZ (\bstau) + C(\bar\omega)} \,
\sum_{\bar\beta \in {\bf CR} (\bar\alpha_\bullet)} 
\sum_{ \bsz \in \bb{Z}_\alpha} 
{\bf F} ( \bstau | \bsz \cdot \bar\mu_{\bar\beta} ) .
\label{gen: construct red F}
\end{equation}
We count the powers of $N_c$ in the same way as \eqref{identity face function} in Section \ref{app:vertex skeleton}.
Namely, we compute $\theta = \theta (\bstau)$ in \eqref{def:theta} and count the number of cycles in $\bar\omega = \nu \theta$. The non-planarity inside open two-points also contribute through $\bZ (\bstau)$ in \eqref{def:bZ tau-all}.

The $n$-point function of single-trace operators is written as 
\begin{equation}
G'_n = \sum_{ \bsl \, \vdash \hL }
\frac{1}{|{\rm Aut} \, V|}
\sum_{\nu \, \in (S_{2 \bL}^{\times \! \times})_{\rm phys} }  \, 
\sum_{\bstau \in S_{\bstau} } \bb{F} ( \bstau |\, \nu ) \,
\delta ( n_{\rm ex} , n )
\delta \Big( \{ L_{{\rm ex},i} \} , \{ L_i \} \Big)
\label{n-pt Sred}
\end{equation}
where the sums over $\bsl$ and $\bstau$ are given in \eqref{def:partition of hL} and \eqref{def:S bstau}, respectively.
We inserted two types of $\delta$-functions from \eqref{physical constraints}.
We sum the face permutation $\nu$ over the range \eqref{skeleton graph coset equivalence}. 
We used $G'_n$ as in \eqref{npt reduced formula vertex} because the last procedure removes two-point functions from $G_n$\,.

The fact that the powers of $N_c$ in \eqref{gen: construct red F} depend on the internal data $\bstau$ causes {\it stratification}, namely the skeleton graph of smaller genera contribute to the $1/N_c$ corrections to the $n$-point functions at a fixed genus \cite{Chekhov:1995cq,Bargheer:2017nne,Bargheer:2018jvq}.

\subsubsection{Sum over graphs}\label{sec:sum over graphs}

We simplify \eqref{n-pt Sred} further in two steps.
The first step is to take the connected part of $n$-point functions. This makes the subtle difference between $G_n$ and $G'_n$ unimportant,
\begin{equation}
( G'_n )_{\rm connected} = ( G_n )_{\rm connected} \,, \qquad (\forall n \ge 3).
\end{equation}
The second step is to interchange the order of summation in \eqref{n-pt Sred}.
Previously, we fixed the bridge length distribution $\bsl \vdash \hL$ first, and summed over the skeleton graph with the edge lengths $\{ \ell_1 \,, \ell_2 \,, \dots , \ell_{\bL} \}$.
Now we fix the skeleton graph first, and distribute the bridge lengths later.
Let us define a set of skeleton graphs with a given set of dimensions by
\begin{equation}
{\bf SG} (L_1,L_2, \dots, L_n) = \Big\{ \text{Skeleton graphs with $n$ vertices} \, \Big| \,
\Delta (\text{$i$-th vertex}) = L_i
\Big \}.
\end{equation}
The elements of ${\bf SG} ( \{ L_i \})$ automatically solve the conditions \eqref{physical constraints}.
The mutually adjacent external Wick-contractions of \eqref{skeleton graph coset equivalence} has disappeared by the definition of skeleton-ness.

Let $\Gamma \in {\bf SG} ( \{ L_i \})$ and define
\begin{equation}
{\rm Face} (\Gamma) \equiv \nu, \qquad
| {\rm Edge} (\Gamma) | = \bL, \qquad
{\rm Vertex} (\Gamma) = \bar\alpha_\bullet \,, \qquad
| {\rm Vertex} (\Gamma)| = n .
\end{equation}
The $n$-point function \eqref{n-pt Sred} becomes
\begin{align}
G'_n &= \sum_{ \Gamma \in {\bf SG} ( \{ L_i \}) } \sum_{\(\text{Distribute $\bsl$}\)}
\frac{1}{|{\rm Aut} \, V \! (\bsl) |}
\sum_{\bstau \in S_{\bstau} } \mathscr{F} \(\bstau | \Gamma \) 
\label{n-pt Sred 2} \\
\mathscr{F} (\bstau | \Gamma) &=
N_c^{\bZ (\bstau) + C(\bar\omega)} \,
\sum_{ \bsz \in \bb{Z}_\alpha} 
\prod_{s=1}^{|{\rm Edge} (\Gamma)|} \bsz \cdot 
h \( \bstau \, \Big| \, \bar\mu_{1,s} \ \bar\mu_{1, s'} \ \dots \, \bar\mu_{n,s} \ \bar\mu_{n,s'} \) 
\label{def:scrF}
\end{align}
where $\bar\omega = {\rm Face} (\Gamma) \cdot \theta (\bstau)$.
The second sum of \eqref{n-pt Sred 2} is taken over the space of assigning a length to each edge of $\Gamma$,
\begin{multline}
\( \text{Distribute $\bsl$} \) = \Bigl\{ 
{\rm Map} :  {\rm Edge} (\Gamma) \, \to \, \bb{Z}_+^{|{\rm Edge} (\Gamma)|} \, \Big| \,
\\
\sum_{q=1}^{\bL_i} \bsl_{i|q} = L_i \,, \ \ 
\bigsqcup_{i=1}^n \bigsqcup_{q=1}^{\bL_i} \, \bsl_{i|q} = [1^{2 \bsr_1} 2^{2 \bsr_2} \dots ] \vdash 2L, \ \ 
\sum_k \bsr_k = |{\rm Edge} (\Gamma)|
\Bigr\} .
\end{multline}
The first sum of \eqref{n-pt Sred 2} is taken over {\it partially labeled} skeleton graphs.\footnote{By a partially labeled skeleton graph, we mean that the $i$-th vertex of the graph has the dimensions $L_i$\,. We do not assign the flavor indices of the original operator $\cO_i$ inside the $i$-th vertex.}
Note that we do distinguish the vertices having the same dimensions, and hence removed the sum over $\bar\beta \in {\bf CR}$ in \eqref{def:scrF}.
Recall that the sum over ${\bf CR}$ was needed in \eqref{abs CR lambda}, because in the group-theory language a permutation $\bar\alpha_\bullet$ cannot specify an ordered list uniquely. 
In the graph-theory language, we may distinguish the $i$-th and $j$-th vertices even if $L_i = L_j$\,.

\bigskip
Let us take the connected part of the $n$-point function and rewrite $( G_n )_{\rm connected}$ further.
We decompose the space ${\bf SG} ( \{ L_i \})$ by the genus of a graph $\Gamma$,
\begin{equation}
2 - 2g = |{\rm Vertex} (\Gamma)| - | {\rm Edge} (\Gamma) |
+ | {\rm Faces} (\Gamma) | .
\end{equation}
We define the gauge theory moduli space $\cM_{g,n}^{\rm gauge} (\{L_i\})$ by the first two sums of  \eqref{n-pt Sred 2},
\begin{equation}
( G_n )_{\rm connected} = \sum_{g \ge 0} \ 
\sum_{ \Gamma ( \bsl ) \in \cM_{g,n}^{\rm gauge} (\{ L_i \}) } \ 
\frac{1}{|{\rm Aut} \, V (\bsl) |}
\sum_{\bstau \in S_{\bstau} } \mathscr{F} \(\bstau | \Gamma \) 
\label{npt reduced formula face}
\end{equation}
where $\Gamma (\bsl)$ is the skeleton graph whose bridge lengths are specified. Note that $g$ is a parameter of $\Gamma (\bsl)$ which is {\it not} the powers of $N_c$ in $( G_n )_{\rm connected}$ due to the stratification.

The skeleton graph with specific bridge lengths describing the connected $n$-point function can be identified as the so-called connected metric ribbon graphs. Using this connection, we will study the properties of $\cM_{g,n}^{\rm gauge} (\{L_i\})$ in Section \ref{sec:geometry}.

\section{Geometry from permutations}\label{sec:geometry}

We start our argument in an abstract way by looking at the equivalence between the space of connected metric ribbon graphs and the decorated moduli space.
Then we define a subset of these spaces by imposing three conditions.
We conjecture that this subspace is equal to the gauge theory moduli space $\cM_{g,n}^{\rm gauge} (\{L_i\})$ introduced above, and discuss some examples.

\subsection{Ribbon graph and Riemann surface}

As explained in Introduction and Appendix \ref{app:quadratic diff}, there is a one-to-one correspondence between the space of connected metric ribbon graphs (CMRG) with the decorated moduli space of Riemann surfaces,
\begin{equation}
{\rm CMRG}_{g,n} = \cM_{g,n} \times \bb{R}_+^n \,.
\label{CMRG Mgn correspondence}
\end{equation}
In this equality, a graph defines a cell decomposition of a Riemann surface, $\Sigma_{g,n} = \sqcup_{i=1}^n R_i$\,. Each region $R_i$ contains a puncture of $\Sigma_{g,n}$, and two regions are separated by the edges of the graph. The edge lengths of the graph $\{ \ell_E \}$ specify the complex structure of $\Sigma_{g,n}$\,.

We denote the sum of edge lengths around the $i$-th puncture by
\begin{equation}
\sum_{E \in \partial R_i} \ell_E = p_i \ \in \ \bb{R}_+ \,.
\label{lengths around a puncture}
\end{equation}
On $\Sigma_{g,n}$\,, LHS is a cycle integral of a differential around the $i$-th puncture, and RHS is the residue.
In fact, for any $\{ p_1 \,, p_2 \,, \dots \,, p_n \} \in \bb{R}_+^n$, there exists a unique quadratic differential $\varphi$ having the residue $p_i$ at the $i$-th puncture, so-called the Jenkins-Strebel (JS) differential. 
The critical graph of $\varphi$ is isomorphic to the metric ribbon graph $\Gamma_{g,n}$\,. See reviews \cite{MP97,Mondello07,Eynard:2016yaa} for details.

\bigskip
Let us inspect the properties of ribbon graphs, constrained by the corresponding Riemann surfaces.
If a ribbon graph completely triangulates a Riemann surface as $\Sigma_{g,n} = \sqcup_{i=1}^n R_i$\,, then the number of vertices, edges and faces are given by
\begin{equation}
V^\bullet_{g,n} = 2n -4+4g, \qquad
E^\bullet_{g,n} = 3n -6+6g, \qquad
F^\bullet_{g,n} = n, \qquad
(\text{all trivalent}).
\label{graph planarity counting}
\end{equation}
Incomplete triangulations also constrains the graph data, as discussed in Appendix \ref{app:geometry graph}.
If we focus on all possible ways of distributing lengths (a positive number) to each edge of the graph, we find
\begin{equation}
{\rm CMRG}_{g,n} \ \simeq \ \bb{R}_+^{3n-6+6g} 
\end{equation}
which is consistent with
\begin{equation}
\dim_{\, \bb{R}} \cM_{g,n} = 2n-6+6g.
\label{dim_R Mgn}
\end{equation}
This relation has been made precise in \cite{Kontsevich:1992ti}, where the integration measure of $\cM_{g,n}$ is expressed by the edge length differentials $\pare{ d \ell_1 \,, d\ell_2 \,, \dots \,, d \ell_{3n-6+6g} }$ under the constraints \eqref{lengths around a puncture}.

By definition, the space ${\rm CMRG}_{g,n}$ splits into the metric part and the ribbon graph part as
\begin{equation}
{\rm CMRG}_{g,n} =
\Bigl( \text{Choice of } \{ \ell_E > 0 \} \Bigr) \times
\Bigl( \text{Choice of ribbon graph $\Gamma_{g,n}$} \Bigr).
\label{split CMRG_gn}
\end{equation}
We can specify the ribbon graph uniquely by permutations.
One way of doing this is
\begin{multline}
\Bigl( \text{Choice of ribbon graph $\Gamma_{g,n}$} \Bigr) 
\\
=
\Bigl( \text{Number of $k$-valent vertices} \Bigr) \times
\Bigl( \text{Cyclic ordering of edges at each vertex} \Bigr) .
\label{split Gamma_gn}
\end{multline}

\subsection{Constraints from gauge theory}\label{sec:gauge constraints}

We impose three constraints on $\cM_{g,n}$ to select the gauge theory moduli space $\cM_{g,n}^{\rm gauge} \subsetneq \cM_{g,n}$.

The first condition is integrality, $\ell_E \in \bb{Z}_+$\,. As a corollary, the residue condition \eqref{lengths around a puncture} becomes
\begin{equation}
\sum_{E \in \partial R_i} \ell_E = L_i \ \in \ \bb{Z}_+ 
\label{lengths around an operator}
\end{equation}
which is equivalent to \eqref{sum lij-rho}.

The second condition is the absence of self-contractions, $\ell_{ii}=0$.
As a corollary, we obtain the extremality condition $\sum_{j \neq i} L_j \ge L_i$ for any $i$.
This can be shown by
\begin{equation}
\sum_{j \neq i} L_j - L_i 
= \sum_{j=1}^n L_j - 2 L_i
= \( \ell_{ii} + 2 \sum_{j \neq i} \ell_{ji} + \sum_{j \neq i} \sum_{k \neq i} \ell_{kj} \) - 2 \sum_{j} \ell_{ji}
= - \ell_{ii} + \sum_{j \neq i} \sum_{k \neq i} \ell_{kj} 
\label{extremal from self-edge}
\end{equation}
which is non-negative if $\ell_{ii}=0$.
Indeed, the correlator of scalar operators in gauge theory vanishes when the extremality condition is violated.
The situation was different in matrix models, because self-contractions are allowed and $\ell_{ii}$ can be non-zero.

\begin{figure}[t]
\begin{center}
\includegraphics[scale=0.7]{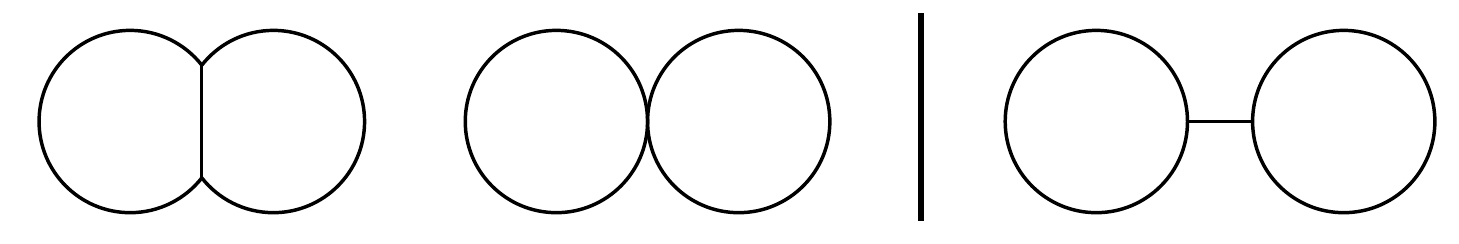}
\vskip 7mm
\includegraphics[scale=0.58]{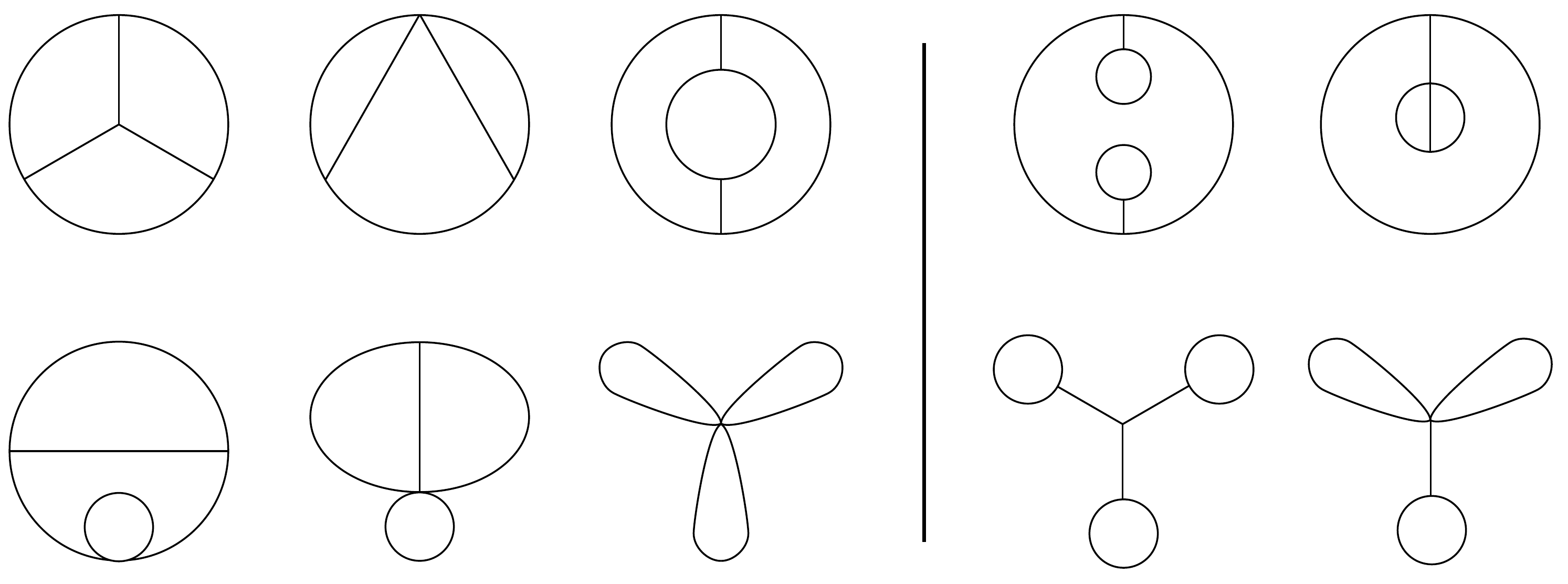}
\caption{Examples of the connected metric ribbon graphs with $(g,n)=(0,3)$ (above) and $(0,4)$ (below). The diagrams on the right half have self-edges, and some diagrams on the left half are extremal.}
\label{fig:CMRG0304}
\end{center}
\end{figure}

The above conditions also constrain the complex structure, which can be argued in two ways.
One argument goes as follows. 
Since the complex structure is induced by decoration in the metric ribbon graph method, any inequality on the decoration parameters selects a proper subset of the moduli space.
Thus we find
\begin{equation}
{\rm CMRG}_{g,n}^\triangle 
= \( \cM_{g,n} \times \bb{Z}_+^n \)^{\triangle}
= \cM_{g,n}^\triangle \times (\bb{Z}_+^n)^{\triangle}
\label{def:CMRG triangle}
\end{equation}
where $( \bb{Z}_+^n)^{\triangle}$ is the subspace of $\bb{Z}_+^n$ satisfying the generalized triangular inequality,
\begin{equation}
( \bb{Z}_+^n)^{\triangle} = \pare{ \(L_1 \,, L_2 \,, \dots \,, L_n \) \in \bb{Z}_+^n \, \Big| \,
\sum_{j \neq i} L_j \ge L_i \quad (i=1,2,\dots, n) } .
\label{def:triangular Z+}
\end{equation}
The moduli space without self-contractions $\cM_{g,n}^\blacktriangle$ is a subspace of $\cM_{g,n}^\triangle$ according to \eqref{extremal from self-edge}.
Another argument is based on the fact that Feynman graphs are dual to the critical graphs on Riemann surfaces, as discussed in Appendix \ref{app:Feynman Riemann}. 
In general, the critical graph of Riemann surface may contain {\it self-edges} \cite{Kontsevich:1992ti}.
The self-edge is a boundary of the same face as shown in Figure \ref{fig:CMRG0304}.
The gauge theory moduli space $\cM_{g,n}^{\rm gauge}$ should not contain self-edges, because they correspond to the self-contractions in the dual Feynman graph \cite{Aharony:2006th}.

The third condition is discrete homotopy, which is stronger than the usual homotopy in classifying edges.
In classical geometry, we call two edges are homotopic if they end on the same vertices and if they wind various one-cycles on $\Sigma_{g,n}$ in the same way.
In our skeleton reduction, two edges are discrete-homotopic as long as they are consecutive and connect the same pair of operators, even if they cross each other.
This condition corresponds to the exclusion of the mutually adjacent pairs discussed in Appendix \ref{app:external Wick space}.
The equivalence under the discrete homotopy selects a subspace of the moduli space as
\begin{equation}
\cM_{g,n}^{\blacktriangle, {\rm disc}} \subset \cM_{g,n}^\blacktriangle \subset \cM_{g,n} \,.
\label{def:Mgn triangle disc}
\end{equation}
Importantly, the discrete homotopy induces a special type of triangulation of Riemann surface.
Since the discrete homotopy is stronger than the usual homotopy, our triangulation is simpler than those used in \cite{Bargheer:2017nne,Bargheer:2018jvq}.\footnote{This simplification may be useful if one is interested only in the contribution from the zero-length bridges in the hexagon methods.}

We conjecture that the restricted moduli space is equal to the gauge theory moduli space defined by the skeleton-reduced $n$-point formula \eqref{npt reduced formula face},\footnote{If we apply the skeleton reduction which assembles only the planar internal Wick-contractions, we should get $\cM_{g,n}^{\triangle} = \cM_{g,n}^{\rm gauge} (\{ L_i \})$.}
\begin{equation}
\cM_{g,n}^{\blacktriangle, {\rm disc}} = \cM_{g,n}^{\rm gauge} (\{ L_i \}).
\end{equation}
We see that LHS indeed depends on $\{ L_i \}$, because the minimal edge length is quantized.
As a result, the length of external operators $\{ L_i \}$ gives an upper bound on $g$.
This property is quite different from the classical moduli space.

\subsubsection*{Cell decomposition}

Since $\cM_{g,n}^{\rm gauge} (\{ L_i \})$ is a set of metric ribbon graphs, it inherits the structure
\eqref{split CMRG_gn} and \eqref{split Gamma_gn}
\begin{multline}
\cM_{g,n}^{\rm gauge} =
\Bigl( \text{Choice of } \{ \ell_E \ge 1 \} \Bigr) \times
\\
\Bigl( \text{Number of $k$-valent vertices} \Bigr) \times
\Bigl( \text{Cyclic ordering of edges at each vertex} \Bigr) .
\label{split M_gn gauge}
\end{multline}
When some of $\ell_E$ become zero, then vertices collide and the graph topology changes. 
In the literature, the classification of $\cM_{g,n}^{\rm gauge}$ by graph topology is called cell decomposition.

Consider the second factor of \eqref{split M_gn gauge} in detail.
Let us denote the valency set of the skeleton graph by $\sfd = [3^{d_3} \, 4^{d_4} \dots ]$, where $d_k$ is the number of $k$-valent vertices. From Appendix \ref{app:geometry graph}, we find that trivalent skeleton graphs have $\sfd = [3^{2n-4+4g}]$, and the extremal skeleton graph with one vertex has $\sfd = [ (2n-2+4g)^1 ]$. The valency set of general skeleton graphs at $(g,n)$ is in between,
\begin{equation}
\sfd \in \sfD_{g,n} = \Bigl\{ [3^{2n-4+4g}],\ [3^{2n-6+4g} \, 4^1], \ \dots ,\ 
[ 3^1 (2n-3+4g)^1 ],\ [ (2n-2+4g)^1 ] \Bigr\}.
\label{def:sfD gn}
\end{equation}
This pattern also constrains the number of edges in the skeleton graphs, denoted by $\bL$ in Section \ref{sec:reduction}. The trivalent graphs have $\bL = 3n-6+6g$ and the extremal graph has $n-1+2g$. 
The number of edges in general skeleton graphs is
\begin{equation}
\bL \in \sfL_{g,n} = \Bigl\{ 3n-6+6g, \ 3n-7+6g, \ \dots, \ 
n+2g, \ n-1+2g \Bigr\}.
\label{def:sfL gn}
\end{equation}
Generally there is no simple correspondence between the entries in \eqref{def:sfL gn} and \eqref{def:sfD gn}. 
As shown in Figure \ref{fig:CMRG0304}, the edges of a non-trivalent graph may end on the same vertex or different vertices, which changes the number of edges.

\subsubsection*{Classical limit of $\cM_{g,n}^{\rm gauge}$}

The classical limit is defined by taking all bridge lengths to be large $\ell_E \gg 1$ with $g,n$ fixed, for the ribbon graphs whose faces are all triangles.
We can neglect the quantization condition \eqref{lengths around an operator} in this limit.
The second line of \eqref{split M_gn gauge} is still a finite set, because the graph data are constrained by $(g,n)$ as in \eqref{def:sfD gn}, \eqref{def:sfL gn}.
From \eqref{def:CMRG triangle} one expects that the dimensions of $\cM_{g,n}^{\rm gauge} ( \{ L_i \} )$ should agree with $\dim_{\bb{R}} \cM_{g,n}$ as
\begin{equation}
\cM_{g,n}^{\rm gauge} ( \{ L_i \} ) \ \sim \ \bb{Z}^{2n-6+6g} .
\label{def:CMRG-gauge dim}
\end{equation}
This equation can be understood in the graph theory language in Appendix \ref{app:count dimensions}.
Owing to the relation \eqref{def:Mgn triangle disc}, the space $\cM_{g,n}^{\rm gauge}$ remains a proper subset of $\cM_{g,n}$ in the classical limit.

We expect that the generalized triangular inequality \eqref{def:triangular Z+} disappears in more general setups.
For example, when we include the covariant derivatives, then there are non-zero three-point functions with $\Delta_1 > \Delta_2 + \Delta_3$ due to the Laplacian $\square^n$.
If we write $\Delta = \tau_s^* + s + 2n$ and study the minimal twist $\tau^{\rm min}$, then a similar triangular inequality was found in \cite{Komargodski:2012ek}.

\subsection{Examples}\label{sec:examples}

We compute the connected $n$-point functions $( G_n )_{\rm connected} (\{L_i\})$ for small $\hL = \frac12 \sum_i L_i$ and recast them into the form of \eqref{npt reduced formula face}. 
This procedure explicitly determines the space of reduced Wick-contractions $\cM_{g,n}^{\rm gauge} ( \{ L_i \} )$ for simple cases.

We expand $( G_n )_{\rm connected}$ in $1/N_c$ and use $\xi=0,1,2,\dots$ to count the powers of $N_c$\,, where $\xi=0$ corresponds to the planar term. 
For a given $\{ L_i \}$, we define
\begin{multline}
\sfN^\circ_{\xi,\lambda} ( \{ L_i \} ) = \Bigl\{ 
\text{Feynman (or ribbon) graphs of genus $\xi$ whose skeleton reduction}
\\
\text{has the face permutation and of cycle type $\lambda$}
\Bigr\}
\end{multline}
Different elements in $\sfN_{\xi,\lambda}^\circ$ have different cyclic orderings in \eqref{split M_gn gauge}, or different $\tau$. We introduce
\begin{equation}
\sfy_{\xi,\lambda} \equiv \frac{| \sfN^\circ_{\xi,\lambda} ( \{ L_i \} ) |}{| \bb{Z}_\alpha |} 
= \frac{| \sfN^\circ_{\xi,\lambda} ( \{ L_i \} ) |}{\prod_{i=1}^n L_i } \ \in \bb{Q}
\end{equation}
which roughly counts the number of skeleton graphs whose face permutation of cycle type $\lambda$.
This counting is not precise, because the action of $\bb{Z}_\alpha$ on Feynman graphs may have non-trivial fixed points, as discussed in Appendix \ref{app:details skeleton}.
We write 
\begin{equation}
\begin{aligned}
\sfN_\xi &\equiv \sum_\lambda \, \sfy_{\xi,\lambda} \times \lambda 
\\[1mm]
&= \sfy_{\xi,a} \, [3^{a_3} \, 4^{a_4} \, \dots ] +
\sfy_{\xi,b} \, [3^{b_3} \, 4^{b_4} \, \dots ] +
\sfy_{\xi,c} \, [3^{c_3} \, 4^{c_4} \, \dots ] + \dots .
\end{aligned}
\end{equation}
It will turn out that the lists $\sfN_\xi$ are perfectly consistent with our earlier classification \eqref{def:sfD gn}.
If $g$ represents the genus of the ribbon graph, $\nu_{\xi}$ may contain graphs of smaller genera, $\nu_{g}$ with $g < \xi$, owing to the stratification.

The details of the computation are explained in Appendix \ref{app:details skeleton} and in the attached {\tt Mathematica} notebook.

\paragraph{Three-point functions.}

\begin{figure}[H]
\begin{center}
\includegraphics[scale=.47]{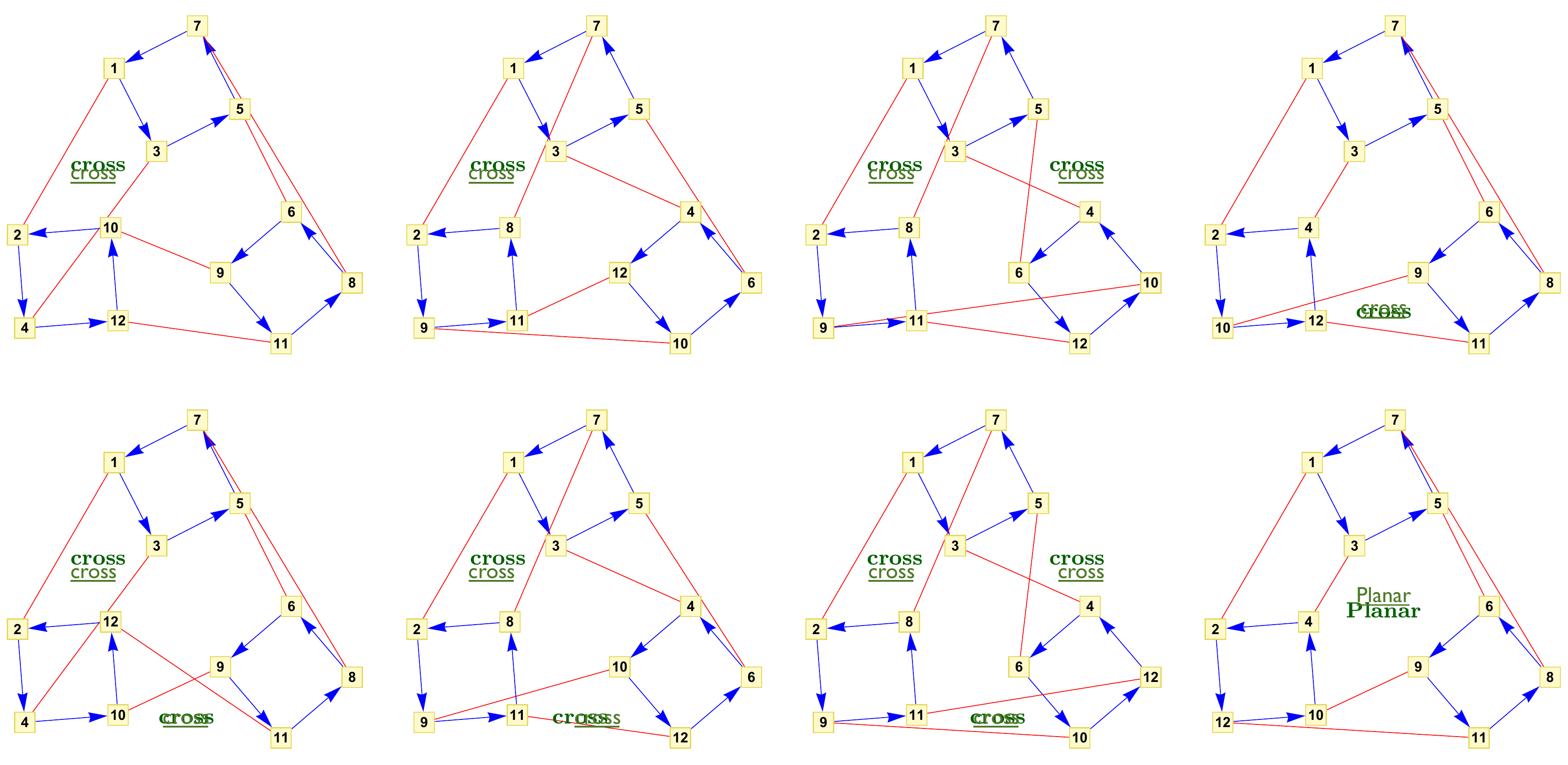}
\caption{Original Feynman graphs corresponding to $7 \times [3^2] \in \sfN_1$ and $[3^2] \in \sfN_0$ in \eqref{sfN list 444}. 
The first 7 graphs are part of the leading non-planar corrections to the three-point function, while the last graph is planar.
All of them gives $\nu \in [3^2]$ after the skeleton reduction of Section \ref{sec:reduction}.}
\label{fig:reduced33}

\vskip 5mm
\includegraphics[scale=.6]{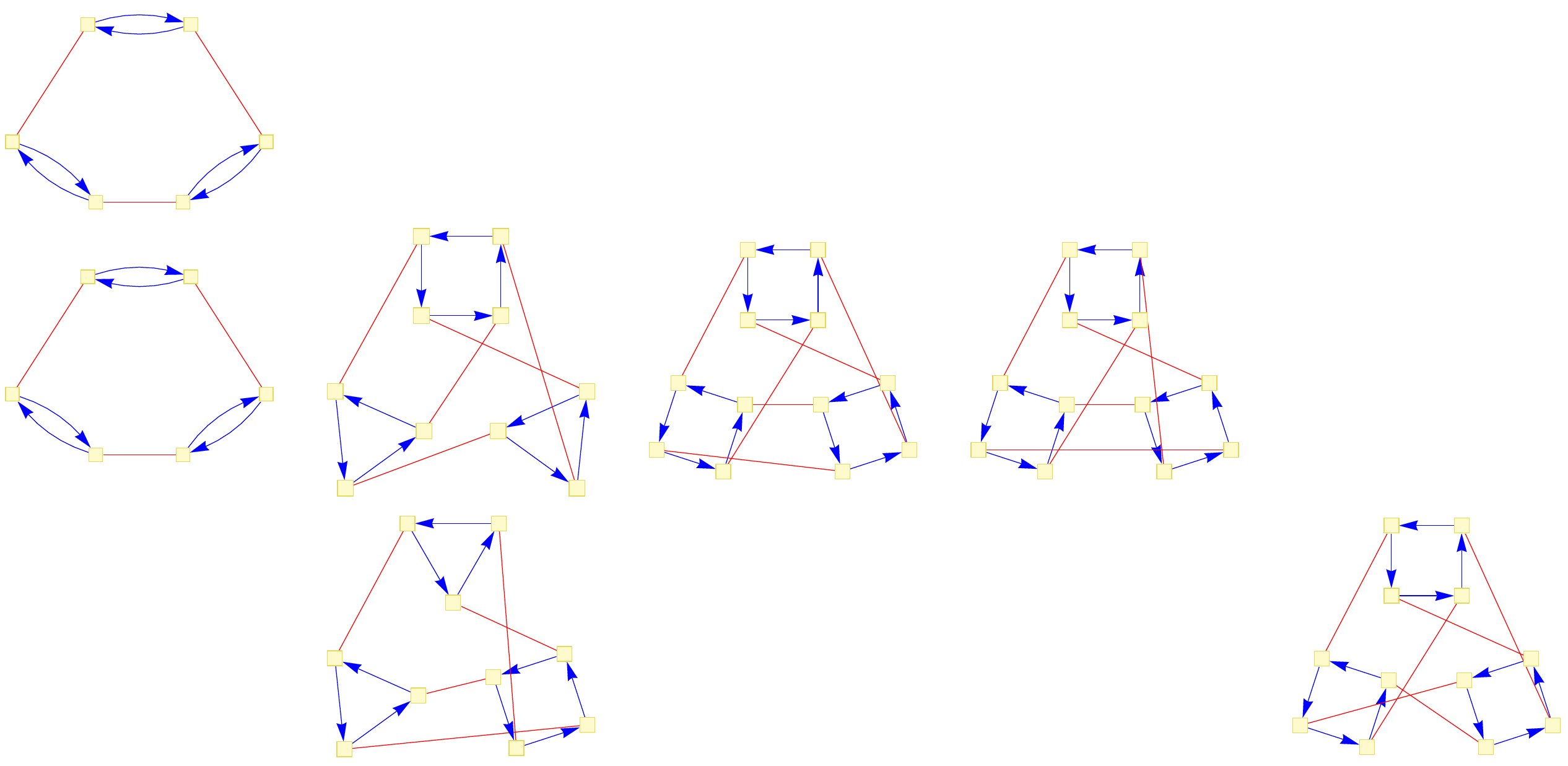}
\caption{Skeleton graphs of $(G_3)_{\rm connected} (4,4,4)$. The top, middle and bottom row corresponds to $\xi=0,1,2$, respectively.}
\label{fig:444}
\end{center}
\end{figure}

\begin{figure}[t]
\begin{center}
\includegraphics[scale=.6]{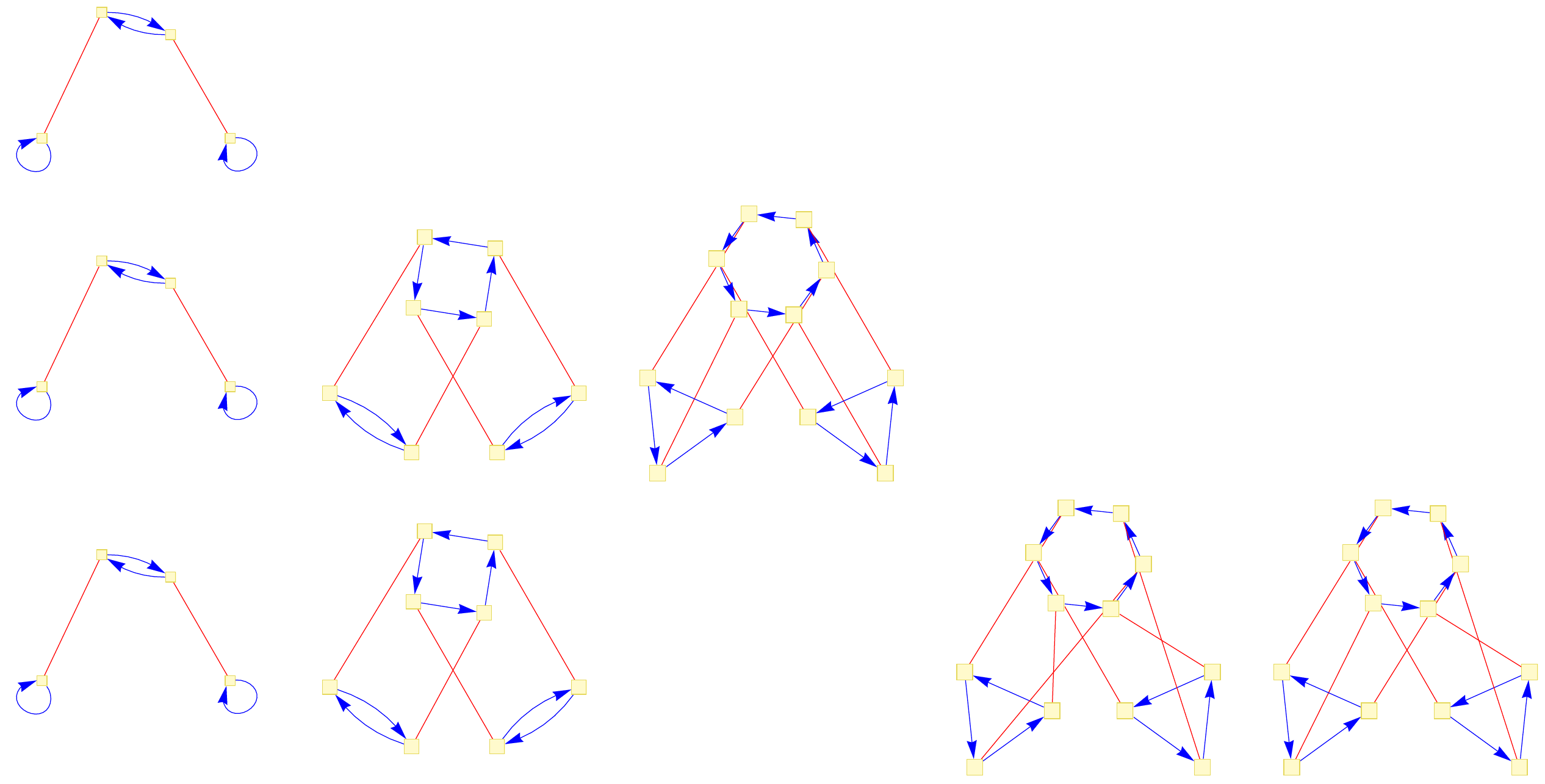}
\caption{Skeleton graphs of $(G_3)_{\rm connected} (6,3,3)$.}
\label{fig:633}

\vskip 5mm
\includegraphics[scale=.5]{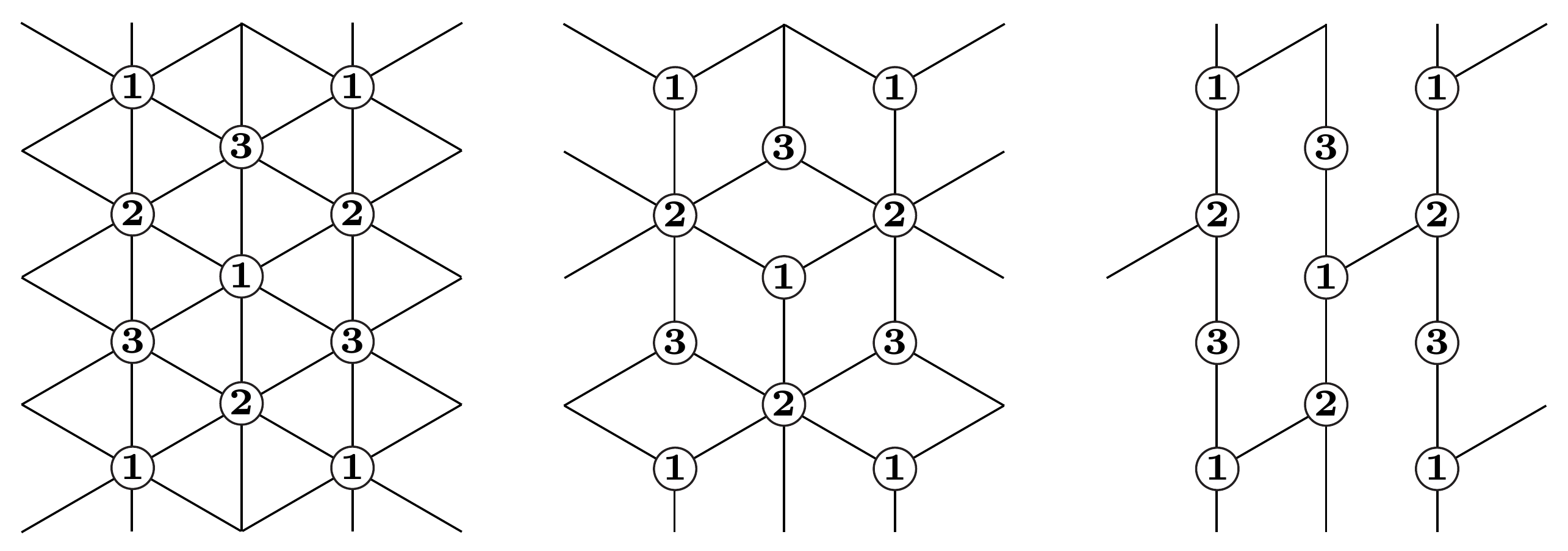}
\caption{Examples of toric diagrams in $\cM_{1,3}^{\rm gauge}$, representing $[\nu] = [3^6], [4^3], [8^1]$. The diagram in the center is extremal and has $\ell_{13}=0$.}
\label{fig:Toric_S13_incomplete}
\end{center}
\end{figure}

We consider two examples, $(L_1,L_2,L_3)=(4,4,4)$ and $(6,3,3)$.
For the first example, the list of the cycle type of face permutations is
\begin{equation}
\sfN_0  = [3^2] , \quad
\sfN_1 = 7 \, [3^2] + 6 \, [3^1 \, 7^1] + 4 \, [3^2 \, 6^1] , \quad
\sfN_2 = 6 \, [3^1 \,7^1] + 3 \, [12^1] .
\label{sfN list 444}
\end{equation}
Figure \ref{fig:reduced33} explains the origin of the first term of $\sfN_0 \,, \sfN_1$ in \eqref{sfN list 444} in view of the original Feynman graphs.
For the second example, the list is
\begin{equation}
\sfN_0 = [4^1] , \quad
\sfN_1 = 2 \, [4^1] + 6 \, [8^1] + \frac13 \, [4^3]  , \quad
\sfN_2 = [4^1] + 2 \, [8^1] + [12^1] .
\end{equation}
The skeleton graphs appearing in $(G_3)_{\rm connected}$ are shown in Figures \ref{fig:444} and \ref{fig:633}.

These lists agree with the classification \eqref{def:sfD gn}. For example, at genus one, we find
\begin{equation}
\sfD_{1,3} = \Bigl\{
[ 3^6 ], \ 
[ 3^4 \, 4^1 ] , \ 
[ 3^3 \, 5^1 ] , \ 
[ 3^2 \, 4^2 ], 
[ 4^3 ] , \ 
[ 3^1 \, 4^1 \, 5^1 ] , \ 
[ 3^2 \, 6^1 ] , \ 
[ 5^2 ] , \ 
[ 4^1 \, 6^1 ] , \ 
[ 3^1 \, 7^1 ] , \ 
[ 8^1 ] 
\Bigr\} .
\label{S13 reduced cycle types}
\end{equation}
Some cycle types are missing in the actual examples, because there is no solution to the constraints on the quantized bridge lengths \eqref{lengths around an operator}. We can represent each entry of $\sfD_{1,3}$ as the toric diagram with three vertices as shown in Figure \ref{fig:Toric_S13_incomplete}.

In general, the planar three-point functions have a unique skeleton graph. This is because the bridge lengths are completely fixed by the external operators
\begin{equation}
\ell_{ij} = \frac{L_i + L_j - L_k}{2}  \,, \qquad \{ i,j,k \} \in \{ 1,2,3 \}
\end{equation}
and the cyclic orderings are fixed uniquely. Thus we have
\begin{equation}
\cM_{0,3}^{\rm gauge} ( L_1 \,, L_2 \,, L_3 ) = \{ pt \}
\end{equation}
up to the change of graph topology in the extremal limit.

Another interesting observation is that the skeleton graph of $(G_3)_{\rm connected} (L,L,2)$ is unique irrespective of the genus. 
By taking first the Wick-contraction of $\cO_3$ with $\cO_1, \cO_2$, we find that the remaining Wick-contractions are same as the two-point function of length $L-1$, whose graph topology is trivial.

\paragraph{Four-point functions.}

\begin{figure}[H]
\begin{center}
\includegraphics[scale=.65]{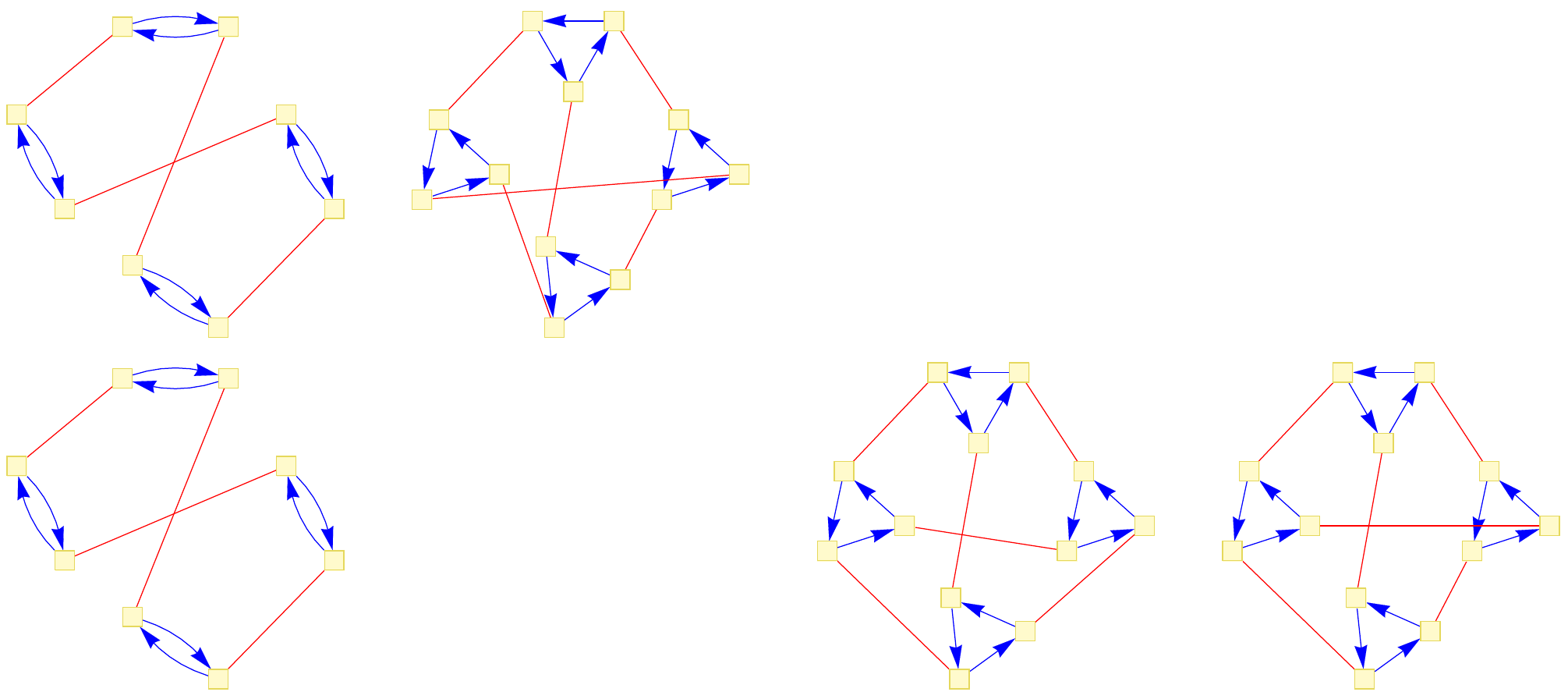}
\caption{Skeleton-reduced diagrams of $(G_4)_{\rm connected} (3,3,3,3)$.}
\label{fig:3333}

\vskip 8mm
\includegraphics[scale=.65]{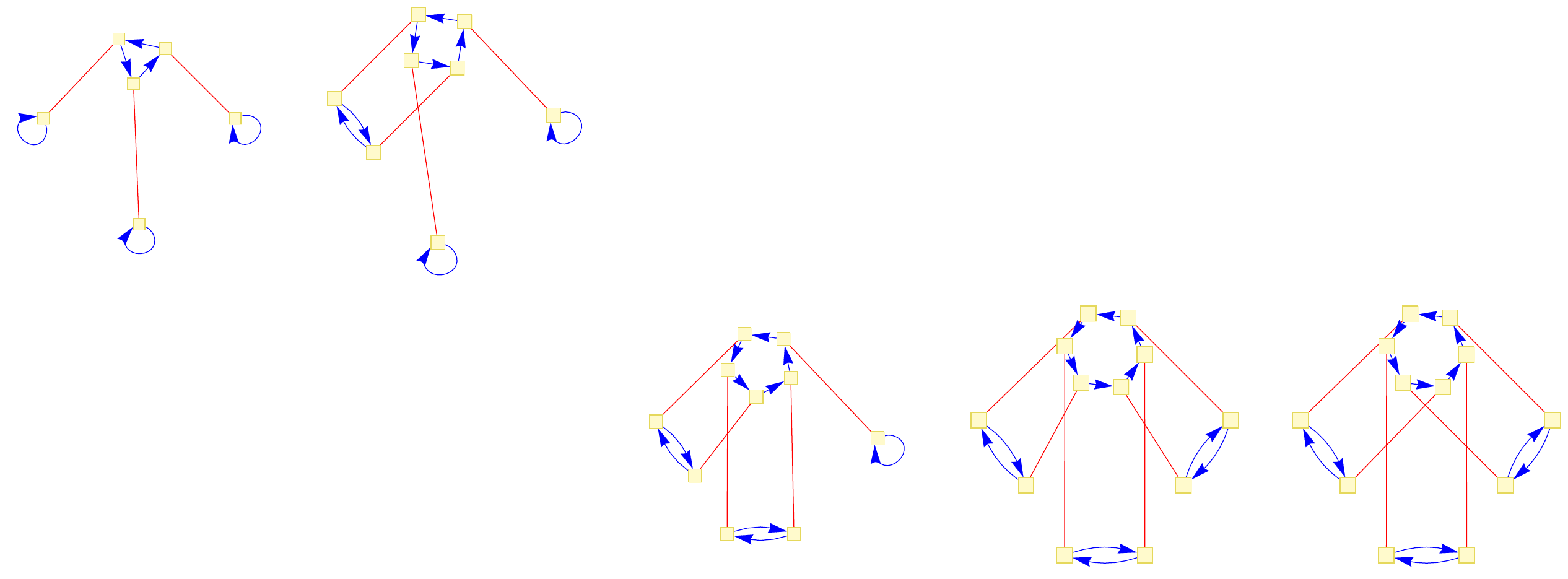}
\caption{Skeleton-reduced diagrams of $(G_4)_{\rm connected} (6,2,2,2)$.}
\label{fig:6222}

\vskip 8mm
\includegraphics[scale=.34]{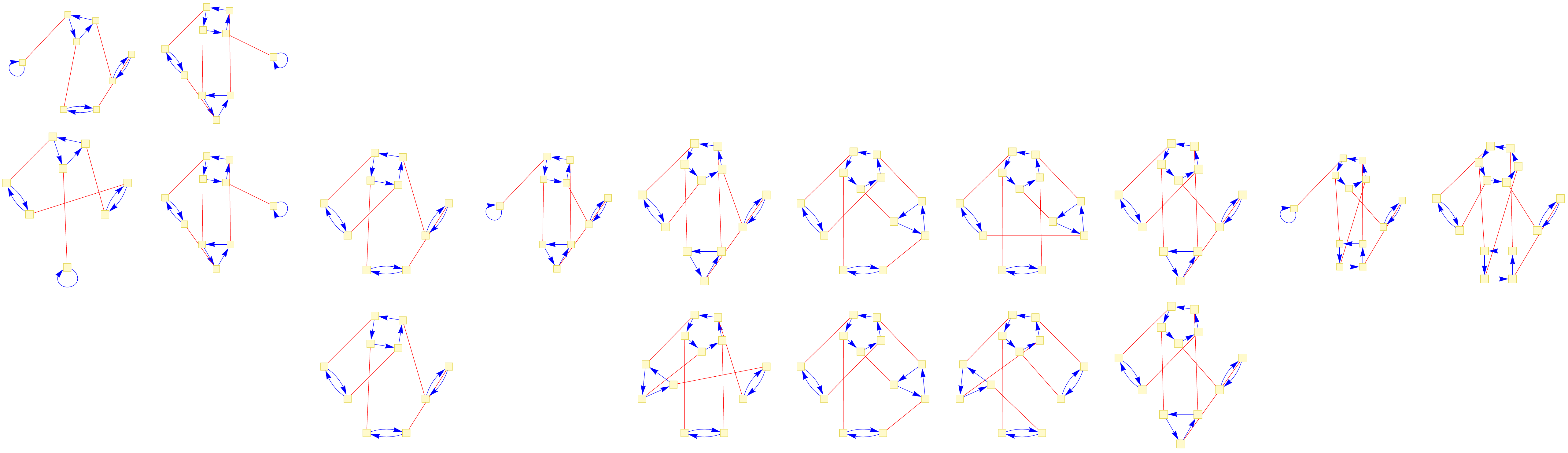}
\caption{Part of the skeleton-reduced diagrams of $(G_4)_{\rm connected} (6,4,2,2)$. There are 28 diagrams in total.}
\label{fig:6422}
\end{center}
\end{figure}

We consider three examples, $(L_1,L_2,L_3,L_4)=(3,3,3,3), (6,2,2,2)$ and $(6,4,2,2)$.
For the first example, the list of the cycle type of face permutations is
\begin{equation}
\sfN_{0} = 6 \, [4^2] + 2 \, [3^4] , \quad
\sfN_{1} = 18 \, [4^2] + 6 \, [4^1 \, 8^1] + 8 \, [3^1 \, 9^1] 
\end{equation}
and for the second example, the list is
\begin{equation}
\sfN_{0} = 2 \, [6^1] + 3 \, [4^2] , \ \quad
\sfN_{1} = 6 \, [10^1] + 3 \, [4^1 \, 8^1] + [6^2] .
\end{equation}
The last example is more complicated,
\begin{equation}
\begin{aligned}
\sfN_{0} &= 6 \, [3^1 \, 5^1] + 4 \, [3^2 \,4^1] ,
\\
\sfN_{1} &= 22 \, [10^1] + 14 \, [5^1 \, 7^1] + 30 \, [3^1 \, 5^1] + 16 \, [3^1 \,9^1]
\\[1mm]
&\hspace{25mm} 
+ 4 \, [3^1 \, 5^1 \, 6^1] + 4 \, [4^1 \, 8^1] + 2 \, [3^2 \, 8^1]
+ 4 \, [3^1 \, 4^1 \, 7^1] + 4 \, [3^2 \, 4^1]   
\\[1mm]
\sfN_{2} &= 12 \, [10^1] + 20 \, [14^1] + 4 \, [5^1 \, 7^1] + 4 \, [3^1 \,9^1]  .
\end{aligned}
\end{equation}
The skeleton graphs appearing in $(G_4)_{\rm connected}$ are shown in Figures \ref{fig:3333}\,-\,\ref{fig:6422}.
Again, these lists are consistent with \eqref{def:sfD gn}. For example, at $g=1$ we get
\begin{equation}
\sfD_{1,4} = \{ [3^8], \dots , [4^4], \dots , [6^2], \dots , [10^1 ] \}.
\end{equation}

The planar four-point functions have several skeleton graphs. The moduli space can be written as
\begin{equation}
\cM_{0,4}^{\rm gauge} ( \{ L_i \} ) = 
\Bigl\{ (\xi_1 \,, \xi_2 ) \in \bb{Z}^2 \, | \, \ell_{ij} \ge 0 \Bigr\} \times
\Bigl\{ \text{Two choices in cyclic ordering} \Bigr\} .
\end{equation}
The first factor comes from the sum over bridge lengths given by
\begin{equation}
\begin{pmatrix}
\ell_{12} & \ell_{13} & \ell_{14} \\
& \ell_{23} & \ell_{24} \\
& & \ell_{34}
\end{pmatrix} \ \mapsto \ 
\begin{pmatrix}
\ell_{12} + \xi_1 & \ell_{13} + \xi_2 & \ell_{14} - \xi_1 - \xi_2 \\
& \ell_{23} - \xi_1 - \xi_2 & \ell_{24} + \xi_2 \\
& & \ell_{34} + \xi_1
\end{pmatrix} , \qquad
(\xi_1 \,, \xi_2 ) \in \bb{Z}^2.
\label{bridge length moduli (0,4)}
\end{equation}
The second factor comes from the ordering of $(\cO_1,\cO_2,\cO_3,\cO_4)$.
Suppose we take the open two-point between $\cO_1 , \cO_2$ as
\begin{equation}
(p \sim p+\ell_{12}-1) \quad {\rm from} \ \ \cO_1 \,, \qquad
(q \sim q+\ell_{12}-1) \quad {\rm from} \ \ \cO_2 
\label{example planar O1O2}
\end{equation}
where $p,q$ can be any number owing to the cyclic symmetry. 
Then, the subsequent color index $p+\ell_{12}$ of $\cO_1$ is contracted either by $\cO_3$ or $\cO_4$,
\begin{equation}
\begin{cases}
(p+\ell_{12} \sim p+\ell_{12}+\ell_{13}-1) \quad {\rm from} \ \ \cO_1 \ \ \text{for contracting} \ \ (\cO_1 , \cO_3)
\\
(p+\ell_{12} \sim p+\ell_{12}+\ell_{14}-1) \quad {\rm from} \ \ \cO_1 \ \ \text{for contracting} \ \ (\cO_1 , \cO_4) .
\end{cases}
\label{example planar O1234}
\end{equation}
This choice determines the skeleton graph completely, from the planarity and cyclic symmetry. 
For example, the first line of \eqref{example planar O1234} means that the color index $q+\ell_{12}$ of $\cO_2$ must be contracted with $\cO_4$\,.
Graphically, $(\cO_1, \cO_2 , \cO_3 , \cO_4)$ are placed at the left, right, top, bottom white circles in Figure \ref{fig:Bipartite g=0} in the counterclockwise ordering.

\section{Metric ribbon graphs in string theory}\label{sec:string}

We briefly discuss how metric ribbon graphs show up in string theory.

In the light-cone SFT, closed string worldsheet is characterized by the Giddings-Wolpert quadratic differential \cite{Giddings:1986rf}, whose analytic structure is classified by Nakamura graphs \cite{Nakamura00}. 
This situation is in parallel with the JS differential and its critical graph.
The moduli space of light-cone worldsheet theory can be regarded as the space of Wick-contractions in the Hermitian matrix model \cite{Freidel:2014aqa}, described by permutations \cite{Garner:2015wxa}. 
The light-cone and conformal methods differ in many ways.
For example, Nakamura graphs consist of quadrivalent vertices (four-point interactions) only.
Furthermore, the Giddings-Wolpert differential introduces a global time coordinate on $\Sigma_{g,n}$, whereas the JS differential does not have such a quantity.

In conformal gauge, we argue that the metric ribbon graphs arise from the analytic properties of the classical energy-momentum tensor, and characterize the permutation of Stokes sectors of string worldsheet on \AdSxS.

\bigskip
As discussed in Appendix \ref{app:geometry and graph}, the JS quadratic differential dual to a Feynman diagram of $n$-point functions has the analytic structure
\begin{equation}
\varphi = - \frac{L_i^2}{4 \pi^2} \frac{dz^2}{(z-z_i)^2} + O(|z-z_i|^{-1}), \qquad (i=1,2, \dots, n).
\end{equation}
It also has $2n-4+4g$ simple zeroes somewhere on $\Sigma_{g,n}$\,.
This behavior is similar to that of the classical worldsheet energy-momentum tensor
\begin{equation}
T(z) \sim \frac{\kappa_i}{(z-z_i)^2} + O(|z-z_i|^{-1}), \qquad (i=1,2, \dots, n).
\end{equation}
Here $\{ \kappa_i \}$ is related to the conformal dimensions of vertex operators.
If we interpret $T(z)$ as the ``JS'' differential, then the parameters $\{ \kappa_i \}$ determine the complex structure of the underlying Riemann surface. 
Physically, we may replace the Strebel condition by other conditions like the equations of motion, as long as the external parameters uniquely fix the quadratic differential and its critical graph.\footnote{Non-Strebel critical graphs are considered in \cite{Hollands:2013qza}.}

In AdS/CFT, the three-point functions of heavy operators in $\cN=4$ SYM at strong coupling have been studied by Pohlmeyer reduction \cite{Kazama:2011cp,Kazama:2012is,Kazama:2013qsa,Kazama:2016cfl}, borrowing the techniques developed for the gluon scattering \cite{Alday:2009ga,Alday:2009yn,Alday:2009dv,Alday:2010vh}.
The Pohlmeyer reduction on $S^3$ gives us the Hitchin system,
\begin{equation}
[ \partial + B_z, \bar \partial + B_{\bz} ] = 0.
\label{def:SL2 Hitchin}
\end{equation}
The $SL(2,\bb{C})$ flat connections $(B_z \,, B_{\bz})$ can be written as \cite{Kazama:2013qsa}
\begin{alignat}{9}
B_z &= \frac{\Phi_z}{\zeta} +A_z
&\ &= \frac{1}{\zeta} \begin{pmatrix}
0 & -\frac{\sqrt{T}}{2}e^{-i\gamma} \\
-\frac{\sqrt{T}}{2}e^{i\gamma} & 0
\end{pmatrix}  
&\ &+
\begin{pmatrix}
-\frac{i\partial \gamma}{2} & \frac{\rho e^{i\gamma}}{\sqrt{T}\sin 2\gamma} \\
\frac{\rho e^{-i\gamma}}{\sqrt{T}\sin 2\gamma} & \frac{i\partial \gamma}{2}
\end{pmatrix}
\notag \\[1mm]
B_{\bz} &= \zeta \Phi_{\bz} + A_{\bz}
& &= \zeta \begin{pmatrix}
0 & \frac{\sqrt{\bar{T}}}{2}e^{i\gamma} \\
\frac{\sqrt{\bar{T}}}{2}e^{-i\gamma} & 0
\end{pmatrix} 
& &+ \begin{pmatrix}
\frac{i\bar\partial \gamma}{2} & \frac{\tilde{\rho} e^{-i\gamma}}{\sqrt{\bar{T}}\sin 2\gamma} \\
\frac{\tilde{\rho} e^{i\gamma}}{\sqrt{\bar{T}}\sin 2\gamma} & -\frac{i\bar\partial \gamma}{2}
\end{pmatrix}
\label{def:Hitchin}
\end{alignat}
where $T (z), \bar T (\bar z)$ are the worldsheet energy-momentum tensors, $\zeta = \frac{1-x}{1+x}$ is the spectral parameter, and $(\gamma, \rho, \tilde \rho)$ are the $SO(4)$-invariant degrees of freedom after the Pohlmeyer reduction.
The flatness condition \eqref{def:SL2 Hitchin} is equivalent to the compatibility of the auxiliary linear problem
\begin{equation}
\(  \partial + B_z \) \psi = \( \bar \partial + B_{\bz} \) \psi = 0.
\end{equation}
These have two linearly independent solutions, called small and large depending on the asymptotics of $|z| \to \infty$ with ${\rm arg} \, (z)$ fixed. When we rotate $z$ the asymptotic behavior of the solutions changes. The region of the common asymptotics is called the Stokes sector.

In the semi-classical analysis, we uniformize the worldsheet by $dw = \sqrt{T(z)} \, dz$, which introduces branch points at the simple zeroes of $T(z)$.\footnote{In the scattering problem of $n$ gluons on ${\rm AdS}_3$\,, one chooses $dw = \sqrt{p_n (z)} \, dz$ where $p_n (z)$ is a polynomial of degree $n-2$. The number of coefficients of $p_n(z)$ is equal to the number of cross-ratios, or $\dim_{\bb{C}} \cM_{0,n} = n-3$. After integration we find $w \sim z^{n/2}$, showing that there are $n$ Stokes sectors at each half-plane.}
If we regard $T(z)$ as a quadratic differential on $\Sigma_{g,n}$\,, then the number of simple zeroes in $T(z)$ is given by \eqref{face degree g,n}.
We can draw a metric ribbon graph by connecting the zeroes and the poles of $T(z)$.
This graph knows how Stokes sectors are permuted around the branch points.
A similar argument has been done in the light-cone string theory in \cite{Freidel:2014aqa,Garner:2015wxa}.

\section{Conclusion and Outlook}

We studied the general $n$-point functions of scalar multi-trace operators in $\cN=4$ SYM with $U(N_c)$ gauge group at tree-level.
We obtained permutation-based formulae valid for any $n$ and to all orders of $1/N_c$\,.
The edge-based formula is interpreted as the topological partition function on $\Sigma_{0,n}$ with defects, which naturally decomposes into pairs of pants.
We applied a new skeleton reduction to find another set of formulae.
The skeleton-reduced Feynman graphs generate metric ribbon graphs, which form a subset of the moduli space of complex structure of Riemann surfaces.
Our skeleton reduction stratifies the moduli space with respect to the genus, whose top component is simpler than the usual skeleton reduction.

We find open/closed duality from the $n$-point functions of gauge-invariant operators.
The pants decomposition resembles closed-string interaction, while the triangulation of $\Sigma_{g,n}$ through the skeleton reduction resembles open-string interaction.

\bigskip
This work hopefully initiates several future directions of research, which we want to revisit in the near future.

The first direction is to generalize our formulae into the full sector of $\cN=4$ SYM, and to include the $g_{\rm YM}$ loop corrections.
Perturbatively, the loop corrections are computed by taking the OPE with the insertion of interaction terms in the Lagrangian, and integrating over the insertion point in spacetime.
The integrand of $n$-point functions at $\ell$-loop can be rewritten roughly as the $(n+\ell)$-point function at tree-level.
Pants decomposition in Section \ref{sec:PF pants} may be useful to deal with such quantities.

The second direction is to compare our results with the BPS correlators derived by the $S^4$ localization \cite{Gerchkovitz:2016gxx,Rodriguez-Gomez:2016ijh,Rodriguez-Gomez:2016cem}. 
Whether one can recast the generating function of BPS correlators as the $\tau$-function of an integrable system is an interesting question.

The third direction is to apply our formalism to the determinant-like operators, or equivalently the LLM geometry \cite{Lin:2004nb}. It has recently been conjectured that certain states on the LLM backgrounds are isomorphic to those of $\cN=4$ SYM, and the correlators are related by the redefinition of $N_c$ \cite{deMelloKoch:2018ert}. This conjecture may be checked by rewriting our results in an appropriate representation basis.

The last direction is to employ our Wick-contractions techniques for studying integrable systems in the free-field representation \cite{Zamolodchikov:1978xm,Faddeev:1980zy}. 
The skeleton reduction in Section \ref{sec:reduction} is analogous to the hexagon program \cite{Eden:2017ozn,Bargheer:2017nne,Bargheer:2018jvq}.
Then, the pants decomposition in Section \ref{sec:PF pants} may give a hint on the three-point functions or related form-factors in terms of Quantum Spectral Curve along the line of \cite{Kim:2017sju,Cavaglia:2018lxi}.

\medskip \noindent
{\bf Acknowledgments}

RS thanks Hiroyuki Fuji and Sanjaye Ramgoolam for discussions and comments on the manuscript.

\appendix

\section{Notation}\label{app:notation}

We denote an indexed set $\alpha_1 \,, \alpha_2 \,, \dots$ collectively by $\{ \alpha_i \}$. 
A multiple sum or product is written as
\begin{equation}
\sum_{\alpha_1} \sum_{\alpha_2} \dots f( \alpha_1, \alpha_2, \dots) = 
\sum_{\{\alpha_i\}} f (\{ \alpha_i\}) .
\end{equation}
The tensor product of groups is written as $\otimes_i G_i$\,, or as $\prod_i G_i$ when any elements of $G_i$ and $G_j$ $(i \neq j)$ commute.

\subsection{Permutations}

\begin{figure}[t]
\begin{center}
\includegraphics[scale=1]{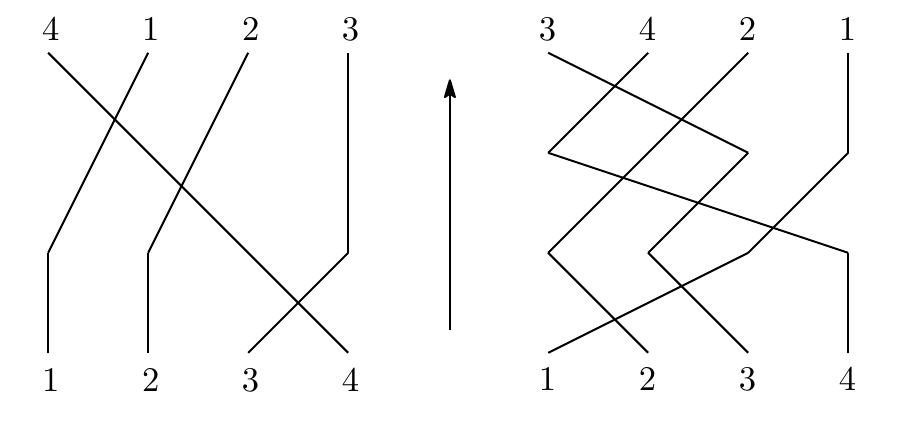}
\caption{Multiplication of permutations represented as diagrams, showing $(123)(34)=(1234)$ and $(123)(1234)(132)=(2314)$.}
\label{fig:perm notation}
\end{center}
\end{figure}

A cyclic permutation is denoted by $(i_1 i_2 \dots i_\ell)$, and $(i)(j) \dots$ represents an identity element.\footnote{Usually we do not put commas inside the bracket for permutation cycles.}
We define the permutation action by
\begin{equation}
\sigma \,:\, \{ v (1), \dots, v (n) \} \ \mapsto \ \{v' (1), \dots , v' (n) \}, \qquad
v'(\sigma(n)) = v(n) .
\end{equation}
For example, we have
\begin{equation}
\sigma = (123) \,:\, v = \{a,b,c\} \ \mapsto \ v' = \{c,a,b\}
\end{equation}
as represented in Figure \ref{fig:perm notation}.
We find the identity
\begin{equation}
\alpha \cdot \beta \( n \) = \alpha ( \beta (n) ), \qquad
\alpha (i_1 i_2 \dots i_\ell) \, \alpha^{-1} = (\alpha(i_1) \alpha(i_2) \dots \alpha (i_\ell)).
\label{perm identities}
\end{equation}
In {\tt Mathematica}, the group multiplication $\alpha \cdot \beta$ is implemented by {\tt PermutationProduct[$\beta$,$\alpha$]}.

The $\delta$-function on the symmetric group is defined by
\begin{equation}
\delta_L (\sigma) = \begin{cases}
1 &\qquad (\sigma = 1 \in S_L) \\
0 &\qquad ({\rm otherwise}) .
\end{cases}
\label{def:deltaL}
\end{equation}
We denote the conjugacy class of $\alpha$ by $[\alpha]$,
\begin{equation}
[\alpha] \equiv \frac{1}{\hL!} \sum_{\sigma \in S_{\hL}} \sigma^{-1} \alpha \sigma \,.
\label{def:conj class}
\end{equation}
Generally, two conjugacy classes commute; $[\alpha_1] [\alpha_2] = [\alpha_2] [\alpha_1]$.
The conjugacy class of $S_L$ is characterized by its cycle type, or equivalently a partition of $L$. We denote the partition of a positive integer by
\begin{equation}
\lambda \equiv [ 1^{\lambda_1} \, 2^{\lambda_2} \, \dots \, L^{\lambda_L} ] \vdash L
\label{def:partition of integer}
\end{equation}
which means
\begin{equation}
\sum_{k=1}^L k \, \lambda_k = L , \qquad
\sum_{k=1}^L \lambda_k \equiv C(\lambda).
\label{def:cycle type}
\end{equation}
Here $C(\lambda)$ counts the number of cycles in $\lambda \in S_L$\,, e.g.
\begin{equation}
C({\rm identity}) = C \Bigl( (1)(2)(3) \dots (L) \Bigr)=L, \qquad
C \Big((12) \Big) = C \Bigl( (12)(3) \dots (L) \Bigr)=L-2 .
\label{example C(perm)}
\end{equation}
Often we use $[\alpha]$ to signify the cycle type of $\alpha$.

The wreath product group is denoted by $S_L [G]$, which is a composition of $G^{\otimes L}$ and $S_L$ permuting the order of the copies of $G$'s. The order of this group is given by
\begin{equation}
\Big| S_L [G] \Big| = L! \, |G|^L .
\end{equation}
For example, the action of $S_L[\bb{Z}_2]$ on $L$ pairs is generated by $( \{ \pi_i \}, \sigma)$, where
\begin{equation}
\bb{Z}_2^{\otimes L} \ni \pi_i \, : \, (a_{2i-1} \,, a_{2i}) \ \mapsto \ (a_{2i} \,, a_{2i-1}), \qquad
S_L \ni \sigma \, : \, (a_{2k-1} \,, a_{2k}) \ \mapsto \ (a_{2\sigma(k)-1} \,, a_{2\sigma(k)}).
\label{app:SLZ2 action}
\end{equation}

The stabilizer subgroup is denoted by
\begin{equation}
{\rm Stab}_G (x) = \pare{ g \in G \mid g(x)=x } \ \subset \ G.
\end{equation}
Here $g(x)$ can be any group action, such as $g(x) = g x g^{-1}$.
By the orbit-stabilizer theorem, the number of inequivalent elements of the group coset is given by $|G|/|{\rm Stab}_G (x)|$.

\subsection{Permutation basis of operators}\label{app:perm basis}

We express a multi-trace operator of length $L$ as the equivalence class in the permutation group $S_L$\,. Let $\cO^{\vec A}_\alpha$ be a scalar multi-trace operator,
\begin{equation}
\cO^{A_1 A_2 \dots A_L}_\alpha = \tr_{L} \( \alpha \, \Phi^{A_1} \Phi^{A_2} \dots \Phi^{A_L} \)
\equiv \sum_{a_1, a_2 \dots, a_L = 1}^{N_c} (\Phi^{A_1})^{a_1}_{a_{\alpha (1)}}
(\Phi^{A_2})^{a_2}_{a_{\alpha (2)}} \dots
(\Phi^{A_L})^{a_L}_{a_{\alpha (L)}} \,.
\label{app:pm basis}
\end{equation}
Here $\alpha \in S_L$ defines the color structure of $\cO^{\vec A}_\alpha$. If $\alpha \in \bb{Z}_L$\,, then $\cO^{\vec A}_\alpha$ becomes a single-trace operator, usually denoted by
\begin{equation}
\cO^{\vec A}_\alpha = \tr \( \Phi^{A_1} \Phi^{A_2} \dots \Phi^{A_L} \), \qquad
(\alpha \in \bb{Z}_L).
\end{equation}
For later purposes, we introduce another notation for the same operator by
\begin{equation}
\cO^{A_1 A_2 \dots A_L}_\alpha =
\( \prod_{p=1}^L \delta^{a_{p'}}_{a_{\alpha (p)}} \) 
\sum_{ \{ a_p , a_{p'} \} } 
(\Phi^{A_1})^{a_1}_{a_{1'}}
(\Phi^{A_2})^{a_2}_{a_{2'}} \dots
(\Phi^{A_L})^{a_L}_{a_{L'}} \,.
\label{def:double pm basis}
\end{equation}
The $\delta$-function prefactor can be represented by another permutation
\begin{equation}
\iota_\alpha \equiv \prod_{p=1}^L (\alpha(p) \, p') \, \in \, \bb{Z}_2^{\otimes L} \,.
\label{def:iota}
\end{equation}

We can reorder the flavor indices in RHS of \eqref{app:pm basis}, which leads to the gauge symmetry,
\begin{equation}
\cO^{A_1 A_2 \dots A_L}_\alpha = \cO^{A_{\gamma(1)} A_{\gamma(2)} \dots A_{\gamma(L)}}_{\gamma^{-1} \alpha \gamma} \,, \qquad
(\forall \gamma \in S_L ). 
\label{def:gauge symm}
\end{equation}
If we neglect the spacetime dependence, the scalar fields obey the $U(N_c)$ Wick-contraction rule,
\begin{equation}
\contraction{(}{\Phi^A}{)_a^b (x_i) \, (}{\Phi^B}
(\Phi^A)_a^b (x_i) \, (\Phi^B)_c^d (x_j) = g^{AB} (x_{ij}) \, \delta^d_a \, \delta^b_c \,, \qquad
g^{AB} (x_{ij}) = \tilde g^{AB} \, |x_i - x_j|^{-2} 
\label{UNc Wick}
\end{equation}
where $\tilde g^{AB} = \delta^{AB}$ if $\Phi^A$ is a real scalar of $\cN=4$ SYM.
If the gauge group is $SU(N_c)$, we replace the scalar fields of $SU(N_c)$ theory by those of $U(N_c)$ theory as
\begin{equation}
(\Phi^{SU(N_c)})_a^b \equiv (\Phi^{U(N_c)})_a^b - \frac{\delta_a^b}{N_c} \, \tr (\Phi^{U(N_c)}) .
\end{equation}

Let us write the two-point functions of scalar operators of $\cN=4$ SYM as
\begin{equation}
G_2 (\alpha_1 , \alpha_2) = \vev{ \cO_{\alpha_1} (x_1) \cO_{\alpha_2} (x_2) } , \qquad
(\alpha_1 \,, \alpha_2 \in S_L ).
\label{def:2pt} 
\end{equation}
The RHS is given by the sum over all possible Wick-contractions,
\begin{equation}
\contraction[2.5ex]{G_{\alpha_1 \alpha_2} = \Big\langle \prod_{p=1}^L (}{\Phi^{A_p}}{)^{a_p}_{a_{\alpha_1 (p)}} (x_1) \prod_{q=1}^L (}{\Phi^{B_q}}
G_{\alpha_1 \alpha_2} = \Big\langle 
\prod_{p=1}^L (\Phi^{A_p})^{a_p}_{a_{\alpha_1 (p)}} (x_1)
\prod_{q=1}^L (\Phi^{B_q})^{b_q}_{b_{\alpha_2 (q)}} (x_2) \Big\rangle
= \sum_{\sigma \in S_L} \prod_{p,q=1}^L \, 
g^{A_p B_q} (x_{12}) \, \delta^{a_p}_{b_{\alpha_2 (q)}} \, 
\delta^{b_q}_{a_{\alpha_1 (p)}} \ \Big|_{q = \sigma^{-1} (p)} \,.
\label{Wick 2pt}
\end{equation}
The product over the color indices can be simplified by relabeling,
\begin{equation}
\prod_{p=1}^L \, 
\delta^{a_p}_{b_{\alpha_2 \sigma^{-1} (p)}} \, 
\delta^{b_{\sigma^{-1} (p)}}_{a_{\alpha_1 (p)}} 
=
\prod_{p=1}^L \, 
\delta^{a_p}_{b_{\alpha_2 \sigma^{-1} (p)}} \, 
\prod_{r=1}^L
\delta^{b_{\sigma^{-1} (r)}}_{a_{\alpha_1 (r)}} 
=
\prod_{p=1}^L \, 
\delta^{a_p}_{a_{\alpha_1 \sigma \alpha_2 \sigma^{-1} (r)}} 
\end{equation}
where we used $\alpha_2 \sigma^{-1} (p) = \sigma^{-1} (r)$.
This gives
\begin{equation}
G_{\alpha_1 \alpha_2} = \sum_{\sigma \in S_L} \( \prod_{p} \, g^{A_{\sigma^{-1} (p)} B_p} \)
\delta_L ( \Omega \, \alpha_1 \sigma^{-1} \alpha_2 \sigma ) 
\label{two-point MT correlator}
\end{equation}
where the $\delta$-function is given in \eqref{def:deltaL}. The symbol $\Omega$ is an element of the group algebra $\bb{C} [S_L]$
\begin{equation}
\Omega \equiv \sum_{\omega\in S_L} N_c^{C(\omega)} \, \omega^{-1} .
\label{def:Omega}
\end{equation}
where $C(\omega)$ is in \eqref{def:cycle type}.
Thanks to the property $C(\omega) = C(\gamma \, \omega \, \gamma^{-1})$, the $\Omega$ is a class function
\begin{equation}
\Omega = \gamma \, \Omega \, \gamma^{-1} \qquad (\forall \gamma \in S_L) .
\end{equation}

\subsection{Feynman diagrams in the double-line notation}\label{app:double line}

The space of Wick-contractions $\cW$ represents the sum over all Feynman diagrams.
In the double-line notation, Feynman diagrams become ribbon graphs owing to the cyclic ordering at each vertex. The graph data of vertices, edges, and faces uniquely specifies a ribbon graph.
Thus we introduce a triple of permutations to represent the graph data:\footnote{The permutations $(\alpha, W, \omega)$ are denoted by $(\sigma_0 , \sigma_1 , \sigma_\infty)$ in \cite{Koch:2010zza}.}
\begin{equation}
\( \text{Vertices, Edges, Faces} \) \quad \leftrightarrow \quad 
\( \alpha, W, \omega \) \, \in \, (S_{2 \hL}, \bb{Z}_2^{\otimes \hL} , S_{2 \hL}) , \quad
\omega = \alpha W.
\label{def:perm triples}
\end{equation}
Here the vertex permutation $\alpha$ knows the color structure of external operators $\{ \cO_i \}$, and the face permutation $\omega$ knows the color-index loops.
By using permutations, we can represent the Feynman graph as the Cayley graph of the composition of $(\alpha,W)$.\footnote{The Cayley graph $\Gamma(G)$ is a colored directed graph, defined canonically from the generators of a finite group $G$.}

\begin{figure}[t]
\begin{center}
\includegraphics[scale=0.75]{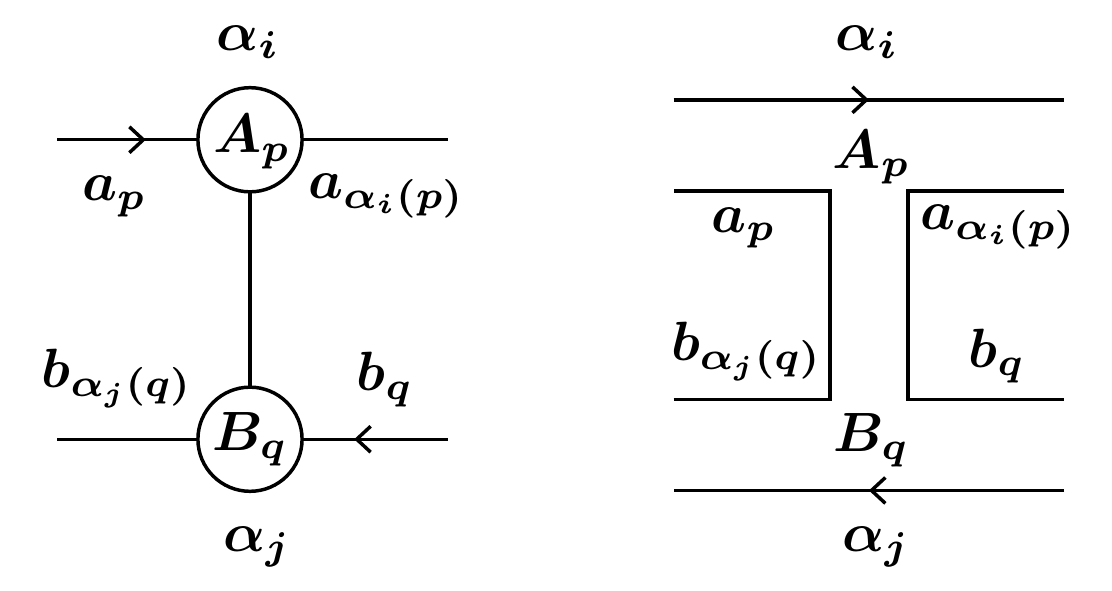}
\caption{(Left) Usual Feynman graph as the Cayley graph of $(\alpha, W)$. (Right) The Feynman graph in the double-line notation as the Cayley graph of $(\iota_\alpha \,, W)$, representing the Wick-contraction $\vev{ (\Phi^{A_p})^{a_p}_{a_{\alpha_i(p)}} \, (\Phi^{B_q})^{b_q}_{b_{\alpha_j(q)}} }$.}
\label{fig:Feynman}
\end{center}
\end{figure}

The Feynman graphs in the double-line notation can be understood as a Cayley graph in the following way.
We rewrite the permutation basis of operators in the double-index notation as in \eqref{def:double pm basis}. Now a scalar field carries one flavor half-edge and two color half-edges. 
All half-edges are paired by the pair of involutions $(\iota_\alpha , W)$ with $\iota_\alpha = \prod_{i=1}^n \iota_{\alpha_i}$ in \eqref{def:iota}.
The Cayley graph of the composition of $(\iota_\alpha , W)$ is the usual Feynman graph in the double-line notation; see Figure \ref{fig:Feynman}.

The Feynman graphs in the double-line notation have the ribbon graph structure.
A ribbon graph is a graph with cyclic ordering at each vertex, which is represented by $\alpha$ or $\omega$ in our setup.

\subsection{Summary of skeleton reduction}\label{app:skeleton}

Let us present an overview of our notation in Section \ref{sec:reduction}, where we introduced various different symbols for making our argument straightforward.

The number of all Wick-contractions between $\cO_i$ and $\cO_j$ is denoted by $\{ \ell_{ij} \}$ in \eqref{Wicknum total n-pt}.
We introduce two more symbols as the refinement of $\{ \ell_{ij} \}$ in \eqref{sum lij-rho} and \eqref{partition of Li}.
The first one gives a partition of $\ell_{ij}$ where each segment represents  the consecutive sets of Wick-edges between $\cO_i$ and $\cO_j$\,,
\begin{equation}
\ell_{ii}^{(\rho)} = 0 , \qquad
\ell_{ij}^{(\rho)} = \ell_{ji}^{(\rho)} , \qquad
\ell_{ij} = \sum_{\rho=1}^{r_{ij}} \ell_{ij}^{(\rho)} , \qquad
\sum_{i<j}^n \sum_{\rho=1}^{r_{ij}} \ell_{ij}^{(\rho)} = \hL .
\end{equation}
The second one defines a partition of $L_i$ and of the operator $\cO_i$\,,
\begin{equation}
\bsl_i \equiv ( \bsl_{i|1} \,, \bsl_{i|2} \,, \dots \,, \bsl_{i|\bL_i} ) \vdash L_i \,, \qquad
L_i = \sum_{q=1}^{\bL_i} \bsl_{i|q} \,.
\end{equation}
Again each segment represents the consecutive Wick-contractions. Hence, both symbols are relabeling of the same object.
The collection of $\{ \bsl_i \}$ gives a partition of $2 \hL$
\begin{equation}
\bigsqcup_{i=1}^n \, \bsl_i = [1^{2 \bsr_1} 2^{2 \bsr_2} \dots ] \vdash 2 \hL \,.
\end{equation}
We take half of this partition and define
\begin{equation}
\[ \bsl \] = [1^{\bsr_1} 2^{\bsr_2} \dots ] \vdash \hL, \qquad
\sum_k \bsr_k = \bL
\end{equation}
where $\bsr_k$ is the number of consecutive Wick-contractions with $k$ fields.
The partition $[\bsl] \vdash \hL$ is related to the skeleton reduction, and should not be confused by the cycle type of $\alpha = \prod_i \alpha_i$ denoted by $\lambda = [1^{\lambda_1} \, 2^{\lambda_2} \dots ] \vdash 2 \hL$.
An example of $\ell_{ij}^{(\rho)}$ is illustrated in Figure \ref{fig:ReducedGraph}, and another example of $\bsl_i$ is given in Figure \ref{fig:length_partitioned}.

\bigskip
One definition of the skeleton reduction is to assemble scalar fields in $\cO_i$ into small groups.
Let us regroup the color (or flavor) indices as
\begin{equation}
p \in \pare{ 1, 2, \dots , L_i } \ = \ 
\Bigl\{ ( \sfp, \bp) \, \Big| \, \bp \in \pare{ 1,2, \dots, \ell_{\sfp} },
\sfp \in \pare{ 1, 2, \dots, \bL_i } \Bigr\}
\end{equation}
where $\sfp$ labels the open end-points and $\bp$ labels the internal indices as
\begin{equation}
\cO_i = \prod_{p=1}^{L_i} \, (\Phi^{A_p})^{a_p}_{a_{\alpha_i (p)}}
\quad \stackrel{{\rm partition}}{\longrightarrow} \quad
\olcO_i 
= \prod_{\sfp=1}^{\bL_i} \, (\bsPhi^{(i)}_p)^{a_{\sfp}}_{a_{\bar \alpha_i (\sfp)}} 
= \prod_{\sfp=1}^{\bL_i} \ 
\prod_{\bp=1}^{\ell_\sfp} (\Phi^{A_{\sfp, \bp}})^{a_{\sfp + \bp -1 }}_{a_{\sfp + \bp} } .
\end{equation}
We call $\olcO_i$ a reduced operator, and $\bsPhi^{(i)}_\sfp$ a sequence or sequential fields. Each $\bsPhi^{(i)}_\sfp$ consists of $\ell_\sfp$ consecutive fields.
There are many ways to create a reduced operator from $\cO_i$\,. We call $\{ \olcO_i \}$ a partition of $\cO_i$\,.

We impose the following rule for a pair of sequential fields,
\begin{equation}
\contraction[1.5ex]{( }{{\bf \Phi}}{^{{\bsA}^{(i)}_{p +1 \sim p+\ell}} )^{a_{p+1}}_{a_{p+\ell+1}} \, 
( }{{\bf \Phi}}
( {\bf \Phi}^{{\bsA}^{(i)}_{p +1 \sim p+\ell}} )^{a_{p+1}}_{a_{p+\ell+1}} \, 
( {\bf \Phi}^{{\bsA}^{(j)}_{q +1 \sim q + \ell'}} )^{b_{q+1}}_{b_{q+\ell'+1}} 
= \delta_{\ell, \ell'} \, \sum_{\bW} \ \prod_{k=1}^\ell \, 
g^{A_{p+k} B_{q+k'}} \, 
\delta^{a_{p+k}}_{b_{q+k'+1}} \, \delta^{b_{q+k'}}_{a_{p+k+1}} \ 
\Big|_{q + k' = \bW^{-1} (p + k)} 
\end{equation}
which we call an open two-point function. 
Using the $\delta$-functions of color indices in the open two-point function, we define two functions $R$ and $Z$ as
\begin{equation}
\prod_{k=1}^{\ell} \, \delta^{a_{p+k}}_{b_{\alpha_j \bW^{-1} (p+k)}} \, \delta^{b_{\bW^{-1}(p+k)}}_{a_{\alpha_i (p+k)}} 
\equiv N_c^{Z (\alpha_i \bW \alpha_j \bW^{-1} | \, a_{p+1} \,, b_{q+1} )} \,
R^{a_{p+1} \, b_{q+1}}_{a_{p+\ell+1} \, b_{q+\ell+1} } \,.
\end{equation}
The properties of these functions are studied at the end of Section \ref{sec:open 2pt}.

After the skeleton reduction, the Wick-contractions are taken in two steps. 
The first step is to take all pairs of sequential fields $( \bsPhi^{(i)}_\sfp, \bsPhi^{(j)}_\sfq)$, and the second step is to take the sum inside open two-points. 
Correspondingly, the pairing map $W \in \bb{Z}_2^{\otimes \hL}$ in \eqref{def:gWginv} can be rewritten as
\begin{equation}
W \, \ni \, (a_p \,, b_q) = (a_{\sfp, \bp} \,, b_{\sfq, \bq} ) \quad \leftrightarrow \quad
(a_\sfp, b_\sfq) \times (\bp, \bq) \, \in \, (V, \tau).
\end{equation}
We call $V$ an external Wick-contraction, and $\tau$ an internal Wick-contraction.

We write the union of the double-index $(\sfp, \bp)$ as
\begin{equation}
P \in \pare{ 1, 2, \dots , 2 \hL } \ = \ 
\Bigl\{ (\sfP, \bp) \, \Big| \, \bp \in \pare{ 1,2, \dots, \ell_{\sfP} },
\sfP \in \pare{ 1, 2, \dots, 2 \bL } \Bigr\} .
\end{equation}
The symbol $\sfP$ describes the partitions of $\{ \cO_i \}$,
\begin{equation}
\pare{1,2,\dots, 2\bL} \ni \sfP 
\qquad \leftrightarrow \qquad
(\sfp_1 \,, \sfp_2 \,, \dots \,, \sfp_n), \quad 
\sfp_i \in \pare{ 1, 2, \dots, \bL_i }, \qquad
\sum_{i=1}^n \bL_i = 2 \bL .
\end{equation}
The skeleton-edges can be represented in two notations, $\sfP=1,2,\dots, 2\bL$ and $E=1,2,\dots, \bL$.
The symbol $E$ is rewriting of $(ij,\rho)$,
\begin{equation}
\sum_{E=1}^{\bL} \ell_E = \hL , \qquad
\sum_{j \neq i} r_{ij} = \bL_i \,, \qquad
\sum_{i < j} r_{ij} = \bL \,.
\end{equation}
The symbol $\sfP$ is a double copy of $E$. For example, we find the identity
\begin{equation}
E \sim (\sfP, \sfQ), \qquad
\tau_\sfP = \tau_\sfQ \quad \Leftrightarrow \quad
(a_\sfP \, a_\sfQ) \in V .
\end{equation}

\section{Another face-based formula}\label{sec:face rim}

In Section \ref{sec:face-based formula} we derived a face-based formula of $n$-point functions.
Below we discuss another formula by using the rims labeled by $\alpha$, as illustrated in Figure \ref{fig:half-edge}.
It turns out that the constraint $\delta_{2 \hL} ( \alpha^{-1} \omega W )$ disappears in this labeling, but another constraint must be imposed.

Let us denote the cycle type of $\omega$ by $[\omega] = [1^{w_1} 2^{w_2} \dots ] \vdash 2 \hL$,
\begin{equation}
C(\omega) = \sum_f w_f \,, \qquad
2 \hL = \sum_f f \, w_f \,.
\end{equation}
The Feynman graph has $w_f$ faces with $f$ edges, and $C(\omega)$ color-index loops. We have $w_1=0$ because the self-contractions are forbidden. Each face defines a cyclic translation by going around the boundary of the face counterclockwise. We can parametrize $\omega$ as a product of cyclic permutations,
\begin{equation}
\omega = \prod_{R=1}^{C(\omega)} \Phi_R 
= \prod_{R=1}^{C(\omega)} ( \phi_{R,1} \phi_{R,2} \dots \phi_{R,f_R} ) 
\label{def:Phi-F}
\end{equation}
or using $\{ w_f \}$ as
\begin{equation}
\omega = (\phi^{(2)}_{1,1} \phi^{(2)}_{1,2}) (\phi^{(2)}_{2,1} \phi^{(2)}_{2,2}) \dots (\phi^{(2)}_{w_2,1} \phi^{(2)}_{w_2,2})
(\phi^{(3)}_{1,1} \phi^{(3)}_{1,2} \phi^{(3)}_{1,3}) \dots
(\phi^{(3)}_{w_3,1} \phi^{(3)}_{w_3,2} \phi^{(3)}_{w_3,3}) \dots .
\label{parametrize omega-phi}
\end{equation}

The face permutation $\omega$ written as sequences of rims define Wick-contractions. For example, 
\begin{equation}
\omega \ni ( \dots, p, q, r \dots ) \quad \Rightarrow \quad
W \ni \(p ,\alpha^{-1}(q) \) \(q , \alpha^{-1}(r) \) .
\end{equation}
For each factor inside $\omega$ in \eqref{def:Phi-F}, we introduce the function on faces by
\begin{equation}
\bb{F} \cdot \Phi_R = N_c \( \prod_{k=1}^{f_R} h^{A_{\phi_{R,k}} , \, A_{\alpha^{-1} (\phi_{R,k+1} )}} \)^{1/2}, \qquad
\Bigl( \alpha = \prod_{i=1}^n \alpha_i \in S_{2\hL} \Bigr)
\label{def:face function}
\end{equation}
and write $n$-point functions as $G_n = \sum_{\omega} \ \bb{F} ( \omega )$.

Let us sum over $\omega$ in two steps. The first sum is over the cycle type $w \vdash 2L$ and the second sum over $\omega$'s at a fixed cycle type. 
The latter sum can be generated by
\begin{equation}
\cW (w) \equiv \pare{ \gamma \, \omega_0 \, \gamma^{-1} \, \Big| \, \gamma \in S_{2 \hL} / {\rm Aut} (\omega_0) }, 
\qquad
{\rm Aut} (\omega_0) = \prod_{f \ge 1} S_{w_f} [\bb{Z}_f ]
\label{def:cWw}
\end{equation}
where $\omega_0$ is a fixed (or representative) permutation with the cycle type $w$.
The division by ${\rm Aut} (\omega_0)$ removes the redundancy of relabeling in \eqref{parametrize omega-phi}.

The space $\cW (w)$ in \eqref{def:cWw} contains various types of unphysical Wick-contractions.
The first type is self-contractions. 
This happens when two consecutive numbers $( \dots, p, q \dots) \in \omega$ come from the same operator, $(p,q) \in \alpha_i$\,, or when $\omega$ contains one-cycles, $(p) \ni \omega$.
We remove the self-contractions by the property of $h^{AB}$ in \eqref{def:hAB}. 
The second type comes from the faces with the wrong orientation.
According to our definition of $\bb{F}$ in \eqref{def:face function}, two cycles which differ by orientation, $( \phi_{1} \phi_{2} \dots \phi_{f} )$ and $( \phi_{f} \phi_{f-1} \dots \phi_{1} )$, produce different flavor factors.
We should choose the orientation of each face so that they are consistent with the global choice. When wrongly oriented faces are glued together, the square-root on $h^{AB}$ in \eqref{def:face function} remains unresolved. 
In terms of permutations, this condition for the consistent orientation can be rephrased as
\begin{equation}
\omega \ \ni \ ( \dots p \,, \alpha(q) \dots ) \quad \Leftrightarrow \quad
\omega \ \ni \ ( \dots q \,, \alpha(p) \dots ), \qquad (\forall p, q)
\label{local consistency omega}
\end{equation}
where the two sequences $\dots p \,, \alpha(q) \dots$ and $\dots q \,, \alpha(p) \dots$ may belong to the same cycle. 
Algebraically, this condition comes from $\omega = \alpha W$ in \eqref{def:dual omega}. If we write $W \ni (p \, q)$, then the equation \eqref{local consistency omega} implies
\begin{equation}
\omega \ \ni \ ( \dots p \,, \alpha W (p) \dots ) \quad \Leftrightarrow \quad
\omega \ \ni \ ( \dots q \,, \alpha W (q) \dots ).
\label{local consistency omega2}
\end{equation}
These two sequences produce the same flavor factor $\sqrt{h^{A_p A_q}}$. Thus, the consistency condition \eqref{local consistency omega} or \eqref{local consistency omega2} is equivalent to the square-root-free condition; namely, the total product $\prod_{R=1}^{C(\omega)} \bb{F} \cdot \Phi_R$ in \eqref{n-pt S*2hL-1} does not have the square root of $h$'s.

Based on the above arguments, we conjecture that the $n$-point function is given by the sum over $\omega \in \cW(w)$ which does not contain $\sqrt{h}$'s:
\begin{equation}
G_n = \sum_{\substack{w \, \vdash 2 \hL \\w_1=0}} \ \prod_f \, \frac{1}{w_f! \, f^{w_f} } \,
\sum_{\gamma \in S_{2 \hL}} \ \bb{F} (\gamma \, \omega_0 \, \gamma^{-1}) \, \Big|_{\sqrt h \to 0} \,.
\label{n-pt S*2hL-2}
\end{equation}
We checked that this equation is consistent with the original $S_{2\hL}$ formula \eqref{npt Wick 2hL extended}, by computing some simple cases by {\tt Mathematica}.

\section{Vertex-based skeleton formula}\label{app:vertex skeleton}

We express $G_n$ by gluing open two-points, following the vertex-based method in Section \ref{sec:S2L formalsim}.

\subsection{Gluing open two-point}\label{sec:gluing open}

\begin{figure}
\begin{center}
\includegraphics[scale=0.65]{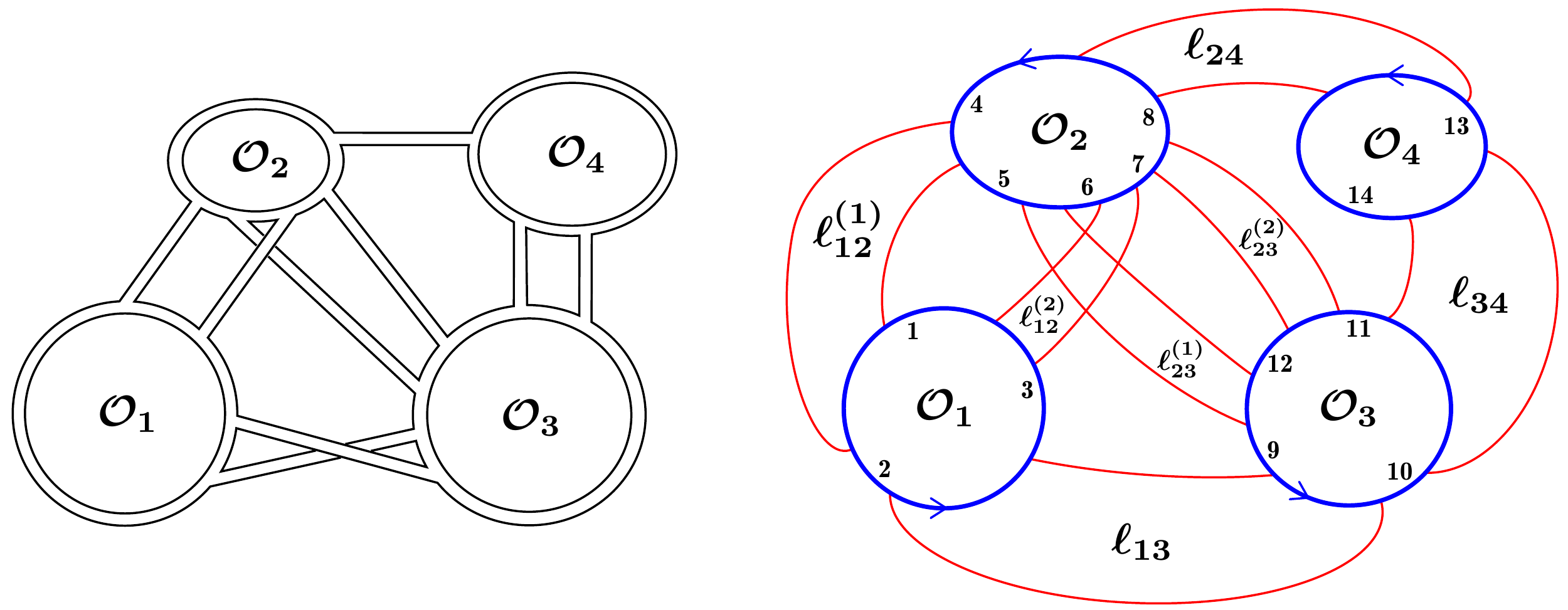}
\caption{(Left) Example of a set of the Wick-contractions in $\Vev{ \cO_1 \, \cO_2 \, \cO_3 \, \cO_4 }$ with $(L_1,L_2,L_3,L_4)=(4,5,6,3)$. 
(Right) Skeleton-reduced graph for the same set of Wick-contractions, whose edges are colored as in Figure \ref{fig:Cayley644}.
This corresponds to $V=(1,4)_1 \, (2,9)_2 \, (3,6)_1 \, (5,12)_1 \, (7,11)_1 \, (8,13)_1 \, (10,14)_2$ with $\bL=7$, where the subscript in $V$ represents the bridge length of each open two-point.}
\label{fig:ReducedGraph}
\end{center}
\end{figure}

Let us unify the color indices of $\{ \cO_i \}$ in \eqref{def:P to sfP} as
\begin{equation}
P \in \pare{ 1, 2, \dots , 2 \hL } \ = \ 
\Bigl\{ (\sfP, \bp) \, \Big| \, \bp \in \pare{ 1,2, \dots, \ell_{\sfP} },
\sfP \in \pare{ 1, 2, \dots, 2 \bL } \Bigr\} .
\end{equation}
We also unify the reduced permutations and the reduced Wick-contractions as
\begin{equation}
\bar \alpha = \prod_{i=1}^n \bar \alpha_i  \in S_{\bL} \,, \qquad
V = \prod_{\sfP=1}^{\bL} \Big(\bar\gamma(a_{2\sfP - 1}) \, \bar\gamma( a_{2 \sfP}) \Big)  \in \bb{Z}_2^{\otimes \bL} \,, \qquad
(\bar \gamma \in S_{2 \bL})
\label{def:reduced alpha V}
\end{equation}
just like \eqref{def:gWginv}.
The Cayley graph of $(V, \bar\alpha)$ is the skeleton-reduced Feynman graph, and a positive integer $\ell_E$ is associated to the $E$-th edge. Figure \ref{fig:ReducedGraph} shows an example of the skeleton reduction applied to a Feynman graph in a four-point function.

In the new notation, the bridge lengths $\{ \ell_{ij}^{(\rho)} \}$ in \eqref{sum lij-rho} simply become $\{ \ell_{\sfP} \}$, and the number of non-zero $\ell_{ij}^{(\rho)}$'s is equal to $\bL$,
\begin{equation}
\sum_{j \neq i} r_{ij} = \bL_i \,, \qquad
\sum_{i < j} r_{ij} = \bL \,, \qquad
\bL = \frac12 \sum_i \bL_i \,.
\end{equation}
Note that the range of $\sfP$ is twice as big as that of $(ij,\rho)$. 
An open two-point function has four endpoints, labeled by $\sfP$. Using the global orientation, we may use two of the four endpoints to specify the same open two-point. It means that
\begin{equation}
\tau_\sfP = \tau_\sfQ \quad \Leftrightarrow \quad
(a_\sfP \, a_\sfQ) \in V .
\label{skeleton-edges identified}
\end{equation}
The same remark applies to the color $R$-matrix, which has the symmetry $R^{ab}_{a'b'}= R^{ba}_{b'a'}$\,.

Let us rewrite $G_n$ in \eqref{npt Wick lij-rho Vtau} as\footnote{By abuse of notation, we use $(a_\sfP \, a_\sfQ)$ and $(\sfP \, \sfQ)$ interchangeably, like $(a_\sfP \, a_\sfQ) \in V$ or $(\sfP \, \sfQ) \in V$. Recall that originally $a_{\sfP}$ is the color index running from $1$ to $N_c$ for the $\sfP$-th vertex.}
\begin{align}
G_n &= \sum_{ \{ \text{partition of $\cO_i$} \} } \,
\sum_{V : \, {\rm external}} \, 
\sum_{\tau : \, {\rm internal}} \, 
\prod_{\sfP=1}^{2 \bL} \Bigg[
\( \prod_{\bp=1}^{\ell_{\sfP} } \, h^{A_{\sfP,\bp} \, A_{V(\sfP),\tau_{\sfP} (\bp)}} \)
N_c^{\bZ (\tau_{\sfP} )} \, 
R^{a_{\sfP} \, a_{V(\sfP)} }_{a_{\sfP+\ell_\sfP } \, a_{V(\sfP)+\ell_\sfP } } 
\Bigg]^{1/2} 
\label{npt open product1} \\[2mm]
&= \sum_{ \{ \text{partition of $\cO_i$} \} } \,
\sum_{V : \, {\rm external}} \, 
\sum_{\tau : \, {\rm internal}} \, 
N_c^{C(\bar\omega) + \frac12 \sum_{\sfP} \bZ (\tau_\sfP) } \, \fH (W) 
\label{npt open product3}
\end{align}
where we introduced new symbols
\begin{align}
\fH (W) &\equiv \( \prod_{\sfP=1}^{2 \bL} \prod_{\bp=1}^{\ell_{\sfP}} \, 
h^{A_{\sfP,\bp} \, A_{V(\sfP),\tau_{\sfP} (\bp)}} \)^{1/2} ,
\label{def:fHW} \\
N_c^{C(\bar\omega)} &\equiv \( \prod_{\sfP=1}^{2 \bL} \prod_{\bp=1}^{\ell_{\sfP}} \, 
R^{a_{\sfP} \, a_{V(\sfP)} }_{a_{\sfP+\ell_\sfP } \, a_{V(\sfP)+\ell_\sfP } } 
\)^{1/2} , \quad (\bar\omega \in S_{2 \bL}).
\label{def:C_bar_omega}
\end{align}
We can think of $\bar\omega$ as a face permutation, because $R$ is a pair of $\delta$-functions as discussed in \eqref{Rab selection rule}.
The equation \eqref{def:C_bar_omega} makes sense because RHS has $4 \bL$ color indices, and the same index always appears as a superscript and a subscript, once for each.
Furthermore, by comparing \eqref{npt Wick 2hL} and \eqref{npt open product3}, we find
\begin{equation}
C(\alpha W) = C(\bar\omega) + \frac12 \sum_{\sfP} \bZ (\tau_\sfP)
= C(\bar\omega) 
+ \sum_{i<j} \sum_{\rho} Z \(\alpha_i \, W_{ij,\rho} \, \alpha_j \, W_{ij,\rho}^{-1} \mid \{\text{end-points}\} \)
\label{identity face function}
\end{equation}
where we used \eqref{def:bZ tau}.
This means that $\bZ$ counts the number of cycles without open end-points, and $C(\bar\omega)$ counts those with open end-points.

\subsection{Face permutations of the skeleton graph}\label{sec:reduced face}

There exist two face permutations for the skeleton graph. The first one is given by
\begin{equation}
\nu \equiv \bar \alpha \, V \ \in \ S_{2 \bL} 
\label{def:nu}
\end{equation}
which generates the faces of the skeleton graph. Since the skeleton graph triangulates a Riemann surface, $\nu$ does not have one- and two-cycles.
The second one is $\bar\omega \in S_{2 \bL}$ introduced in \eqref{def:C_bar_omega}, which counts the color-index cycles and follows from the skeleton reduction of the original face permutation,
\begin{equation}
\text{Reduction} \, : \, \omega \ \mapsto \ \bar\omega .
\label{reduce omega}
\end{equation}
If we consider the ladder-skeleton reduction (assembling only planar ladders), $\nu$ and $\bar\omega$ are identical.

\begin{figure}[t]
\begin{center}
\includegraphics[scale=0.75]{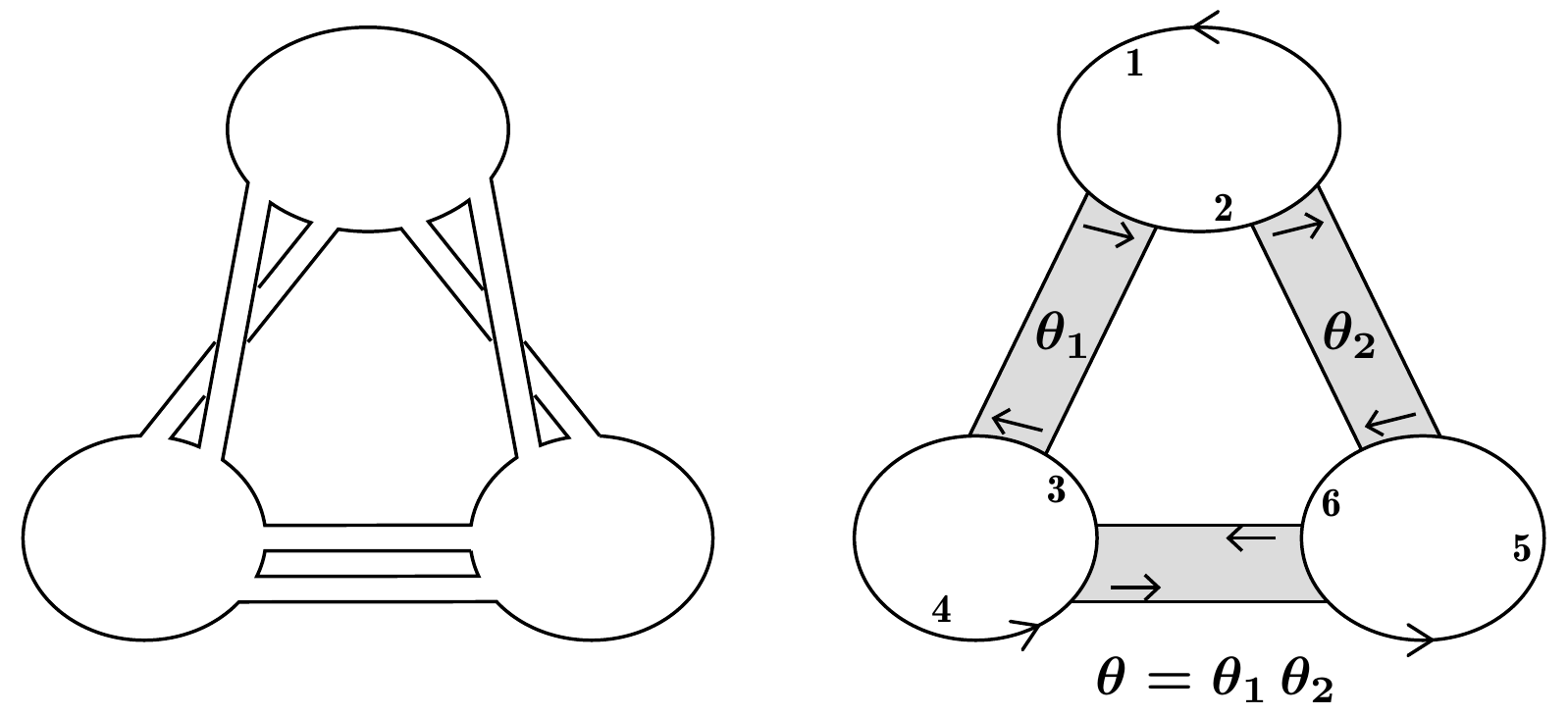}
\caption{(Left) Original Feynman graph, (Right) Skeleton-reduced graph with $\bar\alpha=(12)(34)(56)$ and $V=(13)(25)(46)$. The two face permutations of the skeleton graph are $\nu=(145)(263)$ which neglects the internal Wick-contractions, and $\bar\omega=(12)(3456)$ which remembers the internal structure through $\theta = (13)(25)$.}
\label{fig:theta_nu}
\end{center}
\end{figure}

The two face permutations $\nu$ and $\bar\omega$ agree if the matrix $R^{ab}_{a'b'}$ in \eqref{def:R2pt} takes the planar form, $R^{ab}_{a'b'} = \delta^a_{b'} \, \delta^b_{a'}$\,. For non-planar cases, we can define another involution $\theta \in \bb{Z}_2^{\bL}$ such that
\begin{equation}
\nu \equiv \bar \omega \, \theta, \qquad \theta^2 = 1.
\label{def:theta}
\end{equation}
This $\theta$ also follows from $\tau$ in \eqref{def:bW tau} thanks to the selection rule \eqref{Rab selection rule}. An example of the relation \eqref{def:theta} is illustrated in Figure \ref{fig:theta_nu}.

\subsection{Space of the reduced Wick-contractions}\label{sec:space reduced Wick}

Our next goal is to construct the space of the reduced Wick-contractions in $G_n$, namely the range of three sums in \eqref{npt open product3}.

\subsubsection{Partitions of an operator}\label{app:partitions op}

We generate partitions of the operator $\cO_i$ in \eqref{def:partition of cO_i} as follows: choose an ordered partition of $L_i$\,, 
divide $\cO_i$ according to that partition, and apply $\bb{Z}_{L_i}$ translations.
An ordered partition of $L_i$ is denoted by
\begin{equation}
\bsl_i \equiv ( \bsl_{i|1} \,, \bsl_{i|2} \,, \dots \,, \bsl_{i|\bL_i} ) \vdash L_i \,, \qquad
L_i = \sum_{q=1}^{\bL_i} \bsl_{i|q} \,.
\label{partition of Li}
\end{equation}
We group the sequences of $\bsl_{i|q}$ consecutive fields together as in \eqref{def:sequence fields} to define the reduced operator
\begin{equation}
\olcO_i = \tr_{\bL_i} \( \bar \alpha_i \,  \bsPhi^{(i)}_1 \,  \bsPhi^{(i)}_2 \, \dots \, \bsPhi^{(i)}_{\bL_i} \), \qquad
\bsPhi^{(i)}_q = \bsPhi^{\bsA^{(i)}_{p_q +1 \sim p_q+ \bsl_{i|q} } } \,, \quad 
p_q = \sum_{j=1}^{q-1} \bsl_{i|j} \,,
\label{def:reduced operator}
\end{equation}
with $\bar \alpha_i \in S_{\bL_i}$\,.

For simplicity, we assume that $\bar \alpha_i \in \bb{Z}_{\bL_i}$ and $\olcO_i$ are single-traces.
We should identify the cyclically translated operators as long as the two ordered partitions are equal,
\begin{multline}
\tr \( \bsPhi^{(i)}_1 \,  \bsPhi^{(i)}_2 \, \dots \, \bsPhi^{(i)}_{\bL_i} \)
\sim \tr \( \bsPhi^{(i)}_{m + 1} \dots \, \bsPhi^{(i)}_{\bL_i} \,  \bsPhi^{(i)}_1 \, \dots \, \bsPhi^{(i)}_{m} \)
\\[1mm]
\quad \Leftrightarrow \quad
( \bsl_{i|1} \,, \bsl_{i|2} \,, \dots \,, \bsl_{i|\bL_i} ) \ = \ 
( \bsl_{i|m+1} \,, \dots \,, \bsl_{i|\bL_i} \,, \bsl_{i|1} \,, \dots \,, \bsl_{i|m} ).
\label{cyclic sequential}
\end{multline}
This identification becomes important when $\bsl_i$ has a repeated pattern.\footnote{For example, if $\bsl_i = (3,2,3,2)$, the cyclic translation $\Phi^{A_p^{(i)}} \mapsto \Phi^{A_{p+5}^{(i)}}$ generates the same reduced operator, which corresponds to the identity $\tr ( \bsPhi^{(i)}_1 \,  \bsPhi^{(i)}_2 \, \bsPhi^{(i)}_3 \,  \bsPhi^{(i)}_4 ) = \tr ( \bsPhi^{(i)}_3 \,  \bsPhi^{(i)}_4 \, \bsPhi^{(i)}_1 \,  \bsPhi^{(i)}_2 )$.}
The original $\bb{Z}_{L_i}$ symmetry generates inequivalent reduced operators, unless the equivalence relation \eqref{cyclic sequential} applies. For example, the following two reduced operators are inequivalent:
\begin{multline}
\tr \Big( (\Phi^{A_1} \Phi^{A_2}) \, (\Phi^{A_3} \Phi^{A_4}) \Big) \sim 
\tr \Big( (\Phi^{A_3} \Phi^{A_4}) \, (\Phi^{A_1} \Phi^{A_2}) \Big)
\\
\nsim 
\tr \Big( (\Phi^{A_2} \Phi^{A_3}) \, (\Phi^{A_4} \Phi^{A_1}) \Big) 
\sim \tr \Big( (\Phi^{A_4} \Phi^{A_1}) \, (\Phi^{A_2} \Phi^{A_3}) \Big) .
\label{example cyclic identification}
\end{multline}
We introduce an ordered set of $L_i$ numbers $\varpi_i$ as a function of $z \in \bb{Z}_{L_i}$\,,
\begin{multline}
\varpi_i (z) \equiv
\Big\{ z(1) \,, z(2) \,, \dots \,, z(\bsl_{i|1} ) \Big\}
\Big\{ z(p_2 + 1) \,, z(p_2 + 2) \,, \dots \,, z(p_2 + \bsl_{i|2} ) \Big\}
\dots
\\
\Big\{ z(p_{\bL_i} + 1) \,, z(p_{\bL_i} + 2) \,, \dots \,, z(p_{\bL_i} + \bsl_{i|\bL_i} ) \Big\}
\end{multline}
and denote its equivalence class by $[\varpi_i (z)]$ modulo the relation \eqref{cyclic sequential}.
In general, the number of equivalence classes is given by the orbit-stabilizer theorem, which means
\begin{equation}
\begin{gathered}
\Big( \text{Number of $[\varpi_i (z)]$} \Big) = \frac{|\bb{Z}_{L_i}|}
{|{\rm Stab}_{\, \bb{Z}_{L_i}} (\varpi_i (z_0)) |} 
\\[1mm]
{\rm Stab}_{\,\bb{Z}_{L_i}} (\varpi_i (z_0) ) = \Bigl\{
z \in \bb{Z}_{L_i} \, \Big|\,
\varpi_i (z) \sim \varpi_i (z') \ \text{as in \eqref{cyclic sequential}} \Bigr\}
\end{gathered}
\label{count varpi orbits}
\end{equation}
where $z_0$ is a reference point, e.g. identity.
In summary, the space of the partitions of $\cO_i$ can be written as
\begin{multline}
\( \text{Partition of $\cO_i$} \)
= \Bigl\{ \text{Choice of an ordered list } \, \bsl_i \vdash L_i \Bigr\} \ 
\\
\times
\pare{ \text{Choice of an inequivalent cyclic translation} \, [ \varpi_i (z) ] \, \Big| \, z \in \bb{Z}_{L_i} }.
\label{def:space of partitions of cOi}
\end{multline}

\subsubsection{Space of internal Wick-contractions}\label{app:internal Wick space}

\begin{figure}[t]
\begin{center}
\includegraphics[scale=0.8]{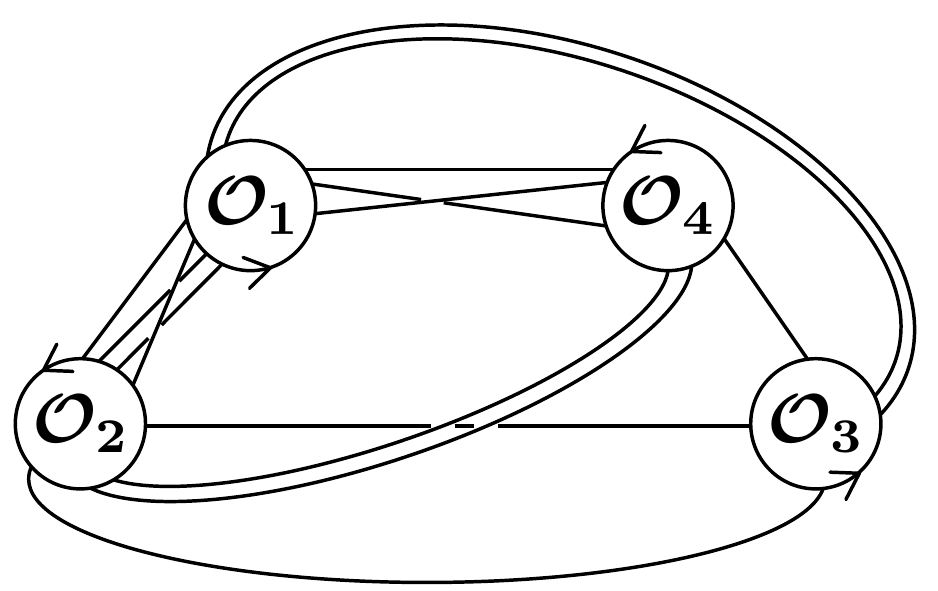}
\caption{This diagram corresponds to $[\bsl]=[1^3 \, 2^2 \, 3^1 \, 4^1], L=14$ and $\bL=7$.
The individual partitions of lengths are $\bsl_1 = (4,3,2), \bsl_2 = (4,1,2,1), \bsl_3 = (2,1,1,1)$ and $\bsl_4=(3,2,1)$, given in the counterclockwise order.
}
\label{fig:length_partitioned}
\end{center}
\end{figure}

Let us introduce a new partition of $\hL$ by\footnote{This $\[ \bsl \]$ is no longer an ordered set.}
\begin{equation}
\[ \bsl \] = [1^{\bsr_1} 2^{\bsr_2} \dots ] \vdash \hL, \qquad
C( [\bsl] ) = \sum_k \bsr_k = \bL
\label{def:partition of hL}
\end{equation}
where $\bsr_k$ is the number of consecutive Wick-contractions with $k$ fields. This partition is related to $\bsl_i = \{ \bsl_{i|q} \} $ in \eqref{partition of Li} by
\begin{equation}
\bigsqcup_{i=1}^n \, \bsl_i = [1^{2 \bsr_1} 2^{2 \bsr_2} \dots ] \vdash 2 \hL \,.
\label{def:partition of 2hL}
\end{equation}
This relation is exemplified in Figure \ref{fig:length_partitioned}.
For each open two-point, we can define the internal Wick-contractions $\tau_E \in S_{\bsl_E}$ by \eqref{def:bW tau}.
The union of $\{ \tau_E \}$ gives
\begin{equation}
\bstau \equiv \( \tau_1 \,, \tau_2 \,, \dots \,, \tau_{\bL} \) \ \in \ 
S_{\bstau} \equiv \bigotimes_{k \ge 1}  S_k^{\otimes \bsr_k} .
\label{def:S bstau}
\end{equation}

\subsubsection{Space of external Wick-contractions}\label{app:external Wick space}

One definition of this space has already been given in \eqref{def:reduced alpha V} as
\begin{equation}
V \in \pare{ \prod_{\sfP=1}^{\bL} \Big(\bar\gamma(a_{2\sfP - 1}) \, \bar\gamma( a_{2 \sfP}) \Big) 
\, \Big| \,
\bar \gamma \in S_{2 \bL} / S_{\bL} [\bb{Z}_2] }.
\end{equation}
We restrict this space, based on a given set of partitions $\{ \olcO_i \}$.
The open two-point is non-zero only if the pair of the sequential fields have the same length \eqref{def:open Wick 2pt}.
We take the external Wick-contractions over all possible pairs of the same length.
Let us write $V = (V_1 \,, V_2 \,, \dots ) \in \bb{Z}_2^{\otimes \bL}$ where $V_k \in \bb{Z}_2^{\otimes \bsr_k}$ is the external Wick-contractions of length $k$.
We sum $V_k$ over the space
\begin{equation}
V_k \in \cV_k = \pare{ \prod_{j=1}^{\bsr_k} \Big( g_k (a^{(k)}_{2j-1}) \, g_k (a^{(k)}_{2j}) \Big)
\, \Big| \, 
g_k \in S_{2 \bsr_k} / S_{\bsr_k} [ \bb{Z}_2 ] } .
\label{def:cV_k}
\end{equation}
The restricted space of external Wick-contractions is defined as the union of $\{ \cV_k \}$,
\begin{equation}
V \ \in \ \bscV \equiv \( \cV_1 \,, \cV_2 \,, \dots \,, \cV_{\bL} \) .
\label{def:cV_all}
\end{equation}

This definition of $\bscV$ still contains two types of unphysical Wick-contractions.
The first type is self-contractions, which can be removed by $h^{AB}$.
The second type is mutually adjacent pairs of open two-points.
As discussed at the beginning of this section, a skeleton graph should not have faces made of two or fewer edges.
As a result, the symbol $\bsPhi^{(i)}_p$ represents the {\it maximal} sequence of consecutive fields appearing in the original Wick-contractions. 
This condition is equivalent to the exclusion of taking external Wick-contractions from the mutually adjacent pairs.
For example, when $( \bsPhi^{(i)}_p \,, \bsPhi^{(j)}_q )$ are contracted, then the four possibilities $( \bsPhi^{(i)}_{p \pm 1} \,, \bsPhi^{(j)}_{q \pm 1} )$ are not allowed.
We define $\bscV_{\rm phys}$ as the subspace of $\bscV$ in \eqref{def:cV_all} having no mutually adjacent pairs.
A necessary but not sufficient condition for getting $\bscV_{\rm phys} \subset \bscV$ is to require that the faces have no two-cycles,\footnote{The one-cycles in $\bar \alpha V$ were removed by $h^{AB}$.}
\begin{equation}
\bscV_{\rm phys} \subsetneq \Bigl\{ V \in \bscV \, \Big| \,
C(\bar \alpha V) = C(\bar \alpha V \bar \alpha V) \Bigr\} .
\label{def:bscV phys} 
\end{equation}
RHS does not have adjacent ladder-type skeleton edges, but still contains adjacent skeleton edges which mutually cross and do not form a two-cycle.
We need to throw away external Wick-contractions of this type.

\subsubsection{Rewriting the $n$-point functions}

As discussed in Section \ref{sec:Wick relabel}, there is a bijection between the original Wick-contraction and $W$ and the triple $\( \{ \olcO_i \} , V, \bstau \)$.
The spaces \eqref{def:space of partitions of cOi}, \eqref{def:bscV phys} and \eqref{def:S bstau} offer an alternative definition of the space of (physical) Wick-contractions given in \eqref{def:moduli cW}.
We write the $n$-point functions \eqref{npt open product3} as
\begin{equation}
G'_n = \sum_{ \{ \text{partition of $\cO_i$} \} } \,
\sum_{ V \in \bscV_{\rm phys} } \, 
\sum_{ \bstau \in S_{\bstau} }
N_c^{\bZ (\bstau) + C(\bar\omega)} \, \fH (W) 
\label{npt reduced formula vertex}
\end{equation}
where
\begin{equation}
\bZ (\bstau) \equiv \frac12 \sum_{\sfP=1}^{2 \bL} \bZ (\tau_\sfP)
= \sum_{E=1}^{\bL} \bZ (\tau_E ).
\label{def:bZ tau-all}
\end{equation}
The symbol $G_n'$ in \eqref{npt reduced formula vertex} means that the disconnected parts involving a two-point function are removed. This is because the skeleton reduction of a two-point function gives a face permutation consisting of a two-cycle. We excluded such terms in \eqref{def:bscV phys}.

Let us compare the vertex-based skeleton formula \eqref{npt reduced formula vertex} and the face-based one \eqref{n-pt Sred}. In the vertex-based formula, we partition the operators $\{ \cO_i \}$ and apply the Wick contractions of open two-point functions. Throughout this process, we keep track the flavor indices of $\{ \cO_i \}$ and avoided the over-counting of the reduced operator as in \eqref{cyclic sequential}.
In contrast, in the face-based formula we sum over the skeleton-graph topologies by using face permutations. The flavor indices of $\cO_i$ are restored only in the end \eqref{gen: construct red F}, and we can sum over the orbits of cyclic translations without bothering about the over-counting.

One may argue that the face-based formula is simpler than the vertex-based one, because the former contains more fictitious degrees of freedom. Roughly speaking, the sum $V \in \bscV_{\rm phys}$ in \eqref{npt reduced formula vertex} should be smaller than $\nu \in \( S_{2 \bL}^{\times \! \times} / {\rm Aut} \, V \)_{\rm phys}$ in \eqref{n-pt Sred}, because $\bscV_{\rm phys} \subset \prod_k S_{2 \bsr_k} \subset S_{2 \bL}$.

\section{More on geometry and graphs}\label{app:geometry and graph}

\subsection{Moduli space}\label{app:moduli}

In the literature there are various definitions of the moduli space of Riemann surfaces denoted by $\cM_{g,n}$\,;
\begin{itemize}[nosep,leftmargin=16mm]
\item Space of the complex structure on $\Sigma_{g,n}$
\item Space of the conformal class of Riemannian metric $g \sim e^{2 \rho} \, g$ on $\Sigma_{g,n}$
\item Space of hyperbolic metrics modulo mapping class groups for $g \ge 2$ \cite{FM12Book}
\item Space of smooth or stable curves up to isomorphism \cite{Eynard18rev}
\item $\cdots$
\end{itemize}
The moduli space $\cM_{g,n}$ has the dimensions
\begin{equation}
\dim_{\,\bb{C}} \cM_{g,n} = n-3+3g 
\label{dim_C Mgn}
\end{equation}
which is parametrized locally by
\begin{equation}
\cM_{g,n} \simeq \begin{cases}
\{ z_4 , \dots, z_n \} \ 
&\qquad (g=0)
\\
\{ z_2 , \dots, z_n \} \ \cup \{ \tau \} \ 
&\qquad (g=1)
\\
\{ z_1 , \dots, z_n \} \ \cup \{ \mu_1 , \dots, \mu_{3g-3} \} \ 
&\qquad (g \ge 2).
\end{cases}
\label{cpx structure moduli}
\end{equation}
One way to identify $\{ \mu_a \}$ is to take pants decomposition by cutting a genus $g$ curve $(3g-3)$ times, and use Fenchel-Nielsen coordinates to fix the moduli\cite{DHoker:1988pdl}.

In general, it is not easy to construct the whole $\cM_{g,n}$ explicitly, because $\cM_{g,n}$ is not compact. 
When one approaches the boundary of $\cM_{g,n}$\,, the Riemann surface $\Sigma_{g,n}$ degenerates, either by pinching the handle or by colliding punctures.
Deligne and Mumford considered compactification of $\cM_{g,n}$ by adding boundary components in the language of algebraic geometry \cite{DM69}.

Harer, Mumford and Thurston suggested studying the moduli space of {\it decorated} Riemann surfaces $\cM_{g,n} \times \bb{R}_+^n$ instead of $\cM_{g,n}$ \cite{Harer86}.
By decorated we mean that the complex structure is induced by the quadratic differential having the double pole with the specific residue $r_i \in \bb{R}_+$ at the puncture $p_i$.
This idea works if there is a bijection between the complex structure and the quadratic differential on $\Sigma_{g,n}$\,.
This is  indeed true for the Jenkins-Strebel quadratic differential discussed in Appendix \ref{app:quadratic diff}.

\subsection{Ribbon graphs}\label{app:geometry graph}

A graph is defined as a set of vertices and edges, and we usually talk of connected unoriented graphs whose vertices are labeled. 
A metric ribbon graph is a graph with the following property:
\begin{enumerate}[nosep]
\item Each vertex has valency at least three,
\item Edges connecting to a vertex are cyclically ordered,
\item A positive real number called length is assigned to each edge.
\end{enumerate}
Any graphs drawn on a Riemann surface without intersection can be ribbonized \cite{Mondello07}.\footnote{A ribbon graph is also called a fat graph in the literature.}
For this purpose, we replace vertices by disks, edges by strips, and glue them together while keeping the cyclic ordering.

\bigskip
A topological cell decomposition of $\Sigma_{g,n}$ defines a ribbon graph, by identifying the punctures as the vertices of the graph. The cell decomposition is called complete (or ideal) triangulation if all faces of the graph are triangles. If some faces are squares or higher polygons, then the triangulation is called incomplete.

Consider a graph which completely triangulates $\Sigma_{g,n}$\,. The number of vertices, edges and faces of the graph is given by
\begin{equation}
V_{g,n} = 2n-4+4g, \quad
E_{g,n} = 3n-6+6g, \quad
F_{g,n} = n, \quad
\chi = V_{g,n}-E_{g,n}+F_{g,n} = 2-2g .
\label{app:graph data counting}
\end{equation}
For example, a planar graph can have at most $3n-6$ non-zero edges, and a toric (genus-one) graph can have $3n$ non-zero edges. 
A graph is called complete if any pairs of faces must be connected by a single edge. A complete graph exists when
\begin{equation}
\binom{n}{2} - \( 3n-6+6g \) \le 0
\label{app:nonzero edge constraints}
\end{equation}
namely, $n \le 4$ at $g=0$ and $n \le 7$ at $g=1$.

Consider the dual ribbon graph $\Gamma_{g,n}$ of an incomplete triangulation of $\Sigma_{g,n}$\,. 
We denote the number of $k$-gons in this graph by $d_k$\,, and define the total face degree of this graph by
\begin{equation}
{\rm Deg} \( \Gamma_{g,n} \) \equiv \sum_{k \ge 3} d_k \( k-2 \).
\label{def:face degree}
\end{equation}
It follows that
\begin{equation}
{\rm Deg} \( \Gamma_{g,n} \) = 2n - 4 + 4g 
\label{face degree g,n}
\end{equation}
for any complete or incomplete triangulations. This is because incomplete triangulations are generated by gluing or splitting of $a$-gons and $b$-gons
\begin{equation}
a \oplus b \quad\ \leftrightarrow \quad (a+k) \oplus (b-k)
\end{equation}
and by removing $2$-gons. This procedure leaves \eqref{def:face degree} invariant. 
The incompleteness of the triangulation can be measured by counting the number of edges,
\begin{equation}
\( \text{Number of edges in } \Gamma_{g,n} \) \le E_{g,n} = 3n - 6 + 6g.
\label{bound on edges in Gamma_gn}
\end{equation}
Since an edge has two endpoints, LHS is also equal to the half of the total valency of the vertices in $\Gamma_{g,n}$\,.

\subsection{Quadratic differentials}\label{app:quadratic diff}

A quadratic differential is a meromorphic function which satisfies the transformation rule
\begin{equation}
\varphi \equiv \phi (z) dz^2 = \phi (w) dw^2 .
\end{equation}
Given a quadratic differential, one can compute the length and area by
\begin{equation}
L(p_0,p_1) = \int_{p_0}^{p_1} \sqrt{ \varphi } , \qquad
A(\cR) = \iint_\cR dz d\bar z \abs{ \phi (z) } .
\label{def:length-area}
\end{equation}
Given a reference point $p_0$\,, the horizontal and vertical trajectories of $\varphi$ are defined by
\begin{equation}
\begin{aligned}
\text{Horizontal trajectory} \ &= \ \pare{ z \in \Sigma \ \Big| \ {\rm Im} \( \int_{p_0}^{z} \sqrt{ \varphi } \) = \text{constant} } ,
\\
\text{Vertical trajectory} \ &= \ \pare{ z \in \Sigma \ \Big| \ {\rm Re} \( \int_{p_0}^{z} \sqrt{ \varphi } \) = \text{constant} } .
\end{aligned}
\end{equation}
Below we consider the Jenkins-Strebel (JS) quadratic differential following \cite{MP97}. The JS differential satisfies a certain minimal-area condition, and it exists uniquely for each punctured Riemann surface \cite{Jenkins57,Strebel84,Zwiebach:1990nh}.

\paragraph{Definition.}

Let $\Sigma_{g,n}$ be a Riemann surface with the punctures at $\{p_1 \,, p_2 \,, \dots \,, p_n \}$. The JS quadratic differential with the residue $\{r_1 \,, r_2 \,, \dots \,, r_n \}$ satisfies the following conditions:
\begin{enumerate}[noitemsep, topsep=0mm, leftmargin=8mm]
\item $\phi$ is holomorphic on $\Sigma_{g,n} \setminus \{ p_i \}$
\item $\phi$ has a double pole at $\{ p_i \}$
\item All horizontal trajectories are compact, except for measure zero subsets of $\Sigma_{g,n}$
\item Each compact trajectory surrounds one of the poles, and the residue at $p_j$ satisfies
\begin{equation}
\oint dz \sqrt{\phi} = r_j \,.
\end{equation}
\end{enumerate}
The branch of the square root is chosen such that $r_j > 0$ for all $j$.\footnote{The indifference to the sign of residue is one of the reasons why ``quadratic'' differential is used.}

By definition, a JS differential does not have poles of order $>2$. We define
\begin{equation}
\text{(Order of $\varphi$)} = 
\sum_{j \ge 1} j \, \text{(Number of the $j$-th zeroes)}
-
\sum_{k=1,2} k \, \text{(Number of the $k$-th poles)} .
\end{equation}
It is known that the order of $\varphi$ for $\Sigma_{g,n}$ is given by\footnote{The order of a quadratic differential is twice of that of a linear differential, whose order is $2g-2$ according to the Riemann-Hurwitz theorem.}
\begin{equation}
\text{(Order of $\varphi$)} = 4g-4.
\label{order JS diff}
\end{equation}
In particular, if $\varphi$ has only simple zeroes and double poles,
\begin{equation}
\cN_{\rm zero} (g,n) \equiv \text{(Number of simple zeroes of $\varphi$)} = -4+4g+2n .
\label{simple-0 JS diff}
\end{equation}
For example, $\cN_{\rm zero} (0,2) = 0$ and $\cN_{\rm zero} (0,3) = 2$.

For a given JS differential $\varphi$, we define its critical graph $\Gamma$ as the collection of all horizontal trajectories. Each zero of $\varphi$ is a vertex of $\Gamma$, and each pole of $\varphi$ is a face of $\Gamma$. Suppose $w$ is a local coordinate around the zero or pole of $\varphi$. From
\begin{equation}
\int dw \sqrt{w} \sim w^{3/2} , \qquad
\int dw w^{-1} \sim \log w
\end{equation}
we find that a simple zero of $\varphi$ gives a trivalent vertex of $\Gamma$, while a face of $\Gamma$ can be surrounded by any number of edges.

The critical graph $\Gamma$ has the structure of metric ribbon graphs.
$\Gamma$ is a ribbon graph because the graph is drawn on an oriented Riemann surface, and $\Gamma$ is a metric graph because we measure the length of the edges by \eqref{def:length-area}.
The structure of $\Gamma$ as a metric ribbon graph is fixed by $\varphi$, which implies that a metric ribbon graph specifies a point of the moduli space $\cM_{g,n}$.
The metric ribbon graphs classify other structures as well, such as the (decorated) Teichm\"uller space \cite{Penner88} and the projective structure \cite{Fock:1993pr}.

\paragraph{Example.}

\begin{figure}[t]
\begin{center}
\includegraphics[scale=0.55]{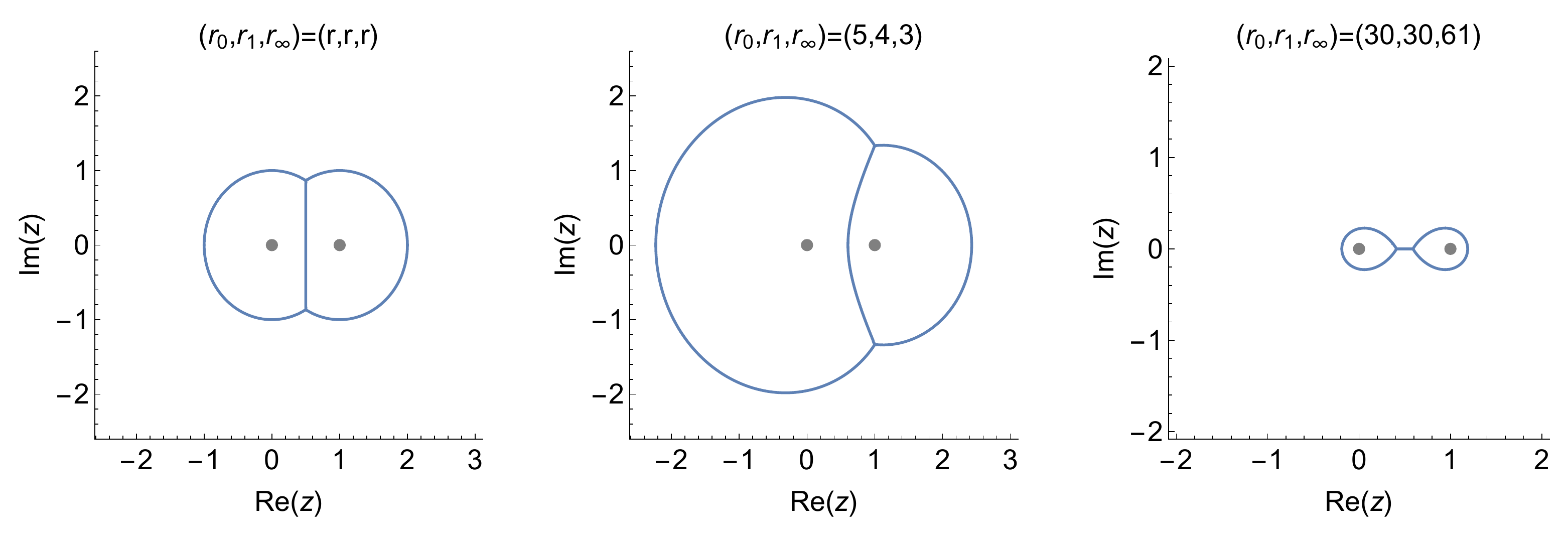}
\caption{Horizontal trajectories of the JS differential of $\Sigma_{0,3}$ in \eqref{example Strebel 3pt}. Three poles are at $z=0,1,\infty$, denoted by gray dots. Two zeroes are at $z=\frac12 \pm i \frac{\sqrt{3}}{2}$ for the left figure, $z=1 \pm i \frac43$ for the middle, and $z=\frac12 \pm \frac{11}{122}$ for the right.}
\label{fig:Strebel 3pt}
\end{center}
\end{figure}

The JS differential of $\Sigma_{0,3}$ takes the form
\begin{equation}
\phi (z) dz^2 = - \frac{1}{4 \pi^2} \pare{ \frac{r_0^2}{z^2} +\frac{r_1^2}{(z-1)^2}
+ \frac{r_0^2 + r_1^2 - r_{\infty}^2}{z(z-1)} }.
\label{example Strebel 3pt}
\end{equation}
Figure \ref{fig:Strebel 3pt} shows the horizontal trajectories for some $(r_0 \,, r_1 \,, r_\infty)$. 
When $(r_0 \,, r_1 \,, r_\infty)$ are not identical, the distance between two zeroes (trivalent vertices) depends on the path one chooses. In particular, there must be a branch cut somewhere on $\Sigma_{g,n}$\,.

Critical graphs may have self-edges as shown in Figure \ref{fig:Strebel 3pt}. If we denote the length of the edges between zeroes of $\phi$ on $\Sigma_{0,3}$ as $(\ell_1 \,, \ell_2 \,, \ell_3)$, then the self-edges show up if three edges do not satisfy the triangular inequality $\ell_i + \ell_j > \ell_k$ for any $(i,j,k)$.

Some examples of the JS differential of genus-one surfaces are given in \cite{MP98}. In general, we need to solve transcendental equations to find the explicit JS differential of given lengths \cite{Ashok:2006du}.

\subsection{Drawing bipartite ribbon graphs}\label{app:drawing bipartite}

According to \eqref{graph planarity counting}, any graph on $\Sigma_{g,n}$ (or ideal triangulation of $\Sigma_{g,n}$) has a fixed number of vertices, edges and faces when all vertices are trivalent. 
From such a graph, we derive a bipartite graph, by drawing a white circle for the face and a black dot for the vertex of the original graph. 
The white circles and black dots should correspond to the double poles and simple zeroes of the Jenkins-Strebel differential on $\Sigma_{g,n}$\,.

\begin{figure}[H]
\begin{center}
\includegraphics[scale=0.6]{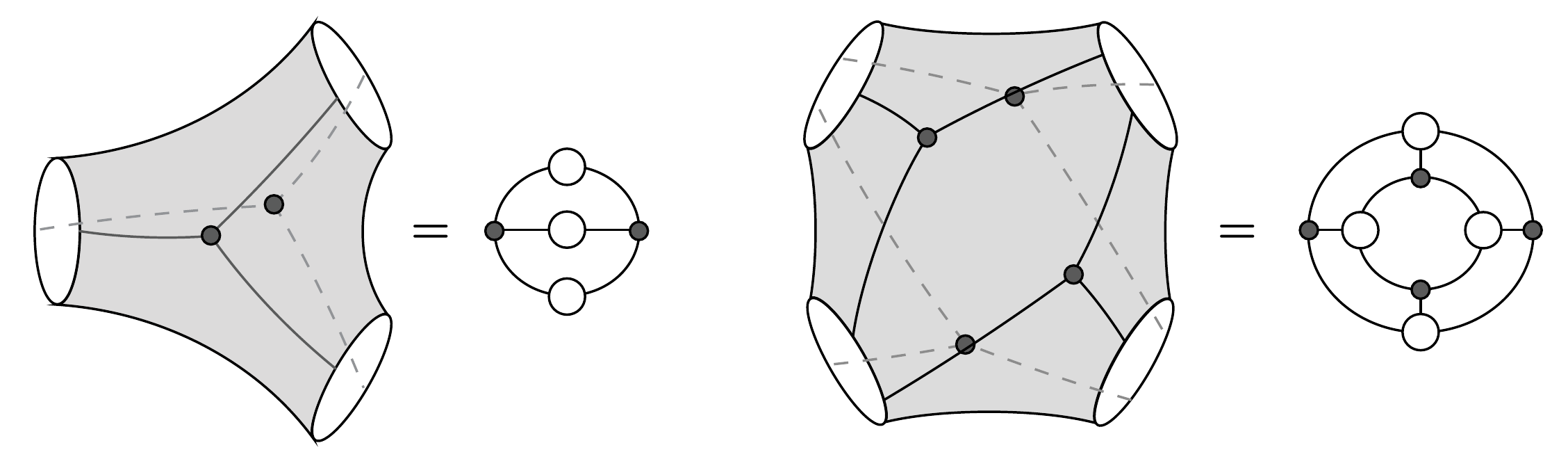}

\vspace{-4mm}
\caption{(Metric) ribbon graph from a triangulation of $\Sigma_{0,3}$ and $\Sigma_{0,4}$.}
\label{fig:Triangulate Sigma Planar2}

\vskip 5mm
\includegraphics[scale=0.6]{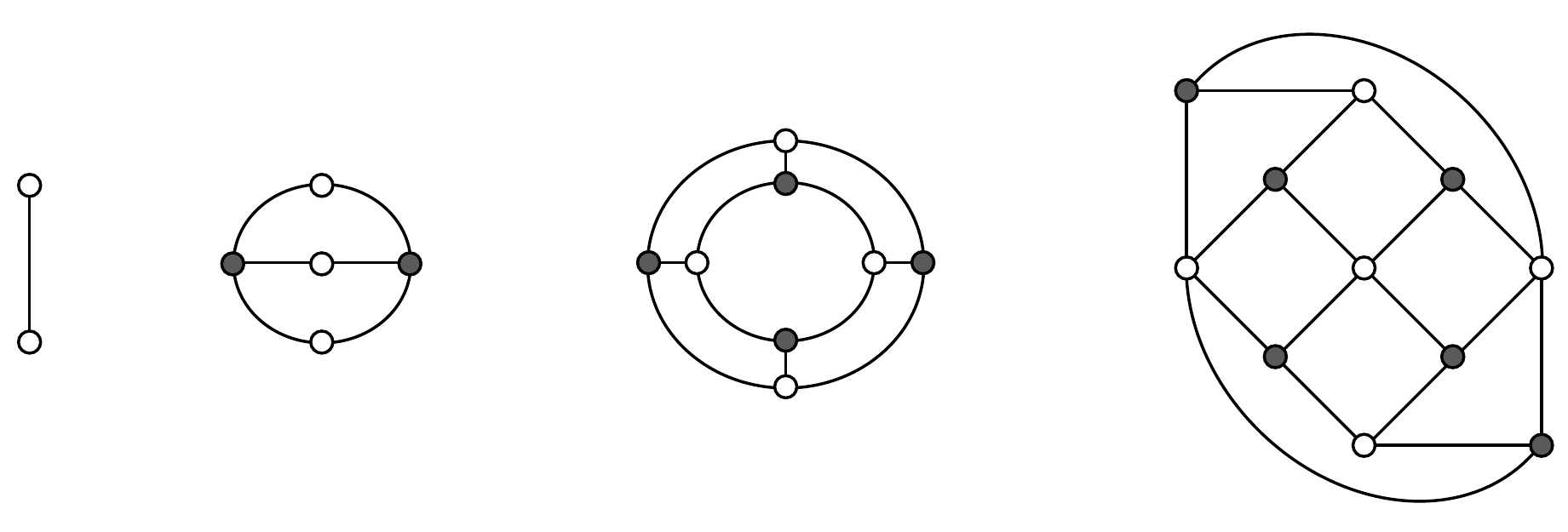}

\vspace{-4mm}
\caption{Examples of the planar bipartite graphs for $n=2,3,4,5$.}
\label{fig:Bipartite g=0}

\vskip 5mm
\includegraphics[scale=.34]{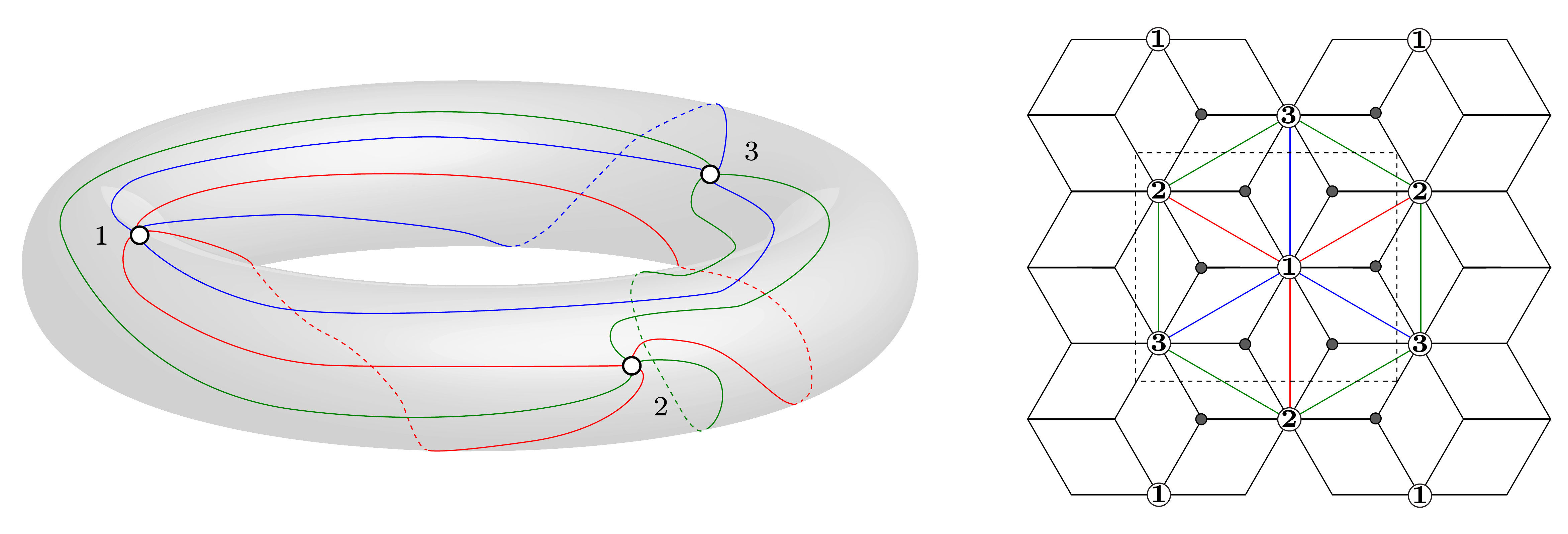}

\vspace{-4mm}
\caption{(Left) Triangulation of $\Sigma_{1,3}$\,. Red, green, blue edges represent $\ell_{12}^{(\rho)}, \ell_{23}^{(\rho)}, \ell_{31}^{(\rho)}$ with $\rho=1,2,3$, respectively.
(Right) The same geometry as a toric bipartite graph. The fundamental region is denoted by the dotted square.}
\label{fig:torus triangulate}

\vskip 5mm
\includegraphics[scale=.45]{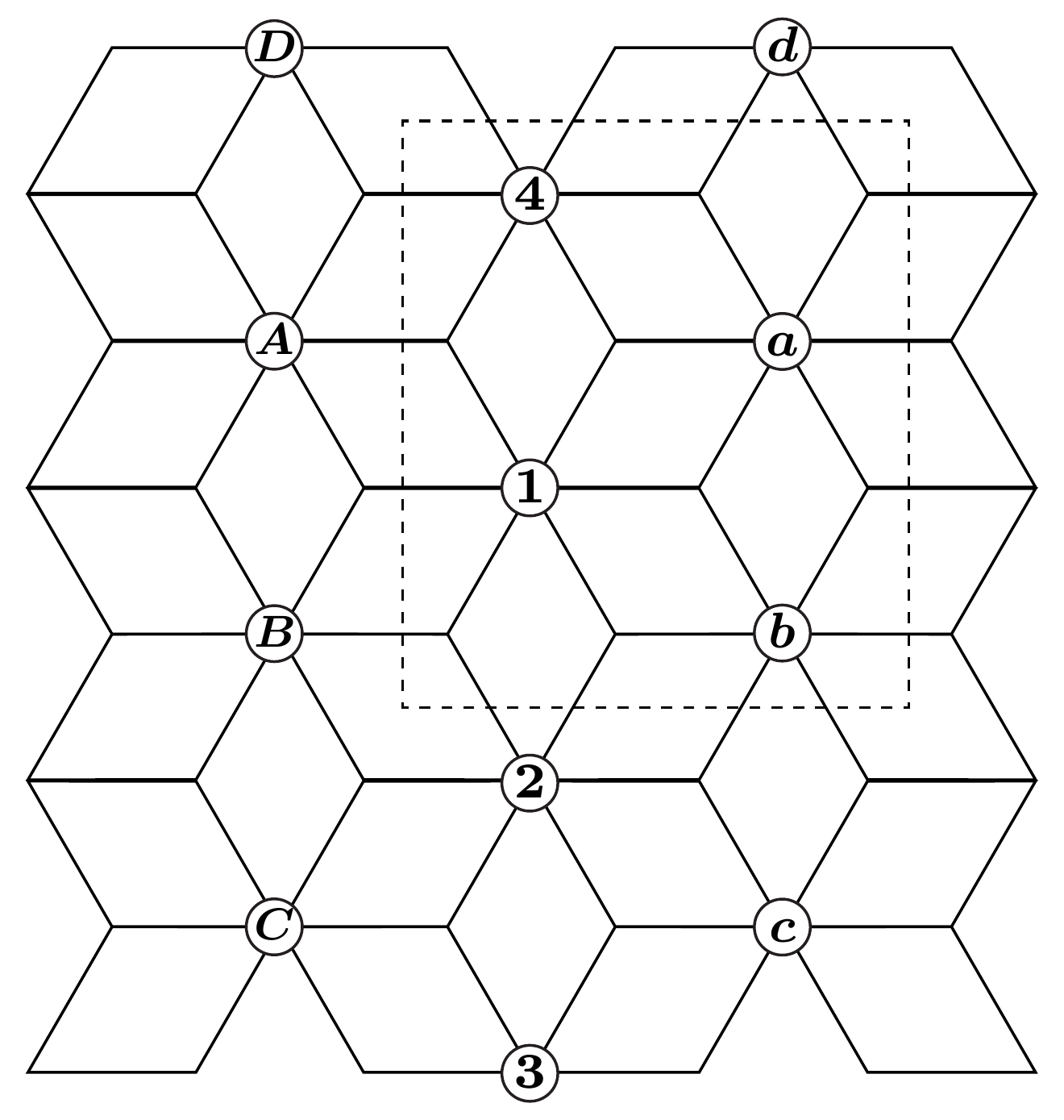}

\vspace{-3mm}
\caption{Example of the bipartite graph from $\Sigma_{1,4}$. The dotted square is the fundamental region.}
\label{fig:toric 4pt}
\end{center}
\end{figure}

Consider examples of the ribbon graph obtained by a triangulation of Riemann surface.
Planar bipartite graphs are shown in Figures \ref{fig:Triangulate Sigma Planar2}\,-\,\ref{fig:Bipartite g=0}, and non-planar graphs in Figures \ref{fig:torus triangulate}\,-\,\ref{fig:toric 4pt}.

Figure \ref{fig:torus triangulate} shows the bipartite graph from a triangulation of $\Sigma_{1,3}$.
This is a toric graph with periodic boundary conditions in both horizontal and vertical directions.
The fundamental region of the bipartite graph contains 6 black dots, 9 edges and 3 white circles, in agreement with \eqref{graph planarity counting}.
Other examples were given in Figure \ref{fig:Toric_S13_incomplete}.
These graphs have the shape of a honeycomb lattice. We can draw similar toric honeycomb lattices for general $n$, satisfying the constraints \eqref{graph planarity counting}.

Figure \ref{fig:toric 4pt} shows the bipartite graph from $\Sigma_{1,4}$\,.
The fundamental region contains 8 black dots, 12 edges and 4 white circles.
Let us label the punctures of this graph. From the periodicity on the torus, we must have
\begin{equation}
(a,b,c,d) = \Bigl\{ (2,3,4,1), \ (3,4,1,2), \ (4,1,2,3),\ (1,2,3,4) \Bigr\}.
\end{equation}
Only the first two choices are allowed to avoid self-contractions. When we fix $(a,b,c,d)$, then the periodicity determines $(A,B,C,D)$ uniquely.

Beware that the case of $(g,n)=(1,2)$ is singular. 
The equation \eqref{graph planarity counting} says that we should have four trivalent vertices.
However, we have only ``bivalent'' vertices because there are only two punctures.
One solution is to introduce fictitious punctures \cite{Eden:2017ozn}. Another solution is to use two quadrivalent vertices.

\clearpage

\subsection{Feynman graphs on a Riemann surface}\label{app:Feynman Riemann}

\begin{figure}[t]
\begin{center}
\includegraphics[scale=0.7]{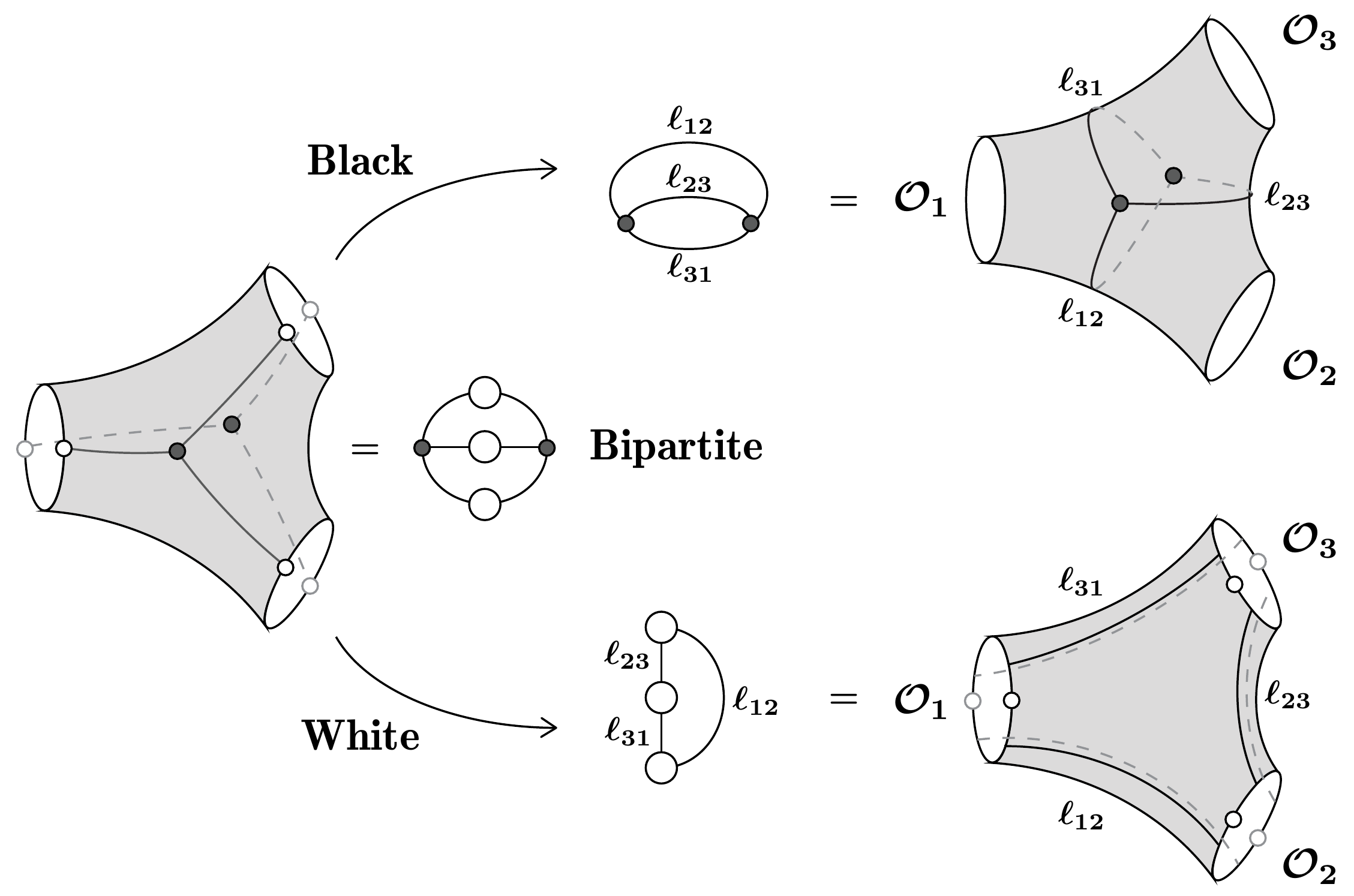}
\caption{Three types of the connected ribbon graph with $(g,n)=(0,3)$; bipartite, black and white. The black and white graphs are dual to each other.}
\label{fig:BipartiteBW}
\end{center}
\end{figure}

Given a bipartite graph in Appendix \ref{app:drawing bipartite}, we can draw two more ribbon graphs, called black and white.
The black graphs are defined by adding edges between the pair of black nodes for each square of the bipartite graph, and then removing all white nodes and the edges connected to them.
The white graphs are defined by the same procedure, interchanging black and white.
The black and white graphs are dual to each other, as shown in Figure \ref{fig:BipartiteBW}.

The black graph represents the complex structure of a Riemann surface.
The white graph represents the Feynman diagram appearing in the connected $n$-point function of single-trace operators.
Thus, the critical graph on a Riemann surface is the dual graph of a Feynman diagram of a gauge theory like $\cN=4$ SYM.
As a Feynman graph, the white circles represent external operators and black dots represent effective three-point interactions.\footnote{They are {\it not} three-point interactions in the $\cN=4$ SYM Lagrangian, because we talk of tree-level correlators.}
The equation \eqref{graph planarity counting} constrains the graph data of a white graph as
\begin{equation}
V^\circ_{g,n} = n,\qquad
E^\circ_{g,n} = 3n -6+6g, \qquad
F^\circ_{g,n} = 2n -4+4g.
\label{dual graph planarity counting}
\end{equation}

Recall that any Feynman diagrams in the double-line notation can be drawn on a Riemann surface.\footnote{In other words, the skeleton reduction is unimportant if we are interested in $\Sigma_{g,n}$ rather than $\cM_{g,n}$\,.}
Thus, the equation \eqref{dual graph planarity counting} constrains the cycle type of the vertex permutation $\omega$ of a general Feynman diagram.
For example, if the graph consists of strips and triangles only, we can write $[ \omega ] = [2^{w_2} \, 3^{w_3}] \vdash 2 \hL$ and impose the equations
\begin{equation}
2 w_2 +3 w_3 = 2 \hL, \qquad w_2+ w_3=C(\omega), \qquad
w_3 = F^\circ_{g,n} 
\end{equation}
where the last equation comes from \eqref{face degree g,n}.
These equations have a unique solution
\begin{equation}
w_2 = \hL - E^\circ_{g,n} = \hL - 3n + 6-6g, \qquad
C(\omega) = \hL- V^\circ_{g,n} +\chi = \hL - n + 2 - 2g.
\label{cycle-type formula gn}
\end{equation}
This value of $C(\omega)$ remains invariant even if we consider more general $[\omega]$ made by gluing triangles under the constraint \eqref{face degree g,n}.
Hence, the formula \eqref{cycle-type formula gn} suggests that higher genus corrections to any $n$-point functions must come in powers of $1/N_c^2$, rather than $1/N_c$\,.\footnote{The $1/N_c$ terms appear when the gauge group is $SO$ or $Sp$ \cite{Caputa:2010ep,Caputa:2013vla,Kemp:2014eca,Lewis-Brown:2018dje}. Also, if the gauge theory has the object in the fundamental representation such as quarks and Wilson loops, the Riemann surface has boundaries \cite{Maldacena:1998im}.}

\subsection{Counting the dimensions of $\cM_{g,n}^{\rm gauge}$}\label{app:count dimensions}

We count the dimensions of $\cM_{g,n}^{\rm gauge}$ by identifying its generators as the edges of the ribbon graph.

\begin{figure}[t]
\begin{center}
\includegraphics[scale=.9]{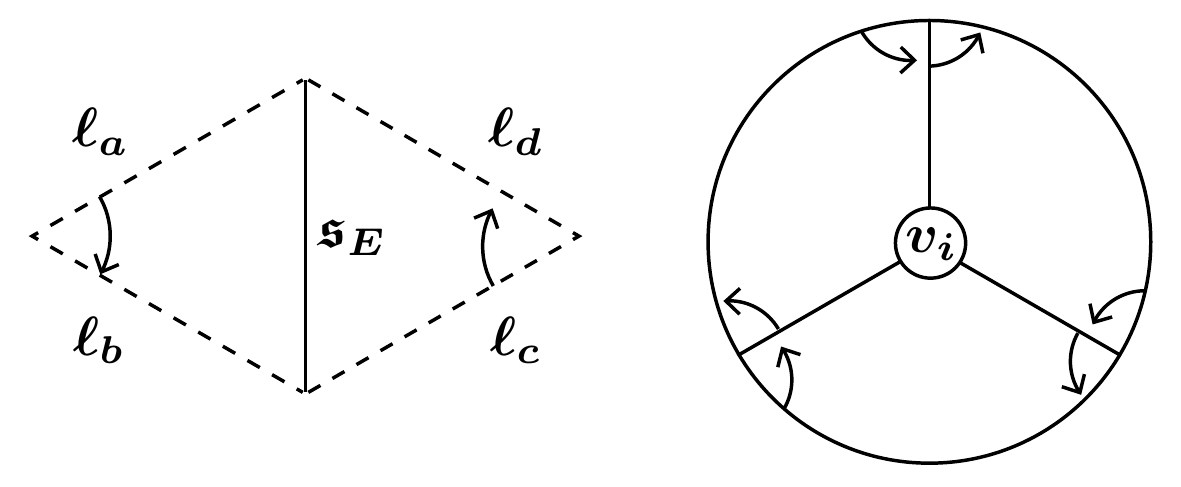}
\caption{(Left) Shift map $\frak{s}_E$ on the $E$-th edge given by \eqref{def:xi-E}. (Right) Simultaneous shift-maps on all edges connected to ${v_i}$ do not change the bridge lengths $\{ \ell_E\}$.}
\label{fig:Xi transform}
\end{center}
\end{figure}

Suppose that a ribbon graph completely triangulates $\Sigma_{g,n}$\,.
Then, any edge is the boundary of two triangles.
Let us make a square by gluing the two triangles, and study the four edges $\{ \ell_a \,, \ell_b \,, \ell_c \,, \ell_d \}$ of the square.
We define the shift map $\frak{s}_E$ on the $E$-th edge as
\begin{equation}
\frak{s}_E \ : \ \{ \ell_a \,, \ell_b \,, \ell_c \,, \ell_d \} \quad \mapsto \quad
\{ \ell_a - 1 \,, \ell_b +1 \,, \ell_c -1 \,, \ell_d +1 \}
\label{def:xi-E}
\end{equation}
which is also shown in Figure \ref{fig:Xi transform}.
If the bridge lengths $\{ \ell_E \}$ solve the constraint \eqref{lengths around an operator}, the shift map $\frak{s}_E$ generates all the other solutions of the constraint.

Denote the $\bb{Z}$\,-module of edge set by
\begin{equation}
\xi \equiv \bigoplus_{E=1}^{3n-6+6g} \, \xi_E \times \( \text{$E$-th edge} \) ,
\qquad (\xi_E \in \bb{Z}) .
\end{equation}
We define the simultaneous shift maps associated to $\xi$, by applying the shift maps $\xi_E$ times for the $E$-th edge,
\begin{equation}
\frak{s} (\xi) \equiv \bigoplus_{E=1}^{3n-6+6g} \xi_E \, \frak{s}_E \,.
\label{def:simultaneous shift maps}
\end{equation}
Let us count the number of the linearly-independent simultaneous shift maps.
Define $\Xi_{v_i}$ as the set of all edges connected to the $i$-th vertex. 
The simultaneous shift-map on this set is always trivial,
\begin{equation}
\frak{s} \( \Xi_{v_i} \) = {\rm identity}, \qquad
\Xi_{v_i} = \bigoplus_{\partial E \, \ni \, v_i} \, \( \text{$E$-th edge} \), \qquad
(i=1,2, \dots n).
\end{equation}
Since there are no more relations among $\frak{s}_E$\,, the number of linearly-independent simultaneous shift maps is equal to $2n-6+6g$. 
This is equal to $\dim_{\, \bb{R}} \cM_{g,n}$ in \eqref{dim_R Mgn}, which implies \eqref{def:CMRG-gauge dim}.


\section{Details of the skeleton reduction}\label{app:details skeleton}

We explain how to compute the skeleton reduction of $G_n$ in order to find $\cM_{g,n}^{\rm gauge} (\{L_i\})$. The attached {\tt Mathematica} files implement the procedures given below.

We need a few preliminary steps.
To begin with, we fix $\{ \cO_i \}$ and compute the connected $n$-point functions of single-trace operators in the standard way, e.g. \cite{PedroCode}. Let us write the result as a function acting on a set of Feynman graphs $\{ \Gamma \}$,
\begin{equation}
(G_n)_{\rm connected} = \sum_{\Gamma} F (\Gamma).
\end{equation}
Since $(G_n)_{\rm connected}$ is long and complicated, we study its cyclic decomposition,
\begin{equation}
\begin{gathered}
(G_n)_{\rm connected} = \sum_{z \in \bb{Z}_\alpha} z \cdot \scr{G}_n \,, \qquad
\scr{G}_n = \sum_{\Gamma_0} \frac{1}{| {\rm Aut}_{\,\bb{Z}_{\alpha}} (\Gamma_0) |} \ 
 F (\Gamma_0) 
\\
{\rm Aut}_{\,\bb{Z}_{\alpha}} (\Gamma_0) =
{\rm Stab}_{\,\bb{Z}_{\alpha}} (\Gamma_0) \equiv
\Bigl\{
\bsz \in \bb{Z}_\alpha \, \Big| \, \bsz \cdot \Gamma_0 = \Gamma_0
\Bigr\}.
\end{gathered}
\end{equation}
Here the action $\bsz \cdot \Gamma_0$ is given in \eqref{action of Zalpha on flavor}.
The division by $| {\rm Aut}_{\,\bb{Z}_{\alpha}} \Gamma_0 |$ is needed to avoid double counting.
For each term of $\scr{G}_n$, we can extract $W \in \bb{Z}_2^{\otimes \hL}$ in the notation of Section \ref{sec:S2L formalsim}. Then we change the labeling so that $W$ becomes $W' = \prod_{p=1}^{\hL} (p \, p')$.
Using the new labeling we compute $\omega = \alpha W$ for each set of Wick-contractions.

In order to apply the skeleton reduction, we need to determine the end-points of open two-point functions.
Once we know the open end-points, we can decompose $W$ into $(V, \bstau)$.
At the same time, we obtain $\bar\alpha$ after relabeling the skeleton-reduced Wick edges.
The set of $\nu = \bar\alpha V$ defines the gauge theory moduli space as in \eqref{npt reduced formula face}.

\bigskip
There are two methods to determine open end-points and perform our skeleton reduction.
The first method is intuitive but slow. 
The second method needs detailed case-studies, but efficient particularly when $\hL$ is small.

\begin{figure}[t]
\begin{center}
\includegraphics[scale=.72]{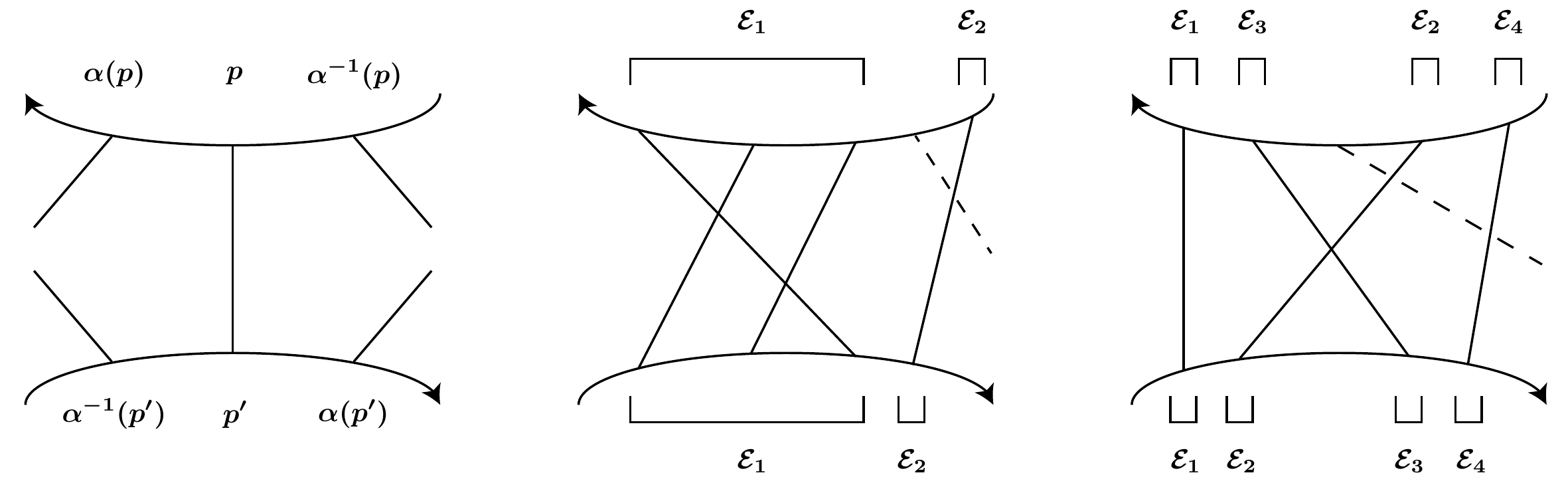}

\vspace{-3mm}
\caption{(Left) Four Wick-edges adjacent to the edge $(p,p')$ given in \eqref{def:4 adj edges}.
(Center) Two maximal Wick-edge groups $\cE_1, \cE_2$.
(Right) Four maximal Wick-edge groups $\cE_1 \sim \cE_4$.
}
\label{fig:edge groups}
\end{center}
\end{figure}

Choose the canonical ordering of $W = \prod (p \,, p')$ and introduce the notation for the Wick-edge,\footnote{The term {\it Wick-edge} is introduced below \eqref{n-pt S*2hL-1}.}
\begin{equation}
{\bf p} = (p , p') .
\end{equation}
We introduce adjacency between Wick-edges, by saying that the Wick-edge ${\bf p}$ is adjacent to the following four Wick-edges,
\begin{equation}
\Big(\alpha (p) , W\alpha (p)\Big) , \quad
\Big(\alpha^{-1} (p) , W\alpha^{-1} (p)\Big), \quad
\Big(W \alpha (p') , \alpha (p')\Big) , \quad
\Big(W \alpha^{-1} (p') , \alpha^{-1} (p')\Big) .
\label{def:4 adj edges}
\end{equation}
A pair of edges can be multiply adjacent.
Let us call a pair of sequences of consecutive fields between two operators a {\it Wick-edge group}.
A Wick-edge group of length $\ell$ consists of a pair of $\ell$ consecutive indices which are Wick-contracted within themselves,
\begin{multline}
\cE = 
\Bigl\{ 
\( p , \alpha(p), \alpha^2 (p), \dots , \alpha^{\ell-1} (p) \) ,
\( q , \alpha(q), \alpha^2 (q), \dots , \alpha^{\ell-1} (q) \) \, \Big|
\\
\text{any $\alpha^m (p)$ for $0 \le m < \ell$ is Wick-contracted with some $\alpha^{n} (q)$ for $0 \le n < \ell$}
\Bigr\}.
\label{def:edge group}
\end{multline}
We also use the notation $\cE = [ {\bf p}_1 \, {\bf p}_2 \, \dots \, {\bf p}_\ell ]$ using edge labels.
We look for the maximal Wick-edge group containing ${\bf p}$, assuming that ${\bf p}$ connects $\cO_{i_p}$ and $\cO_{j_p}$\,. 
Figure \ref{fig:edge groups} shows examples.

\paragraph{First method.}

Define the range of adjacency
\begin{equation}
\begin{aligned}
\cR_1 &\equiv 
\Bigl\{ [ - s, t] \, \Big| \,
\Big(W \alpha^n (p') , \alpha^n (p')\Big) \ \text{connects $\cO_{i_p}$ and $\cO_{j_p}$ for all} \ 
-s \le n \le t \Bigr\} 
\\[1mm]
\cR'_1 &\equiv
\Bigl\{ [ - s', t'] \, \Big| \, \Big(\alpha^n (p) , W\alpha^n (p)\Big)  \ \text{connects $\cO_{i_p}$ and $\cO_{j_p}$ for all} \ 
-s' \le n \le t' \Bigr\} .
\end{aligned}
\label{range of adjacency-1}
\end{equation}
We study the other side of Wick-contractions starting from $n \in \cR_1$.
Since both $p$ and $W \alpha^n (p')$ belong to $\cO_{i_p}$\,, we can define an integer $u_n$ such that
$W \alpha^n (p') = \alpha^{u_n} (p)$. Similarly, we can define $W\alpha^{n} (p) = \alpha^{u'_n} (p')$ for $n \in \cR'_1$\,. 
They are summarized as
\begin{equation}
\begin{aligned}
\Big(W \alpha^n (p') , \alpha^n (p')\Big) 
&\equiv \Big( \alpha^{u_n} (p) , \alpha^n (p')\Big) 
= \Big( \alpha^{u_n} (p) , W \alpha^{u_n} (p) \Big) , \qquad (n \in \cR_1)
\\[1mm]
\Big(\alpha^{n} (p) , W\alpha^{n} (p)\Big) 
&\equiv \Big(\alpha^{n} (p) , \alpha^{u'_n} (p') \Big) 
= \Big( W \alpha^{u'_n} (p'), \alpha^{u'_n} (p') \Big) , \qquad (n \in \cR'_1).
\end{aligned}
\end{equation}
We want to find the ranges $(n , n') \in (\cR_2 , \cR'_2)$ such that
\begin{equation}
\begin{aligned}
\cR_2 &\equiv \Bigl\{ \Big( \alpha^m (p) , W \alpha^m (p) \Big) \ \text{belongs to $\cR_1$ for all} \ \ 
0 \le m \le u_n \ \ 
( {\rm or} \ u_n \le m \le 0) \Bigr\}
\\
\cR'_2 &\equiv \Bigl\{ \Big( W \alpha^{m} (p'), \alpha^{m} (p') \Big) \ \text{belongs to $\cR'_1$ for all} \ \ 
0 \le m \le u'_{n'} \ \ 
({\rm or} \ u'_{n'} \le m \le 0) \Bigr\} .
\end{aligned}
\label{range of adjacency-2}
\end{equation}
Clearly $\cR_2 \subset \cR_1$ and $\cR'_2 \subset \cR'_1$\,. 
If $\cR_2 \,,\cR'_2$ are the two maximal Wick-edge groups \eqref{def:edge group}, the two sets should be identical.

Conversely said, if either of the conditions \eqref{range of adjacency-1}, \eqref{range of adjacency-2} is violated, then we should increase the $\rho$ of $\ell_{ij}^{(\rho)}$ and regard them as different Wick-edges of the skeleton graph.

\paragraph{Second method.}

\begin{figure}[H]
\begin{center}
\includegraphics[scale=.8]{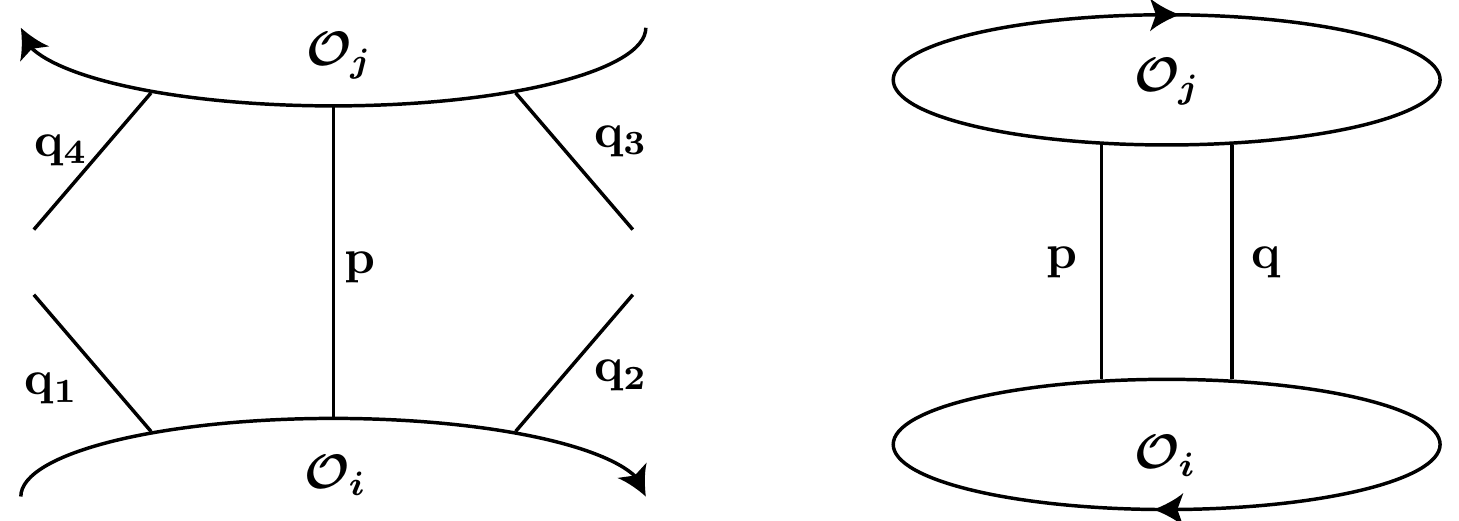}

\vspace{-2mm}
\caption{(Left) $J_{\bf pr} = \sum_{k=1}^4 \delta (\bf{r}, \bf{q}_k)$. Here $\bf{p}$ connects $\cO_i$ and $\cO_j$\,. (Right) Example of $J_{\bf pq} = 4$.
}
\label{fig:edge-adjacency}

\vskip 10mm
\includegraphics[scale=.8]{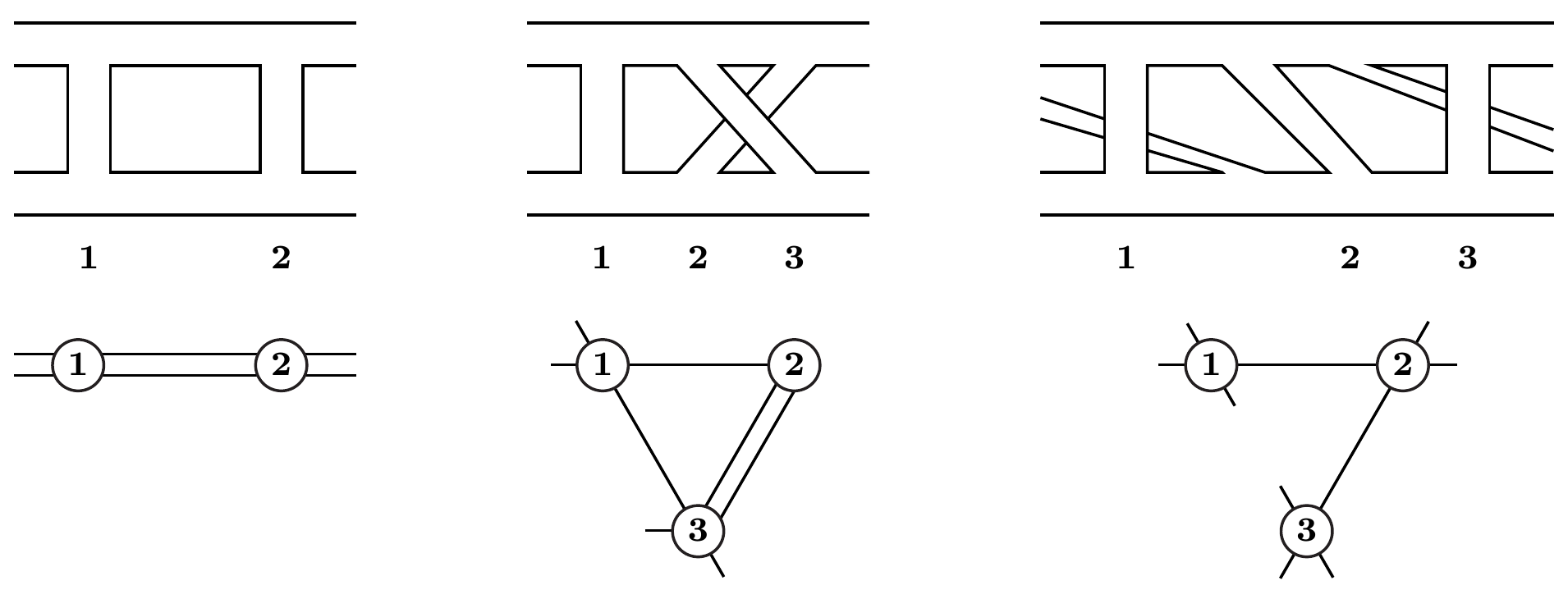}

\vspace{-2mm}
\caption{Open two-points (up) and edge-adjacency graphs (down), respectively corresponding to \eqref{example e-adj matrix}. In the edge-adjacency graphs, \textcircled{$p$} represents a Wick-edge. The edge (or edges) between \textcircled{$p$} and \textcircled{$q$} means they are (multiply) adjacent.}
\label{fig:edge adjacency}

\vskip 10mm
\includegraphics[scale=.8]{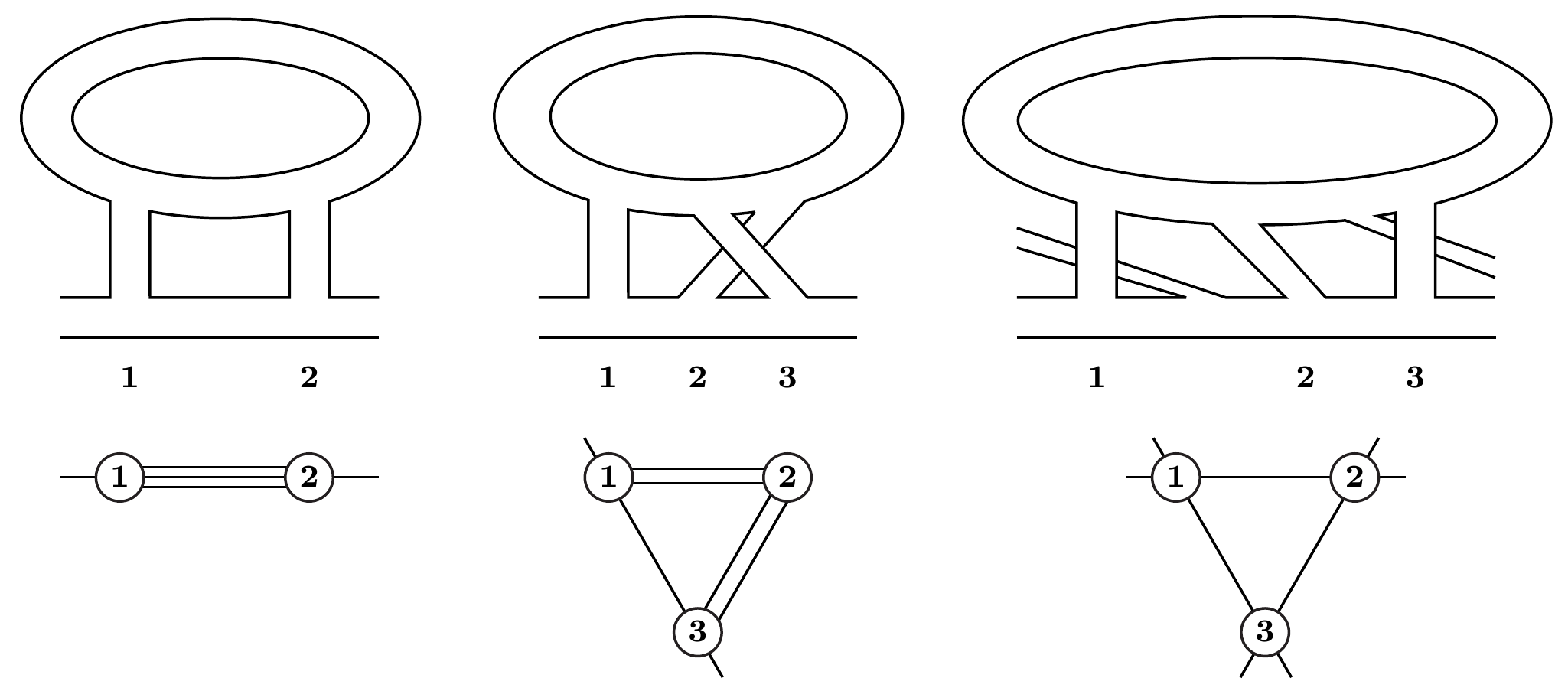}

\vspace{-3mm}
\caption{Open two-points and edge-adjacency matrices represented as adjacency graphs for periodic cases.}
\label{fig:edge adjacency ext}
\end{center}
\end{figure}

Define the edge-adjacency matrix by
\begin{equation}
\begin{gathered}
J \equiv \{ J_{\bf pq} \}_{1 \le {\bf p}, {\bf q} \le \bL} \,,
\\[2mm]
J_{\bf pq} = (\text{Adjacent multiplicity between the Wick-edges ${\bf p}$ and ${\bf q}$}), \qquad
0 \le J_{\bf pq} \le 4.
\end{gathered}
\end{equation}
If $J_{\bf pq}=0$, ${\bf p}$ and ${\bf q}$ are not adjacent. Some examples of $J_{\bf pq}$ are shown in Figure \ref{fig:edge-adjacency}.
Let us denote a pair of operators connected by the edge ${\bf p}$ as $\cO_{i_p}$ and $\cO_{j_p}$.
We have the block decomposition of the matrix $J$ as
\begin{equation}
J = \sum_{i < j} \sum_{k < l} J^{(i j) \, (kl)} \,.
\end{equation}
We call the elements with $(ij) = (kl)$ diagonal blocks, and the rest off-diagonal blocks. 
The diagonal part $J^{(i j) \, (i j)}$ knows how the adjacency inside the Wick-contractions between $\cO_i$ and $\cO_j$\,, somewhat in a cryptic way.

From adjacency relations, we draw the associated adjacency graph.\footnote{Do not confuse the adjacency graph with the Feynman graphs.}
If $J^{(i j) \, (i j)}$ is block-diagonal with more than one blocks, then the associated adjacency graph has more than one connected components. 
Consider the examples. 
Assuming that the left and right ends of the open two-points are not periodically identified as in Figure \ref{fig:edge adjacency}, the following edge-adjacency matrices corresponds to the Wick-edge groups $\pare{ [12] , [123], [1][2][3] }$,
\begin{equation}
J^{(ij) (ij)} \ \sim \ \Biggl\{
\begin{pmatrix}
0 & 2 \\
2 & 0
\end{pmatrix}, \quad
\begin{pmatrix}
0 & 1 & 1 \\
1 & 0 & 2 \\
1 & 2 & 0
\end{pmatrix}, \quad
\begin{pmatrix}
0 & 1 & 0 \\
1 & 0 & 1 \\
0 & 1 & 0
\end{pmatrix}
\Biggr\}.
\label{example e-adj matrix}
\end{equation}
In contrast, if we identify the left and right ends of the above operator as in Figure \ref{fig:edge adjacency ext}, then the following edge-adjacency matrices corresponds to the Wick-edge groups $\pare{ [12] , [123], [1][2][3] }$,
\begin{equation}
J^{(ij) (ij)} \ \sim \ \Biggl\{
\begin{pmatrix}
0 & 3 \\
3 & 0
\end{pmatrix}, \quad
\begin{pmatrix}
0 & 2 & 1 \\
2 & 0 & 2 \\
1 & 2 & 0
\end{pmatrix}, \quad
\begin{pmatrix}
0 & 1 & 1 \\
1 & 0 & 1 \\
1 & 1 & 0
\end{pmatrix}
\Biggr\}.
\label{example e-adj matrix ext}
\end{equation}

\clearpage
Let us call a length-$\ell$ open two-point {\it periodic} if $\ell = {\rm min} (L_i \,, L_j)$.
In the periodic cases, we need to identify both ends of either $\cO_i$ or $\cO_j$ as in Figure \ref{fig:edge adjacency ext}.

Notice that two edges go outside the open two-point in the rightmost graphs of Figures \ref{fig:edge adjacency}\,-\,\ref{fig:edge adjacency ext}. 
We call them external Wick-edges, because they do not connect $\cO_i$ and $\cO_j$ while sitting inside the open two-point between $\cO_i$ and $\cO_j$\,.
In non-periodic cases, there are at most four external adjacent Wick-edges.
In periodic cases, this number decreases by two (or more).

We can count the number of external Wick-edges adjacent to a given single Wick-edge group, based on $J^{(ij) (ij)}_{{\bf pq}}$ and the periodicity.
Suppose there is a single Wick-edge group. We divide this group by inserting an external Wick-edge.
This reduces the adjacent multiplicity of two edges which belonged to the single Wick-edge group.
Therefore, the sum of $J^{(ij) (ij)}_{{\bf pq}}$ over all components ${\bf p}, {\bf q}$ should satisfy
\begin{equation}
\sum_{{\bf p}, {\bf q}=1}^{\ell} J^{(ij) (ij)}_{{\bf pq}} = 4 \ell - 4
+ 2 \( \text{\# of periodic operators} \)
- 2 \( \text{\# of external Wick-edges} \) .
\label{single edge group condition}
\end{equation}
As a corollary, if the adjacency matrix satisfies $\sum_{{\bf p}, {\bf q}=1}^{\ell} J^{(ij) (ij)}_{{\bf pq}} \ge 4 \ell - 4 + 2 \( \text{\# of periodic operators} \)$, then no external Wick-edges are inserted, and all Wick-edges must belong to the single edge-group $\cE$.
If this inequality is violated, we need detailed case-studies in order to see how edges are grouped.
Some examples are shown in Tables \ref{tab:edge groups}\,-\,\ref{tab:edge groups ext}.
\footnote{There are two underlying assumptions in writing this table. First, we consider only those matrices which correspond to a set (or sets) of physically possible Wick-contractions $W$. Second, if several sets of Wick-contractions have the same $J$ and the same periodicity, then they have the same Wick-edge group $\cE$. 
If these assumptions are correct (which seems to be the case), an edge-adjacency matrix $J$ gives the unique corresponding Wick-edge group.}

\begin{table}
\begin{tabular}{c|ccccccc}
\hline \\[-4mm]
$J$ & 
$\begin{pmatrix} 0 \end{pmatrix}$ &
$\begin{pmatrix} 0 & 1 \\ 1 & 0 \end{pmatrix}$ & 
$\begin{pmatrix} 0 & 2 \\ 2 & 0 \end{pmatrix}$ & 
$\begin{pmatrix} 0 & 1 & 0 \\ 1 & 0 & 1 \\ 0 & 1 & 0 \end{pmatrix}$ & 
$\begin{pmatrix} 0 & 1 & 1 \\ 1 & 0 & 1 \\ 1 & 1 & 0 \end{pmatrix}$ & 
$\begin{pmatrix} 0 & 1 & 0 \\ 1 & 0 & 2 \\ 0 & 2 & 0 \end{pmatrix}$ &
$\begin{pmatrix} 0 & 0 & 1 \\ 0 & 0 & 2 \\ 1 & 2 & 0 \end{pmatrix}$ 
\\[11mm]
Edge groups & 
[1] & [1][2] & [12] & [1][2][3] & [1][2][3] & [1][23] & [1][23] 
\end{tabular}

\medskip
\begin{tabular}{c|ccccc}
\hline \\[-4mm]
$\quad$ & 
$\begin{pmatrix} 0 & 1 & 0 & 0 \\ 1 & 0 & 1 & 0 \\ 0 & 1 & 0 & 1 \\ 0 & 0 & 1 & 0 \end{pmatrix}$ &
$\begin{pmatrix} 0 & 1 & 1 & 1 \\ 1 & 0 & 0 & 1 \\ 1 & 0 & 0 & 1 \\ 1 & 1 & 1 & 0 \end{pmatrix}$ &
$\begin{pmatrix} 0 & 1 & 1 & 0 \\ 1 & 0 & 1 & 0 \\ 1 & 1 & 0 & 2 \\ 0 & 0 & 2 & 0 \end{pmatrix}$ &
$\begin{pmatrix} 0 & 2 & 0 & 0 \\ 2 & 0 & 1 & 0 \\ 0 & 1 & 0 & 2 \\ 0 & 0 & 2 & 0 \end{pmatrix}$ &
$\begin{pmatrix} 0 & 1 & 0 & 1 & 1 \\ 1 & 0 & 1 & 1 & 0 \\ 0 & 1 & 0 & 1 & 0 \\
1 & 1 & 1 & 0 & 1 \\ 1 & 0 & 0 & 1 & 0 \end{pmatrix}$ 
\\[13mm]
$\quad$ & 
[1][2][3][4] & [1][2][3][4] & [1][2][34] & [12][34] & [1][2][3][4][5]
\end{tabular}
\caption{Edge-adjacency matrices and the corresponding Wick-edge groups for non-periodic cases.}
\label{tab:edge groups}

\vskip 7mm
\begin{tabular}{c|cccccc}
\hline \\[-4mm]
$J$ & 
$\begin{pmatrix} 0 & 2 \\ 2 & 0 \end{pmatrix}$ & 
$\begin{pmatrix} 0 & 3 \\ 3 & 0 \end{pmatrix}$ & 
$\begin{pmatrix} 0 & 1 & 1 \\ 1 & 0 & 1 \\ 1 & 1 & 0 \end{pmatrix}$ & 
$\begin{pmatrix} 0 & 1 & 1 \\ 1 & 0 & 2 \\ 1 & 2 & 0 \end{pmatrix}$ &
$\begin{pmatrix} 0 & 2 & 0 & 1 \\ 2 & 0 & 1 & 0 \\ 0 & 1 & 0 & 2 \\ 1 & 0 & 2 & 0 \end{pmatrix}$ &
$\begin{pmatrix} 0 & 1 & 1 & 1 \\ 1 & 0 & 1 & 1 \\ 1 & 1 & 0 & 1 \\ 1 & 1 & 1 & 0 \end{pmatrix}$ 
\\[11mm]
Edge groups & 
[1][2] & [12] & [1][2][3] & [1][23] & [12][34] & [1][2][3][4]
\end{tabular}
\caption{Edge-adjacency matrices and the corresponding Wick-edge groups for periodic cases.}
\label{tab:edge groups ext}
\end{table}

\clearpage

\bibliographystyle{utphys}
\bibliography{bibmix}{}

\end{document}